\documentclass[aps,prx,reprint,
superscriptaddress,floatfix,longbibliography,footinbib]{revtex4-1}
\usepackage{mathtools}
\usepackage{amsfonts}
\usepackage{amsmath}
\usepackage{physics}
\usepackage{float}
\usepackage{bbold}
\usepackage[version=4]{mhchem}
\usepackage{hyperref}
\hypersetup{
    colorlinks=true,
    linkcolor=blue,
    filecolor=blue, 
    citecolor=blue,
    urlcolor=blue,
}
\usepackage[doipre={DOI:~}]{uri}

\newcommand{\transfield}{\omega_z}
\newcommand{\spinfield}{h}
\newcommand{\localfield}{\ell}
\newcommand{\longfield}{\varphi}
\newcommand{\ham}{H_{\mathrm{Hopfield}}}
\newcommand{\dE}{\delta\epsilon}
\newcommand{\dEi}{\delta\epsilon_i}
\newcommand{\detune}{\Delta_C}

\newcommand{\spinswidth}{ w}
\newcommand{\noise}{\chi}
\newcommand{\waist}{w_0}
\newcommand{\cutoff}{\omega_c}
\newcommand{\geo}{\beta}
\newcommand{\patt}{\mathbf{\Xi}}
\newcommand{\cavmin}{\tilde{\mathbf{\Phi}}}
\newcommand{\enc}{\mathcal{M}}
\newcommand{\encmat}{\mathbf{M}}
\newcommand{\dec}{\mathcal{M}^{-1}}
\newcommand{\decmat}{\mathbf{M}^{-1}}
\newcommand{\prob}{p}

\begin{document}

\title{Enhancing associative memory recall and storage capacity using confocal cavity QED}

\author{Brendan P.~Marsh}
\affiliation{Department of Applied Physics, Stanford University, Stanford CA 94305, USA}
\affiliation{E.~L.~Ginzton Laboratory, Stanford University, Stanford, CA 94305, USA}
\author{Yudan Guo}
\affiliation{E.~L.~Ginzton Laboratory, Stanford University, Stanford, CA 94305, USA}
\affiliation{Department of Physics, Stanford University, Stanford CA 94305, USA}
\author{Ronen M.~Kroeze}
\affiliation{E.~L.~Ginzton Laboratory, Stanford University, Stanford, CA 94305, USA}
\affiliation{Department of Physics, Stanford University, Stanford CA 94305, USA}
\author{Sarang Gopalakrishnan}
\affiliation{Department of Engineering Science and Physics, CUNY College of Staten Island, Staten Island, NY 10314, USA}
\author{Surya Ganguli}
\affiliation{Department of Applied Physics, Stanford University, Stanford CA 94305, USA}
\author{\\Jonathan Keeling} 
\affiliation{SUPA, School of Physics and Astronomy, University of St. Andrews, St. Andrews KY16 9SS, United Kingdom}
\author{Benjamin L. Lev}
\affiliation{Department of Applied Physics, Stanford University, Stanford CA 94305, USA}
\affiliation{E.~L.~Ginzton Laboratory, Stanford University, Stanford, CA 94305, USA}
\affiliation{Department of Physics, Stanford University, Stanford CA 94305, USA}

\date{\today}

\begin{abstract}
We introduce a near-term experimental platform for realizing an associative memory.  It can simultaneously store many memories by using spinful bosons coupled to a degenerate multimode optical cavity. The associative memory is realized by a confocal cavity QED neural network, with the cavity modes serving as the synapses, connecting a network of superradiant atomic spin ensembles, which serve as the neurons.  Memories are encoded in the connectivity matrix between the spins, and can be accessed through the input and output of patterns of light. Each aspect of the scheme is based on recently demonstrated technology using a confocal cavity and Bose-condensed atoms. Our scheme has two conceptually novel elements. First, it introduces a new form of random spin system that interpolates between a ferromagnetic and a spin-glass regime as a physical parameter is tuned---the positions of ensembles within the cavity. Second, and more importantly, the spins relax via deterministic steepest-descent dynamics, rather than Glauber dynamics. We show that this nonequilibrium quantum-optical scheme has significant advantages for associative memory over Glauber dynamics:   These dynamics can enhance the network's ability to store and recall memories beyond that of the standard Hopfield model. Surprisingly, the cavity QED dynamics can retrieve memories even when the system is in the spin glass phase. Thus, the experimental platform provides a novel physical instantiation of associative memories and spin glasses as well as provides an unusual form of relaxational dynamics that is conducive to memory recall even in regimes where it was thought to be impossible.

\end{abstract}

\maketitle

\section{Introduction}

Five hundred million years of vertebrate brain evolution have produced biological information-processing architectures so powerful that simply emulating them, in the form of artificial neural networks, has lead to breakthroughs in classical computing~\cite{LeCun2015-pp,Bahri2020-mi}. Indeed, neuromorphic computation currently achieves state-of-the-art performance in image and speech recognition, machine translation, and even out-performs the best humans in ancient games like Go~\cite{Silver:2016hl}. Meanwhile, a revolution in our ability to control and harness the quantum world is promising technological breakthroughs for quantum information processing~\cite{nielsen2002quantum,arute2019quantum} and sensing~\cite{Degen2017}.  Thus, combining the algorithmic principles of robust parallel neural computation, discovered by biological evolution, with the nontrivial quantum dynamics of interacting light and matter naturally offered to us by the physical world, may open up a new design space of quantum-optics-based neural networks. Such networks could potentially achieve computational feats beyond anything biological or silicon-based machines could do alone. 

We present an initial step along this path by theoretically showing how a network composed of atomic spins coupled by photons in a multimode cavity can naturally realize associative memory, which is a prototypical function of neural networks.   Moreover, we find that including the effects of drive and dissipation, the naturally arising nonequilibrium dynamics of the cavity QED system enhances its ability to store and recall multiple memory patterns, even in a spin glass phase.  

Despite the biologically inspired name, artificial neural networks can refer to any network of nonlinear elements (e.g., spins) whose state depends on signals (e.g., magnetic fields) received from other elements~\cite{stein2013spin,hertz1991introduction,FisherHertz}.  They provide a distributed computational architecture alternative to the sequential gate-based von Neumann model of computing widely used in everyday devices~\cite{Sompolinsky:1988ez} and employed in traditional quantum computing schemes~\cite{nielsen2002quantum}. 
Rather than being programmed as a sequence of operations, the neural network connectivity encodes the problem to be solved as a cost function, and the solution corresponds to the final steady-state configuration obtained by minimizing this cost function through a nonlinear dynamical evolution of the individual elements. Specifically, the random, frustrated Ising spin glass is an archetypal mathematical setting for exploring neural networks~\cite{stein2013spin}.  Finding the ground state of an Ising spin glass is known to be an NP-hard problem~\cite{Barahona:1982gj,Lucas:2014eb} and so different choices of the spin connectivity may therefore encode many different combinatorial optimization problems of broad technological relevance~\cite{moore2011nature,Mehta:2018tt}. Much of the excitement in modern technological and scientific computing revolves around developing faster, more efficient `heuristic' optimization solvers that provide `good-enough' solutions.  Physical systems capable of realizing an Ising spin glass may play such a role. In this spirit, we present a thorough theoretical investigation of a quantum-optics-based heuristic neural-network optimization solver in the context of the associative memory problem. 

Using notions from statistical mechanics, Hopfield showed how a spin glass of randomly connected Ising spins can be capable of associative memory, a prototypical brain-like neural network function~\cite{Hopfield:1986bw}.   Associative memory is able to store multiple patterns (memories) as local minima of an energy landscape.  Moreover, recall of any individual memory is possible even if mistakes are made when addressing the memory to be recalled:  if the network is initialized by an external stimulus in a state that is not too different from the stored memory, then the network dynamics will flow towards an internal representation of the original stored memory, using an energy minimizing process called pattern completion that corrects errors in the initial state. Such networks exhibit a trade-off between capacity (number of memories stored) and robustness (the size of the basins of attraction of each memory under pattern completion).  Once too many memories are stored, the basins of attraction cease to be extensive, and the model transitions to a spin-glass regime with exponentially many spurious memories (with subextensive basins of attraction) that are nowhere near the desired memories~\cite{Amit1985}.   

From a hardware perspective, most modern neural networks are implemented in CMOS devices based on electronic von-Neumann architectures. In contrast, early work aiming to use classical, optics-based spin representations sought to take advantage of the natural parallelism of light propagation~\cite{Psaltis:1990dw,Anderson:1994fu,Psaltis:1998ka}; such work continues in the context of silicon photonic integrated circuits and other coupled classical oscillator systems~\cite{Shen:2017hb,Csaba:2020eh}.  The use of atomic spins coupled via light promises additional advantages: atom-photon interactions can be strong even on the single-photon level~\cite{Kimble1998}, providing the ability to process small signals and exploit manifestly quantum effects and dynamics.

\begin{figure}
    \centering
    \includegraphics[width=\columnwidth]{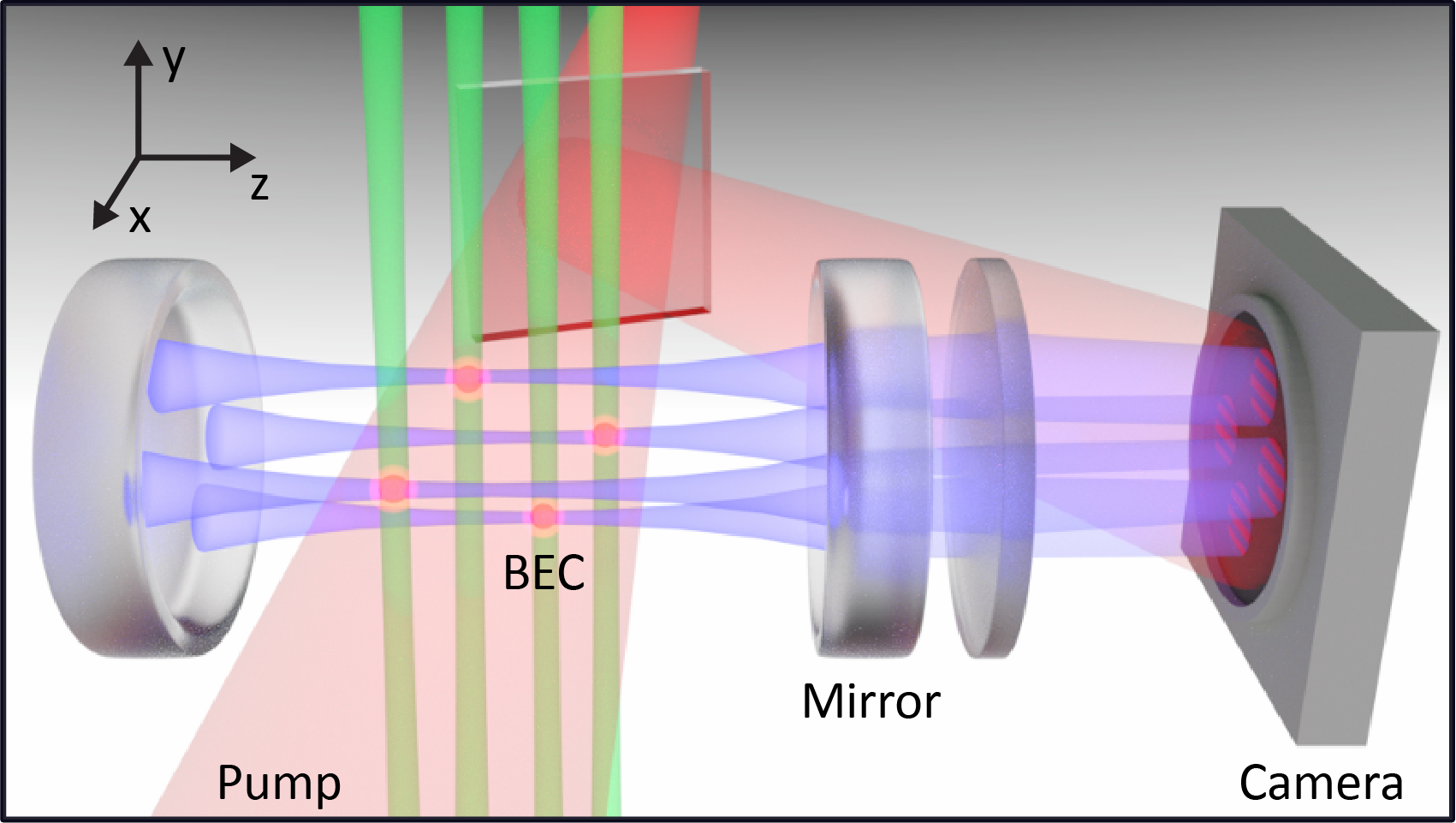}
    \caption{Sketch of the confocal cavity QED system with atomic spin ensembles confined by optical tweezers.  Shown are the cavity mirrors (gray), superpositions of the cavity modes (blue), BECs (red), optical tweezer beams (green), transverse pump beam (red), and interference on a CCD camera for holographic imaging of spin states.}
    \label{fig:cQEDblender}
\end{figure}

Previous theoretical work has sketched how a spin glass and neural network may be realized using ultracold atoms strongly coupled via photons confined within a multimode optical cavity~\cite{Gopalakrishnan:2011jx,Gopalakrishnan:2012cf}. The cavity modes serve as the synaptic connections of the network, mediating Ising couplings between atomic spins. The arrangement of atoms within the cavity determines the specific connection strengths of the network, which in turn determine the stored patterns. The atoms may be reproducibly trapped in a given connectivity configuration using optical tweezer arrays~\cite{Endres:2016fk,deMello:2019jw}.  Subsequent studies have provided additional theory support to the notion that related quantum-optical systems can implement neural networks~\cite{PietroRotondo:2015iq,Rotondo:2018,Torggler:2017hw,Fiorelli:2020kd}. However, all these works, including Refs.~\cite{Gopalakrishnan:2011jx,Gopalakrishnan:2012cf}, left significant aspects of implementation and capability unexplored.  

In the present theoretical study, we introduce the first practicable scheme for a quantum-optical associative memory by explicitly treating photonic dissipation and ensuring that the physical system does indeed behave similarly to a Hopfield neural network. All physical resources invoked in this treatment have been demonstrated in recent experiments~\cite{Kollar:2014us,Kollar2017sm,Vaidya2018,kroeze2018spinor,Guo2019Sign,Guo2019Emergent,Kroeze:2019ex}. Specifically, we show that suitable network connectivity is provided by optical photons in  the confocal cavity, a type of degenerate multimode resonator~\cite{siegman}. The photons are scattered into the cavity from atoms pumped by a coherent field oriented transverse to the cavity axis,  as illustrated in Fig.~\ref{fig:cQEDblender}.  The atoms undergo a transition to a superradiant state above a threshold in pump strength. In this state, the spins effectively realize a system of rigid (easy-axis) Ising spins with rapid spin evolution, ensuring that memory retrieval can take place before heating by spontaneous emission can play a detrimental role.  We moreover find the cavity QED system naturally leads to a discrete analog of ``steepest descent'' dynamics, which we show provides an enhanced memory capacity compared to standard Glauber dynamics~\cite{glauber63}.  Finally, 
the spin configuration can be read-out by holographic imaging of the emitted cavity light, as recently demonstrated~\cite{Guo2019Sign}. That is, the degenerate cavity provides high-numerical-aperture spatial resolving capability, and may be construed as an \textit{active} quantum gas microscope, in analogy to apparatuses employing high-NA lenses near optically trapped atoms~\cite{bakr2009quantum}.  

Our main results are as follows: 
\begin{enumerate}

    \item Superradiant scattering of photons by ensembles of atomic spins plus cavity dissipation naturally realizes a form of zero-temperature dynamics in a physical setting: discrete steepest descent (SD) dynamics. This is because the bath structure dictates that large energy-lowering spin flips occur most rapidly.  This is distinct from the typical zero temperature limit of Glauber~\cite{glauber63} or Zero Temperature Metropolis-Hastings~\cite{Metropolis53,Hastings70} (0TMH) dynamics typically considered in Hopfield neural networks~\cite{hopfield1982neural}.  
    
    \item  The confocal cavity can naturally provide a dense (all-to-all) spin-spin connectivity that is tunable between (1) a ferromagnetic regime, (2) a regime with many large basins of  attraction suitable for associative memory, and (3) a regime in which the connectivity describes a Sherrington--Kirkpatrick (SK) spin glass. This sequence of regimes is characteristic of Hopfield model behavior.
    
    \item  Surprisingly, standard limits on memory capacity and robustness are exceeded under SD dynamics.  This enhancement  is because SD  dynamics enlarge the basins of attraction---i.e., 0TMH can lead to errant fixed points in regimes where SD always leads to the correct one. This is true not just of the cavity QED system, but also for the basic Hopfield model with Hebbian and other learning rules. Moreover, the enhancement persists into the SK spin glass regime, wherein basins of metastable states expand from zero size under 0TMH dynamics to an extensively scaling size under SD.
   
    \item While simulating the SD dynamics requires $\mathcal{O}(N^2)$ numerical operations to determine the optimal energy-lowering spin flip, the physical cavity QED system naturally flips the optimal spin due to the different rates experienced by different spins.  Thus, the real dynamics drives the spins to converge to a fixed point configuration (memory) more efficiently than numerical SD or 0TMH dynamics, assuming similar spin-flip rates.
    
    \item We introduce a pattern storage method that allows one to program associative memories in the cavity QED system. The memory capacity of the cavity under this scheme can be as large as $N$ patterns. Encoding states requires only a linear transformation and a threshold operation on the input and output fields, which can be implemented in an optical setting via spatial light modulators.  While the standard Hopfield model does not require encoding, it cannot naturally be realized in a multimode cavity QED system.  Thus, the physical cavity QED system enjoys roughly an order-of-magnitude greater memory capacity at the expense of an encoder. 
    
    \item Overall, our storage and recall scheme points to a novel paradigm for memory systems in which new stimuli are remembered by translating them into already intrinsically stable emergent patterns of the native network dynamics.

\end{enumerate}

The remainder of this paper is organized as follows.
We first describe the physical confocal cavity QED (CCQED) system in Sec.~\ref{sec:ccQED}, before introducing the Hopfield model in Sec.~\ref{sec:HModel}.  Next, we analyze the regimes of spin-spin connectivity provided by the confocal cavity in Sec.~\ref{sec:connectivity}.  We then discuss in Sec.~\ref{sec:SD} the SD dynamics manifest in a transversely pumped confocal cavity above the superradiant transition threshold. Section~\ref{sec:SDHN} discusses how SD dynamics enhances associative memory capacity and robustness.  A learning rule that maps free-space light patterns into stored memory is presented in Sec.~\ref{sec:learning}. 

Last, in Sec.~\ref{sec:conclusion},  we conclude and frame our work in a wider context.
In this discussion, we speculate about how the quantum dynamics of the superradiant transition might enhance solution finding. This is in contrast to all the rest of this paper, where we consider a semiclassical regime well above any quantum critical dynamics at the transition itself.  Embedding neural networks  in systems employing ion traps, optical lattices and optical-parametric-oscillators has been explored~\cite{Nixon:2013fd,Inagaki:2016eb,McMahon:2016fy,Berloff:2017gs}, and comparisons of the latter to our scheme is discussed here also.  This section also provides concluding remarks regarding how the study of this physical system may provide new perspectives on the problem of how memory arises in biological systems~\cite{Krotov:2020uo}.

Appendices~\ref{app:RamanScheme}--\ref{app:mean-field} present the following:~\ref{app:RamanScheme}, the Raman coupling scheme and effective Hamiltonian;~\ref{app:nonlocal}, the convolution of the confocal connectivity with the finite spatial extent of the spin ensembles; \ref{app:random-coupling}, the derivation of the confocal connectivity probability distribution; \ref{app:classicalbath}, the spin-flip dynamics in the presence of a classical bath with ohmic noise spectrum; \ref{app:quantumbath}, the derivation of the spin-flip rate and dynamics in the presence of a quantum bath; and \ref{app:mean-field}, the derivation of the mean-field ensemble dynamics.


\section{Confocal cavity QED}
\label{sec:ccQED}

As illustrated in Fig.~\ref{fig:cQEDblender}, we consider a configuration of $N$ spatially separated BECs placed in a confocal cavity.  In a confocal cavity,  the cavity length $L$ is equal to the radius of the curvature of the mirrors $R$, which leads to degenerate optical modes~\cite{siegman}. More specifically, modes form degenerate families, where each family consists of a complete set of
transverse electromagnetic TEM$_\mathsf{lm}$ modes with $\mathsf{l}+\mathsf{m}$ either of even or odd parity. Recent experiments have demonstrated coupling between compactly confined Bose-Einstein condensates (BECs) of ultracold $^{87}$Rb atoms and a high-finesse, multimode (confocal) optical cavity with $L=R=1$~cm~\cite{Kollar:2014us,Kollar2017sm,Vaidya2018,kroeze2018spinor,Guo2019Sign,Guo2019Emergent}.

\begin{figure}[t!]
    \centering
    \includegraphics[width=\columnwidth]{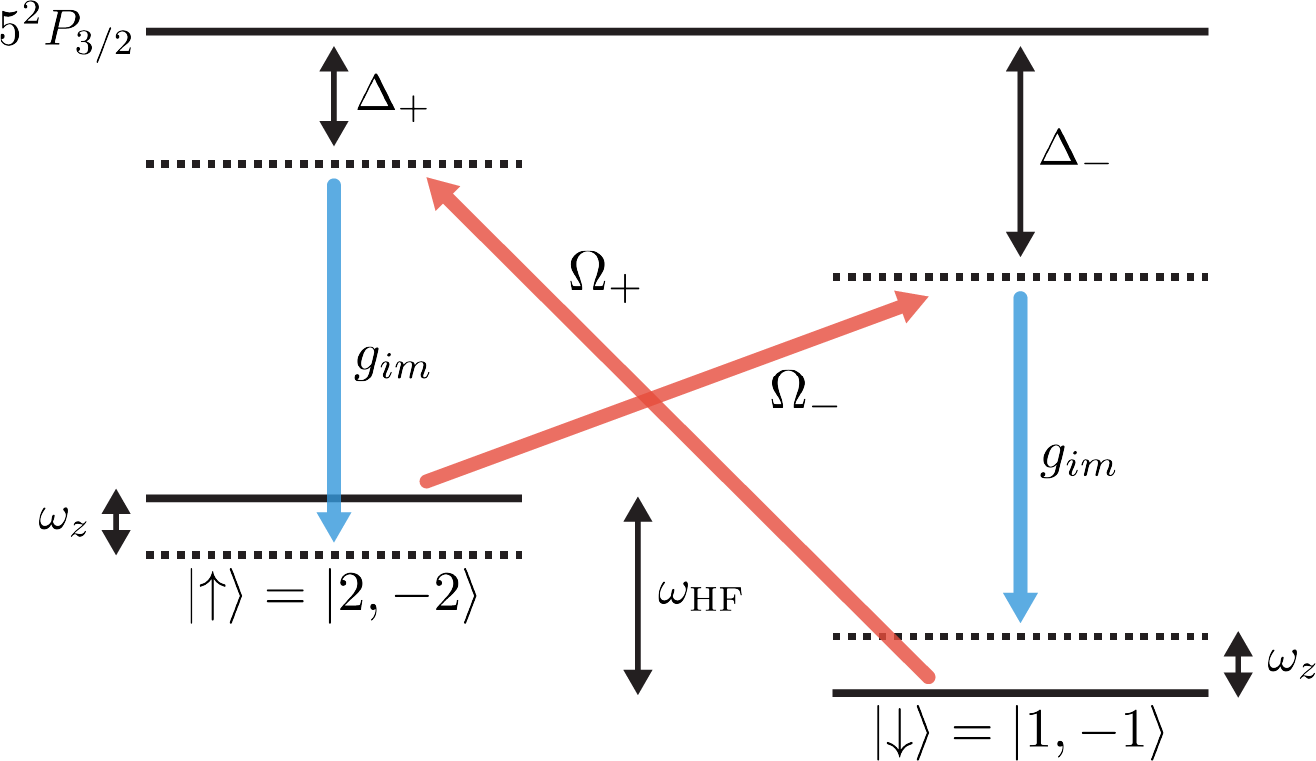}
    \caption{Double Raman atomic coupling.  The scheme provides a two-level system, corresponding to the two hyperfine states of $^{87}$Rb. Two transversely oriented pump beams are used to generate cavity-assisted two-photon processes coupling the states $\ket{F,m_F}=\ket{2,-2}\equiv\ket{\uparrow}$ and $\ket{F,m_F}=\ket{1,-1}\equiv\ket{\downarrow}$.}
    \label{fig:doubleRaman}
\end{figure}

The coupling between the BECs and the cavity occurs via a double Raman pumping scheme illustrated in Fig.~\ref{fig:doubleRaman}. The $|F,m_F\rangle = |1,-1\rangle$ and $|2,-2\rangle$  states are coupled via two-photon processes, involving pump lasers oriented transverse to the cavity axis and the cavity fields~\footnote{Spin-changing collisions in $^{87}$Rb are negligible on the timescale of the experiment, and so contact interactions play little role within each spin ensemble.}.   The motion of atoms may be suppressed by introducing a deep 3D optical lattice  into the cavity~\footnote{Three-dimensional (static) optical lattices inside single mode cavities have been demonstrated~\cite{Klinder2015,Landig2016}. Ultracold thermal atoms may serve as well, provided their temperature is far below the lattice trap depth.}.  When atoms are thus trapped, each BEC can be described as an ensemble of pseudospin-1/2 particles---corresponding to the two states discussed above---with no motional degrees of freedom. The system then realizes a nonequilibrium Hepp--Lieb--Dicke model~\cite{Carmichael07}, exhibiting a superradiant phase transition~\cite{Kollar2017sm,Vaidya2018} when the pump laser intensity is sufficiently large.

In the following, we will assume parameter values similar to those realized in the CCQED experiments of Refs.~\cite{Kollar:2014us,Kollar2017sm,Vaidya2018,kroeze2018spinor,Guo2019Sign,Guo2019Emergent}:  Specifically, we take a single-atom--to--cavity coupling rate of $g_0 = 2\pi\times1.5$~MHz~\footnote{This is the coupling rate to the maximum field of the TEM$_{00}$ mode.}, a cavity field decay rate of $\kappa = 2\pi\times150$~kHz, a pump-cavity detuning of $\detune = - 2\pi\times3$~MHz, and a detuning from the atomic transition of $\Delta_A = -2\pi\times 100$~GHz.  The spontaneous emission rate is approximately $\Gamma \Omega^2 / \Delta_A^2$, where $\Gamma=2\pi \times 6.065(9)$~MHz is the linewidth of the D$_2$ transition in $^{87}$Rb and $\Omega^2$ is proportional to the intensity of the transverse pump laser. A large $\Delta_A$ ensures that the spontaneous emission rate is far slower than the inverse lifetime $\Gamma$ of the Rb excited state. A typical value of $\Omega^2$ is set by the pump strength required to enter the superradiant regime; with 10$^7$ atoms~\footnote{While BECs in the CCQED system are more on the order of 10$^6$ atoms in population, ultracold, but thermal, gases of 10$^7$ atoms may be used since atomic coherence plays little role.}, $\Omega^2$ can be low enough to achieve a spontaneous decay timescale on the order of 100~ms. Hereafter, we will not explicitly include spontaneous emission, but will consider it to set an upper time limit on the duration of experiments.

To control the position of the BECs, one may use optical tweezer arrays~\cite{Endres:2016fk,deMello:2019jw}.  
Experiments have already demonstrated the simultaneous trapping of several ensembles in the confocal cavity using such an approach~\cite{Vaidya2018}.  Extending to hundreds of ensembles is  within the capabilities of tweezer array technology~\cite{deMello:2019jw}.  As we discuss in Sec.~\ref{sec:ensembles}, collective enhancement of the dynamical spin flip rate occurs, depending on the number of atoms in each ensemble. This enhancement is needed so that the pseudospin dynamics is faster than the spontaneous emission timescale.  
Only a few thousand atoms per ensemble are needed to reach this limit, while the maximum number of ultracold atoms in the cavity can reach $10^7$. 
Thus, current laser cooling and cavity QED technology provides the ability to support roughly $10^3$ network nodes and have them evolve for a few decades in timescale.  This number of nodes is similar to state-of-the-art classical spin glass numerical simulation~\cite{Barzegar:2018ez}. 

Improvements to cavity technology can allow the size of the spin ensembles to shrink further, ultimately reaching the single-atom level. Moreover, Raman cooling~\cite{Kaufman:2012ft} within the tight tweezer traps can help mitigate heating effects, allowing atoms to be confined for up to 10 s, limited only by the background gas collisions in the vacuum chamber.   By shrinking the size of the ensembles, it may be possible for quantum entanglement among the nodes to then persist deep into the superradiant regime---we return to this possibility in our concluding discussion in Sec.~\ref{sec:conclusion}.

\begin{figure*}[t!]
    \centering
    \includegraphics[width=\textwidth]{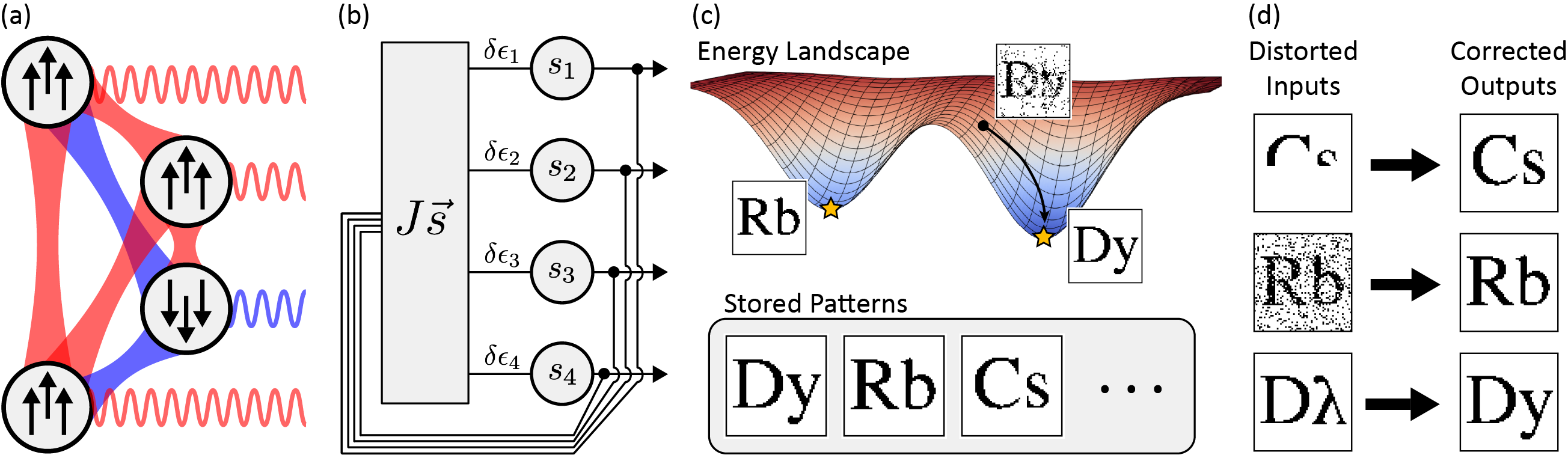}
    \caption{(a) All-to-all, sign-changing connectivity  between spin ensembles  is achieved via photons propagating in superpositions of cavity modes. Blue and red indicate ferromagnetic versus antiferromagnetic $J_{ij}$ links. Only four nodes are depicted.  Individual spins align within an ensemble due to superradiance, while ensembles organize with respect to each other due to $J_{ij}$ coupling. Cavity emission allows for holographic reconstruction of the spin state:  red and blue fields are $\pi$ out of phase, allowing  discrimination between up and down spin ensembles. (b) The spin ensembles realize a Hopfield neural network: a single layer network of binary neurons $s_i=\pm1$ that are recurrently fed back then subjected to a linear transform $J$ and threshold operation at each neuron. (c) The Hopfield model exhibits an energy landscape with many metastable states.  Each local minimum encodes a memory spin configuration (pattern) surrounded by a larger basin of attraction of similar spin states. Energy minimizing dynamics drive sufficiently similar spin configurations to the stored local minimum. There is a phase transition from the associative memory to a spin glass once there are too many memories; i.e., when so many minima exist that basins of attraction vanish. (d) Schematic of the associative memory problem.  Multiple stored patterns (e.g., images of element symbols) may be recalled by pattern completion of distorted input images. }
    \label{fig:cQEDcartoon}
\end{figure*}

The confocal cavity realizes a photon-mediated interaction among the spin ensembles.  This interaction is described by a matrix $J_{ij}$, denoting the coupling between ensembles $i$ and  $j$. As derived in Sec.~\ref{sec:SD}, this interaction involves a sum over all relevant cavity modes, $m$, and takes a form $J_{ij}=-\sum_{m} {\Delta_m g_{im} g_{jm}}/(\Delta_m^2+\kappa^2)$.  Here, $g_{im}$ is the coupling between cavity mode $m$ and ensemble $i$---which depends on the positions of atoms and spatial profiles of modes---while $\Delta_m$ is the detuning of the pump laser from cavity mode $m$.  
For simplicity in writing this equation, we will restrict the BEC positions $\mathbf{r}_i$  to lie within the transverse plane at the center of the cavity. Following refs.~\cite{Vaidya2018,Guo2019Sign,Guo2019Emergent},  in the confocal limit $\Delta_m=\detune$, the interaction then takes the form:
\begin{equation}
\label{eqn:JijForm}
    J_{ij}=\frac{-\tilde g_0^2\detune}{\pi(\detune^2 + \kappa^2)}\left[\geo \delta_{ij}+ \cos\left(2\frac{\mathbf{r}_i\cdot\mathbf{r}_j}{\waist^2} \right) \right].
\end{equation}  
Here, $\tilde g_0=\Omega g_0/\Delta_A$ denotes an effective coupling strength in terms of the transverse pump strength $\Omega$, single-atom--to--cavity coupling $g_0$ and atomic detuning $\Delta_A$.  The length scale $w_0$ is the width (radius) of the Gaussian TEM$_{00}$ mode. The term $\geo$ is a geometric factor determined by the shape of the BEC. For a Gaussian atomic profile of width $\sigma_A$ in the transverse plane, $\geo=w_0^2/8\sigma^2_A$, which is typically $\sim$10.

The first term  $\geo \delta_{ij}$ is a local interaction that is  present only for spins within the same ensemble. This term arises from the light in a confocal cavity being perfectly refocused after two round trips. The  effect of this term is to align spins within the same ensemble, or in other words, to induce the superradiant phase transition of that ensemble.  In practice, imperfect mode degeneracy broadens the refocussing into a local interaction of finite range; Refs.~\cite{Vaidya2018,Guo2019Sign,Guo2019Emergent} discuss this effect.  The range of this interaction is controlled by the ratio between the pump-cavity detuning $\detune$ and the spread of the cavity mode frequencies.  At sufficiently large $\detune$, the interaction range can become much smaller than the spacing between BECs.  Because a confocal cavity resonance contains only odd or even modes, there is also refocussing at the mirror image position; we can ignore this by assuming all BECs are in the same half-plane~\cite{Vaidya2018,Guo2019Sign,Guo2019Emergent}.  

The nonlocal second term arises from the propagation of light in between the refocussing points.  Intuitively, this interaction arises from the fact that each confocal cavity images the Fourier transform of objects in the central plane back onto the objects in that same plane. Thus, photons scattered by atoms in local wavepackets are reflected back onto the atoms as delocalized wavepackets with cosine modulation---the Fourier transform of a spot.  Formally, it arises due to Gouy phase shifts between the different degenerate modes, see Refs.~\cite{Vaidya2018,Guo2019Sign,Guo2019Emergent} for a derivation and experimental demonstrations. This interaction is both nonlocal and nontranslation invariant.  It can generate frustration between spin ensembles due to its sign-changing nature, as discussed in Sec.~\ref{sec:connectivity} below.   The structure of the matrix $J_{ij}$ is quite different from those appearing traditionally in Hopfield models, either under Hebbian or pseudoinverse coupling rules, as discussed in Sec.~\ref{sec:connectivity}.  We note that while the finite spatial extent of the BEC is important for rendering the local interaction finite,  we show in Appendix~\ref{app:nonlocal} that it does not significantly modify the nonlocal interaction in the experimental regime discussed here.


\section{Hopfield model of associative memory} \label{sec:HModel}

The Hopfield associative memory is a model neural network that can store memories in the form of distributed patterns of neural activity ~\cite{Little74, hopfield1982neural, Nishimori01}. In the simplest instantiation of this class of networks, each neuron $i$ has an activity level $s_i$ that can take one of two values: $+1$, corresponding to an active neuron, or $-1$, corresponding to an inactive one.  The entire state of a network of $N$ neurons is then specified by one of $2^N$ possible distributed activity patterns.  This state evolves according to the discrete time dynamics
\begin{equation}\label{update}
s_i(t+1) = \mathrm{sgn}\left(\sum_{j \neq i} J_{ij} s_j(t) - h_i\right),
\end{equation}  
where $J_{ij}$ is a real valued number that can be thought of as the strength of a synaptic connection from neuron $j$ to neuron $i$.  Intuitively, this dynamics computes a total input $h^{\mathrm{eff}}_i = \sum_{j \neq i} J_{ij} s_j(t)$ to each neuron $i$ and compares it to a threshold $h_i$.  If the total input is greater (less) than this threshold, then neuron $i$ at the next timestep is active (inactive).  One could implement this dynamics in parallel, in which case Eq.~\eqref{update} is applied to all $N$ neurons simultaneously.  Alternatively, for reasons discussed below, it is common to implement a serial version of this dynamics in which a neuron $i$ is selected at random, and then Eq.~\eqref{update} is applied to that neuron alone, before another neuron is chosen at random to update. 

The nature of the dynamics in Eq.~\eqref{update} depends crucially on the structure of the synaptic connectivity matrix $J_{ij}$.  For arbitrary $J_{ij}$, and large system sizes $N$, the long-time asymptotic behavior of $s_i(t)$ could exhibit three possibilities:  (1) flow to a fixed point; (2) flow to a limit cycle; or (3) exhibit chaotic evolution.  On the other hand, with a symmetry constraint in which $J_{ij} = J_{ji}$, the serial version of the dynamics in Eq.~\eqref{update} monotonically decreases an energy function 
\begin{equation}
    \label{eqn:H_Hopfield}
    H = -\sum_{i,j=1}^N J_{ij} s_i s_j
        + \sum_{i=1}^N h_i s_i
    .
\end{equation}
In particular, under the update in Eq.~\eqref{update} for a single spin, it is straightforward to see that $H(t+1) \leq H(t)$ with equality if and only if $s_i(t+1) = s_i(t)$.  Indeed, the serial version of the update in Eq.~\eqref{update} corresponds exactly to Zero Temperature Metropolis--Hastings (0TMH)~\cite{Metropolis53,Hastings70} or Glauber dynamics~\cite{glauber63} applied to the energy function in \eqref{eqn:H_Hopfield}.  The existence of a monotonically decreasing energy function rules out the possibility of limit cycles, and every neural activity pattern thus flows to a fixed point, which corresponds to a local minimum of the energy function. A local minimum is by definition a neural activity pattern in which flipping {\it any} neuron's activity state would increase the energy. 

One of Hopfield's key insights was that we could think of neural memories as fixed points or local minima in an energy landscape over the space of neural activity patterns; these are also sometimes known as metastable states or attractors of the dynamics.  Each such fixed point has a basin of attraction, corresponding to the set of neural activity patterns that flow under Eq.~\eqref{update} to that fixed point.  The process of successful memory retrieval can then be thought of in terms of a pattern completion process.  In particular, an external stimulus may initialize the neural network with a neural activity pattern corresponding to a corrupted or partial version of the fixed point memory.  Then, as long as this corrupted version still lies within the basin of attraction of the fixed point, the flow towards the fixed point completes or cleans up the initial corrupted pattern, triggering full memory recall.  This is an example of content addressable associative memory, where partial content of the desired memory can trigger recall of all facts associated with that partial content. A classic example might be recalling a friend who has gotten a haircut. Figure~\ref{fig:cQEDcartoon} illustrates this pattern completion based memory retrieval process.  

In this framework, the set of stored memories, or fixed points, are encoded entirely in the connectivity matrix $J_{ij}$; for simplicity, we set the thresholds $h_i=0$.  Therefore, if we wish to store a prescribed set of $P$ memory patterns $\boldsymbol{\xi}^\mu = (\xi_1^\mu,\ldots,\xi_N^\mu)$ for $\mu = 1,\ldots,P$, where each $\mathbf{\xi}^\mu_i=\pm 1$, we need a learning rule for converting a set of given memories $\{\boldsymbol{\xi}^\mu \}_{\mu=1}^P$ into a connectivity matrix $\mathbf{J}$.  Ideally, this connectivity matrix should instantiate fixed points under the dynamics in Eq.~\eqref{update} that are close to the desired memories $\boldsymbol{\xi}^\mu$, with large basins of attraction, enabling robust pattern completion of partial, corrupted inputs. Of course, in any learning rule, one should expect a trade-off between capacity (the number of memories that can be stored) and robustness (the size of the basin of attraction of each memory, which is related to the fraction of errors that can be reliability corrected in a pattern completion process). 

When the desired memories $\boldsymbol{\xi}^\mu$ are unstructured and random, a common choice is the Hopfield connectivity, which corresponds to a Hebbian learning rule~\cite{hopfield1982neural, Nishimori01}:
\begin{equation}\label{heb}
\mathbf{J}_\mathrm{Hebbian} =  \frac{1}{N} \sum_{\mu = 1}^P \boldsymbol{\xi}^\mu (\boldsymbol{\xi}^\mu)^T.
\end{equation}
We may note that in the magnetism literature, such a model is known as the multicomponent Mattis model~\cite{mattis1976solvable}.
The properties of the energy landscape associated with the dynamics in Eq.~\eqref{update} under this connectivity have been analyzed extensively in the thermodynamic limit $N \to \infty$~\cite{Amit87,amit1992modeling}. When $P \lesssim 0.05 N$ the lowest energy minima are in one to one correspondence with the $P$ desired memories.  For $0.05 \lesssim P/N \lesssim 0.138$, the $P$ memories correspond to  metastable local minima. However, pattern completion is still possible; an initial pattern corresponding to a corrupted memory, with a small but extensive number of errors proportional to $N$, will still flow towards the desired memory.  However, at larger $P$, the network undergoes a spin-glass transition in which there can be exponentially many energy minima with small basins of attraction, none of which are close to the desired memories.  Thus, pattern completion is not possible: an initial pattern corresponding to a corrupted memory, with even a small but extensive number of errors proportional to $N$, is not guaranteed to flow towards the desired memory. This spin glass transition occurs because the large number of memories start to interfere with each other. In essence, the addition of each new memory modifies the existing local minima associated with previous memories.  When too many memories are stored, it is not possible, under the Hebbian rule in Eq.~\eqref{heb} and the dynamics of Eq.~\eqref{update}, to ensure the existence of local minima, with large basins, close to any desired memory. 

Thus, we have noted the Hopfield model undergoes a spin glass transition above a critical pattern loading $P/N \approx 0.138$.  In the exposition below, it will be useful to also consider a prototypical example of a spin glass, namely the 
Sherrington--Kirkpatrick (SK) model~\cite{SherringtonKirkpatrick}.  In the SK model, the matrix elements of the symmetric matrix  $\mathbf{J}_{SK}$ are chosen i.i.d.~from a zero mean Gaussian distribution with variance $\sigma^2 / N$.  At low temperature, such a model also has a spin-glass phase with exponentially many energy minima, and as we shall see below, our CCQED system exhibits a transition from a memory retrieval phase to an SK-like spin-glass phase as the positions of spin ensembles spread out within the cavity.  

Numerous improvements to the Hebbian learning rule have been  introduced~\cite{Storkey1997,Storkey1999} that sacrifice the simple outer product structure of the Hebbian connectivity in Eq.~\eqref{heb} for improved capacity. Notable among them is the pseudoinverse rule, in which $P$ may be as large as $N$.  This large capacity comes at the cost of being a nonlocal learning rule: updating any of the weights requires full knowledge of all existing $J_{ij}$ weights, unlike the Hebbian learning rule.  The $J_{ij}$ matrix for the pseudoinverse learning rule is given by:
\begin{equation}
    \label{pseudoheb}
    \mathbf{J}_{\mathrm{Pseudo}} = \frac{1}{N} \sum_{\mu,\nu=1}^P \boldsymbol{\xi}^\mu \mathbf{C}^{-1}(\boldsymbol{\xi}^\nu)^T,
\end{equation}
where the matrix $C_{\mu\nu} = \frac{1}{N} \boldsymbol{\xi}^\mu\cdot\boldsymbol{\xi}^\nu$ stores the inner products of the patterns. This learning rule ensures that the desired memories $\boldsymbol{\xi}^\nu$ become eigenvectors of the learned connectivity matrix  $\mathbf{J}_{\mathrm{Pseudo}}$ with eigenvalue $1$, thereby ensuring that each desired memory corresponds to a fixed point, or equivalently a local energy minimum, under the dynamics in Eq.~\eqref{update}.  While the pseudoinverse rule does guarantee each desired memory will be a local energy minimum, further analysis is required to check whether such minima have large basins. The basin size will generically depend on the structure of $C_{\mu\nu}$, with potentially small basins arising for pairs of memories that are very similar to each other.  Finally, we note that  the Hebbian learning rule in Eq.~\eqref{heb} is in fact a special case of the pseudoinverse learning rule in Eq.~\eqref{pseudoheb} when the patterns $\boldsymbol{\xi}^\mu$ are all mutually orthogonal, with $C_{\mu \nu} = \delta_{\mu \nu}$. We include the pseudoinverse rule in our comparisons below to demonstrate the generality of results we present, and to apply them to something known to surpass the original Hebbian scheme.

While the simple Hebbian rule and the more powerful pseudoinverse rule are hard to directly realize in a confocal cavity, we show in Sec.~\ref{sec:connectivity} that the connectivity naturally provided by the confocal cavity is sufficiently high rank to support a multitude of local minima. We further analyze the dynamics of the cavity in Sec.~\ref{sec:SD}, and demonstrate that this dynamics endows this multitude of local minima with large basins of attraction in Sec.~\ref{sec:SDHN}.  Thus, the confocal cavity provides a physical substrate for high capacity, robust memory retrieval.  Of course, the desired memories we wish to store may not coincide with the naturally occurring emergent local minima of the confocal cavity.  However, any such mismatch can be solved by mapping the desired memories we wish to store into the patterns that are naturally stored by the cavity (and vice-versa).  We will show in Sec.~\ref{sec:learning} that such a mapping is possible, and further, that it is practicable using optical devices. 

Taken together, the next few sections demonstrate that the CCQED system possesses three critical desiderata of an associative memory: (1) high memory capacity due to the presence of many local energy minima (Sec.~\ref{sec:connectivity}); (2) robust memory retrieval with pattern completion of an extensive number of initial errors (Sec.~\ref{sec:SD} and \ref{sec:SDHN}); and (3) programmability or content addressability of any desired memory patterns (Sec.~\ref{sec:learning}). 

\section{Connectivity regimes in a confocal cavity}
\label{sec:connectivity}

As noted at the end of Sec.~\ref{sec:ccQED}, the form of connectivity $J_{ij}$ in Eq.~\eqref{eqn:JijForm}, which arises naturally for the confocal cavity, is quite different from both the forms arising in the Hebbian (Eq.~\eqref{heb}) and pseudoinverse (Eq.~\eqref{pseudoheb}) learning rules discussed in the previous section.  Therefore, in this section we analyze the statistical properties and energy landscape of the novel random connectivity matrix  arising from the confocal connectivity, taking the ensemble positions $\mathbf{r}_i$ to be randomly chosen from a Gaussian distribution in the cavity transverse midplane.  Specifically, we choose a distribution of positions with zero mean (i.e., centered about the cavity axis) and tunable standard deviation $\spinswidth$. As the interaction in Eq.~\eqref{eqn:JijForm} is symmetric under $\mathbf{r}_i\to-\mathbf{r}_i$, the distribution may also be restricted to a single half plane to avoid the mirror interaction with no adverse affect to local and nonlocal interactions. 

\begin{figure}[t!]
    \centering
    \includegraphics[width=\columnwidth]{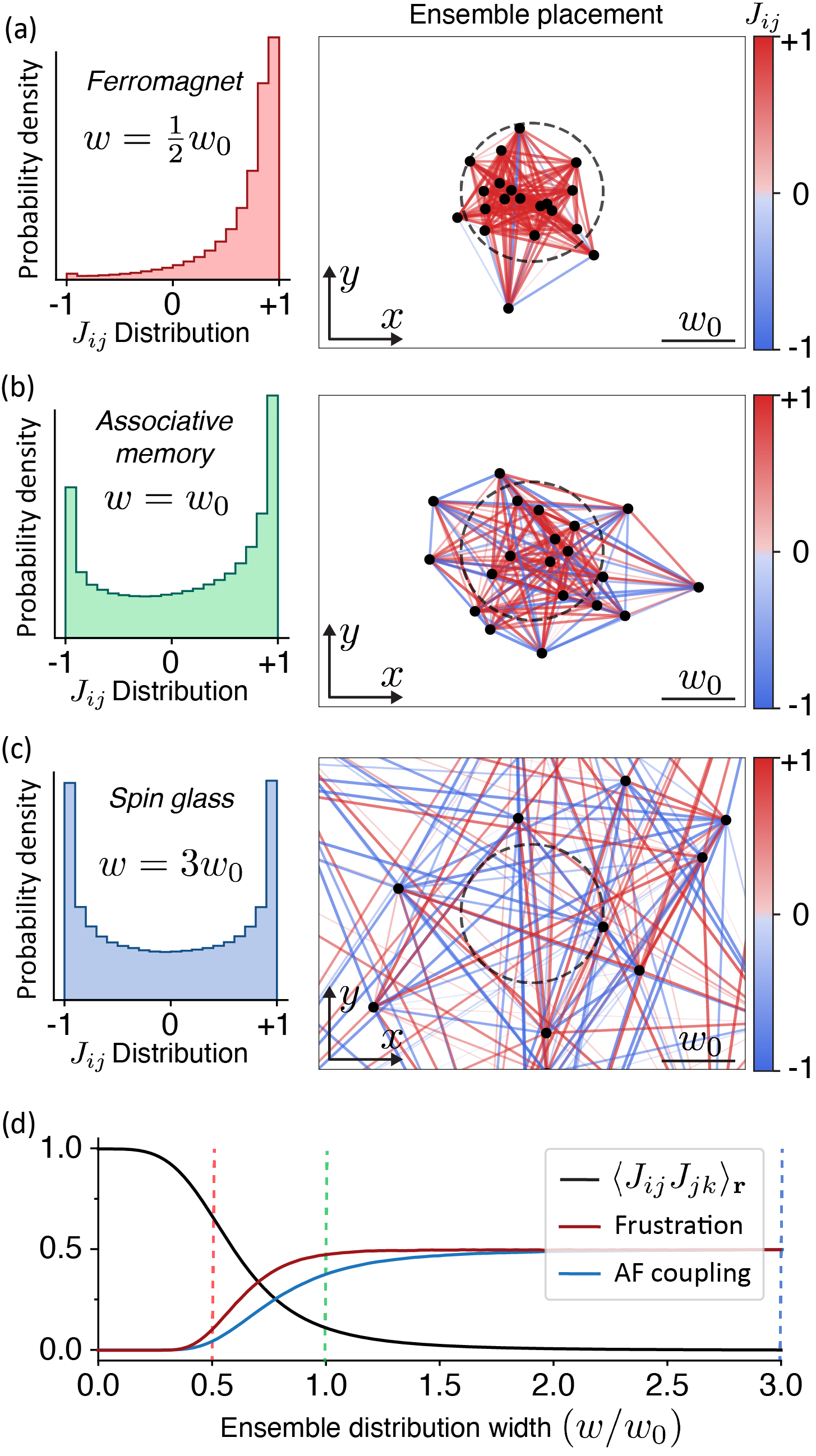}
     \caption{Confocal cavity connectivities.  Distribution of normalized $J_{ij}$ couplings shown for widths (a) $w=w_0/2$, (b) $w=w_0$, and (c) $w=4w_0$. Representative $J_{ij}$ connectivity graphs are shown to the right, with red (blue) links indicating ferromagnetic $J_{ij}>0$ (antiferromagnetic $J_{ij}<0$) coupling.  The radius $w_0$ is the Gaussian width of the TEM$_{00}$ mode in the transverse plane and is indicated by the dashed circle. (The modes of a multimode cavity extend far beyond the characteristic length scale $w_0$.) (d) Statistical properties of the $J_{ij}$ versus $w$. The correlation function in Eq.~\eqref{eqn:Corr_Jij}  between different $J_{ij}$ (black line)   shows that they become uncorrelated at large $w$. The probability of generating frustration between three randomly chosen spins (red line) and the probability of generating an antiferromagnetic coupling between two randomly chosen spins (blue line) are also plotted.  The dashed lines correspond to the widths used in panels (a) through (c). }
    \label{fig:Jij-regimes}
\end{figure}

In the limit $\spinswidth\to0$, so that all $\mathbf{r}_i=0$, all off-diagonal elements of the confocal connectivity matrix $J_{ij}$ become identical and positive.  This describes a ferromagnetic coupling.  As the width increases, some elements of $J_{ij}$ become negative, as illustrated in Fig.~\ref{fig:Jij-regimes}. This can lead to frustration in the network of coupled spins, where the product of couplings around a closed loop is negative, i.e., where  $J_{ij}J_{jk}J_{ki}<0$ for spins $i$, $j$, and $k$.  In such cases, it is not possible to minimize the energy of all pairwise interactions simultaneously.
Such frustration can in principle lead to a proliferation of metastable states, a key prerequisite for the construction of an associative memory.  
As we shall see, tuning the width $\spinswidth$ allows one to tune the degree of frustration, and thus the number of metastable states in the confocal cavity.
In the remainder of this section, we will consider only the normalized nonlocal couplings $\tilde J_{ij}=\cos(2\mathbf{r}_i\cdot\mathbf{r}_j/w_0^2)\in[-1,1]$. The local part of the interaction plays no role in the properties we discuss in this section, but does affect the dynamics as discussed in Sec.~\ref{sec:SD}.

To characterize the statistical properties of the random matrix $J_{ij}$, and its dependence on $\spinswidth$, we begin by considering the marginal probability distribution for individual matrix elements $J_{ij}$. Details of the derivation of this distribution are given in Appendix~\ref{app:random-coupling}; the result is
\begin{equation}
\label{eqn:P_Jij}
        \prob(\tilde J_{ij};w)=\frac{\mathrm{csch}\left(\pi/(2\tilde{w}^2)\right)}{2\tilde{w}^2\sqrt{1-\tilde J_{ij}^2}}\cosh\left[\frac{\pi-\arccos{\tilde J_{ij}}}{2\tilde{w}^2}\right],
\end{equation}
where $\tilde{w}\equiv w/w_0$.  Figure~\ref{fig:Jij-regimes}(a--c) illustrates the evolution of this marginal distribution for increasing width. For small $\tilde{w}$, the distribution is tightly peaked around $\tilde J_{ij}=+1$,
corresponding to an unfrustrated all-to-all ferromagnetic model, with only a single global minimum (up to $\mathbb{Z}_2$ symmetry). As $\tilde{w}$ increases, negative (antiferromagnet) elements of $\tilde J_{ij}$ become increasingly probable. Also plotted in Fig.~\ref{fig:Jij-regimes}(d) are the fractions of $J_{ij}$ links that are antiferromagnetic as well as the fraction that realize frustrated triples of spin connectivity, $J_{ij}J_{jk}J_{ki}<0$. The probability of antiferromagnetic coupling is analytically calculated in Appendix~\ref{app:random-coupling}, while the probability of frustrated triples is evaluated numerically.

If the different matrix elements $J_{ij}$ were uncorrelated, then one could anticipate that, as in the SK model, frustration would occur once the probability of negative $J_{ij}$ becomes sufficiently large.  However, when correlations exist, the presence of many negative elements is not sufficient to guarantee significant levels of frustration and the consequent proliferation of metastable local energy minima.  For example, the rank $1$ connectivity $J_{ij}=\xi_i \xi_j$, for a random vector $\boldsymbol{\xi}$, can have have an equal fraction of positive and negative elements while remaining unfrustrated~\cite{mattis1976solvable}.
In general, we expect the couplings $J_{ij}$ and $J_{jk}$ should be correlated, as they both depend on the common position $\mathbf{r}_j$.
As discussed  in App.~\ref{app:random-coupling}, this correlation can be computed analytically as a function of the width:
\begin{equation}
\label{eqn:Corr_Jij}
    \langle J_{ij}J_{jk}\rangle_{\mathbf{r}}=\frac{1}{1+8\tilde{w}^4},
\end{equation}
where $\langle \cdot \rangle_{\mathbf{r}}$ denotes an average over realizations of the random placement of spin ensembles. Although correlations exist, we see from this expression that they decay like $1/\tilde{w}^4$, so that at large $\tilde{w}$, the correlations are weak; see Figure~\ref{fig:Jij-regimes}(d). Of course, even weak correlations in a large number of $O(N^2)$ off-diagonal elements can, in principle, dramatically modify important emergent properties of the random matrix, such as the induced multiplicity of metastable states and the statistical structure of the eigenvalue spectrum.  Thus, we examine the properties of the actual correlated random matrix ensemble arising from the confocal cavity rather than adopt known results.  

\begin{figure}[t!]
    \centering
    \includegraphics[width=\columnwidth]{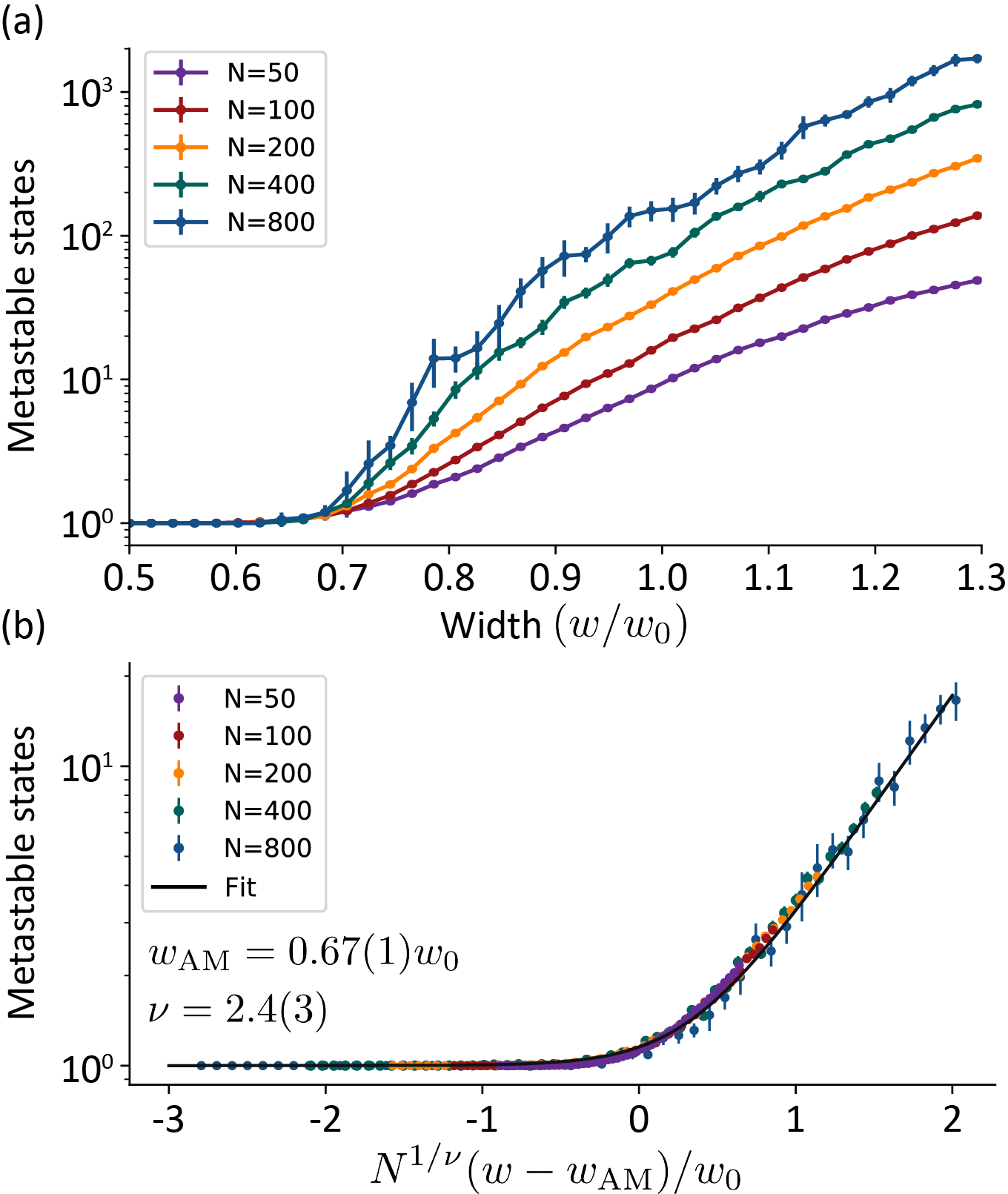}
    \caption{(a) Number of metastable states versus distribution width for various system sizes $N$, as indicated.  The average number of metastable states increases with $N$ above the ferromagnetic--to--associative memory transition at $w_{\mathrm{AM}}=0.67(1)w_0$. (b) Scaling collapse of the above in the region $0.5w_0<w<0.8w_0$, with denser sampling. The x-axis is rescaled as $N^{1/\nu}(w-w_{\mathrm{AM}})/w_0$, while the y-axis is unchanged. The parameters $\nu$ and $w_{\mathrm{AM}}$ are determined by fitting the collapsed data to an exponential form $\sqrt{1+Ae^{Bx}}$, where $x = N^{1/\nu} (w-w_{\mathrm{AM}})/w_0$. The uncertainties are one standard error.}
    \label{fig:Jij-transitions}
\end{figure}

Figure~\ref{fig:Jij-transitions} shows a numerical estimation of the number of metastable states as a function of the width $w$. This number is estimated by initializing a large number of random initial states and allowing those states to relax via 0TMH dynamics until a metastable local energy minimum is found.  This routine is performed for many realizations of the connectivity $J_{ij}$, then averaged over realizations to produce the number plotted in Fig.~\ref{fig:Jij-transitions}.  We regard configurations which are related by an overall spin flip as equivalent. A single metastable state is found at small $\tilde{w}$, as expected for ferromagnetic coupling. A transition to multiple metastable states occurs as $\tilde{w}$ increases; this transition becomes increasingly sharp at larger system sizes.  Finite size scaling analysis of the transition yields a critical point of  $w_{\mathrm{AM}}\approx 0.67 w_0$.  Only one minima exists below this value, while multiple minima emerge above.  
The number of metastable states, shown in Fig.~\ref{fig:Jij-transitions}(b), increases rapidly for $w>w_{\mathrm{AM}}$.  In particular, in the range of $w$ and $N$ that we explored, we find that the following fits the simulations: $\mathcal{N} =  \sqrt{1 + Ae^{Bx}}$, where $x = N^{1/\nu} (w-w_{\mathrm{AM}})/w_0$ is the rescaled width, and $A=0.33$,  $B=3.4$, and $\nu=2.4$ are the fit parameters.  Thus,  the number of metastable states scales with $N$ and $w$ as $O(e^{w N^{0.4}})$ just above $w_{\mathrm{AM}}$. At still larger $w$, the numerical estimation of this number becomes less reliable due to the increasing prevalence of metastable states with small basins of attraction under 0TMH dynamics~\footnote{One might expect this could be overcome by a ``capture--recapture'' approach to estimate the true number of metastable states by determining how often each state is found.  However, as the distribution of basin sizes is non-uniform, a naive application of this approach would be biased toward underestimating the true number of metastable states.}. 

The existence of multiple (metastable) local energy minima is a critical prerequisite for associative memory storage.  An additional requirement, as discussed in Sec.~\ref{sec:HModel}, is that these local energy minima should possess large enough basins of attraction to enable robust pattern completion of partial or corrupted initial states, through the intrinsic dynamics of the CCQED system.  As of yet, we have made no statement about the basins of attraction of the metastable states we have found; we will examine this issue in the next two sections which will focus on the CCQED dynamics.  For now, we simply note that at a fixed width of $\tilde w$ an $O(1)$ amount above the transition, the number of metastable states $\mathcal N$ grows with system size $N$ as $ O(e^{N^{0.4}})$, while the total configuration space grows with exponential scaling $2^N$.   Given that every configuration must flow to one of these energy minima, it seems reasonable to expect that any energy minimizing dynamics should endow these minima with sufficiently large basins of attraction to enable robust pattern completion, assuming these basins are all approximately similar in size.   

At fixed $N$, the growth in the number of energy minima as a function of $\tilde{w}$, depicted in Fig.~\ref{fig:Jij-transitions}, suggests the possibility of a spin glass phase wherein exponentially many metastable local energy minima emerge.  Based on the analysis of the marginal Eq.~\eqref{eqn:P_Jij} and pairwise Eq.~\eqref{eqn:Corr_Jij} statistics of the matrix elements, we expect that the spin glass phase should be like that of an SK model at large $\tilde w$.  To determine if such a state exists, we further analyze the connectivity in this large $\tilde{w}$ regime by comparing properties of the CCQED connectivity to those of the SK spin glass connectivity.  In the limit of large $\tilde{w}$, the probability density for the $J_{ij}$ given in Eq.~\eqref{eqn:P_Jij} takes the limiting form $\prob(\tilde J_{ij})=(1-\tilde J_{ij}^2)^{-1/2}/\pi$.  This functional form differs from the SK spin glass model in which the probability density of the couplings is Gaussian, with only the first two moments nonzero. However, it is known~\cite{panchenko2013sherrington} that the SK model free energy depends on only the first two moments of the marginal distribution in the thermodynamic limit, as long as the third-order cumulant of the distribution is bounded. This is also true for the CCQED connectivity.  Moreover, these first two moments  in the CCQED connectivity can be computed analytically.  The mean $\mu_J$ and standard deviation $\sigma_J$ as a function of width are
\begin{equation}
\mu_J  = \frac{1}{1+4\tilde{w}^4},\quad \sigma_J=\frac{4\tilde{w}^4}{1+4\tilde{w}^4} \sqrt{ \frac{(5+8\tilde{w}^4)}{1+16\tilde{w}^4}}.
\end{equation} 
Thus, at large width, the mean is negligible compared to the standard deviation, which is required for a spin glass (as opposed to ferromagnetic state). 

\begin{figure}
    \centering
    \includegraphics[width=\columnwidth]{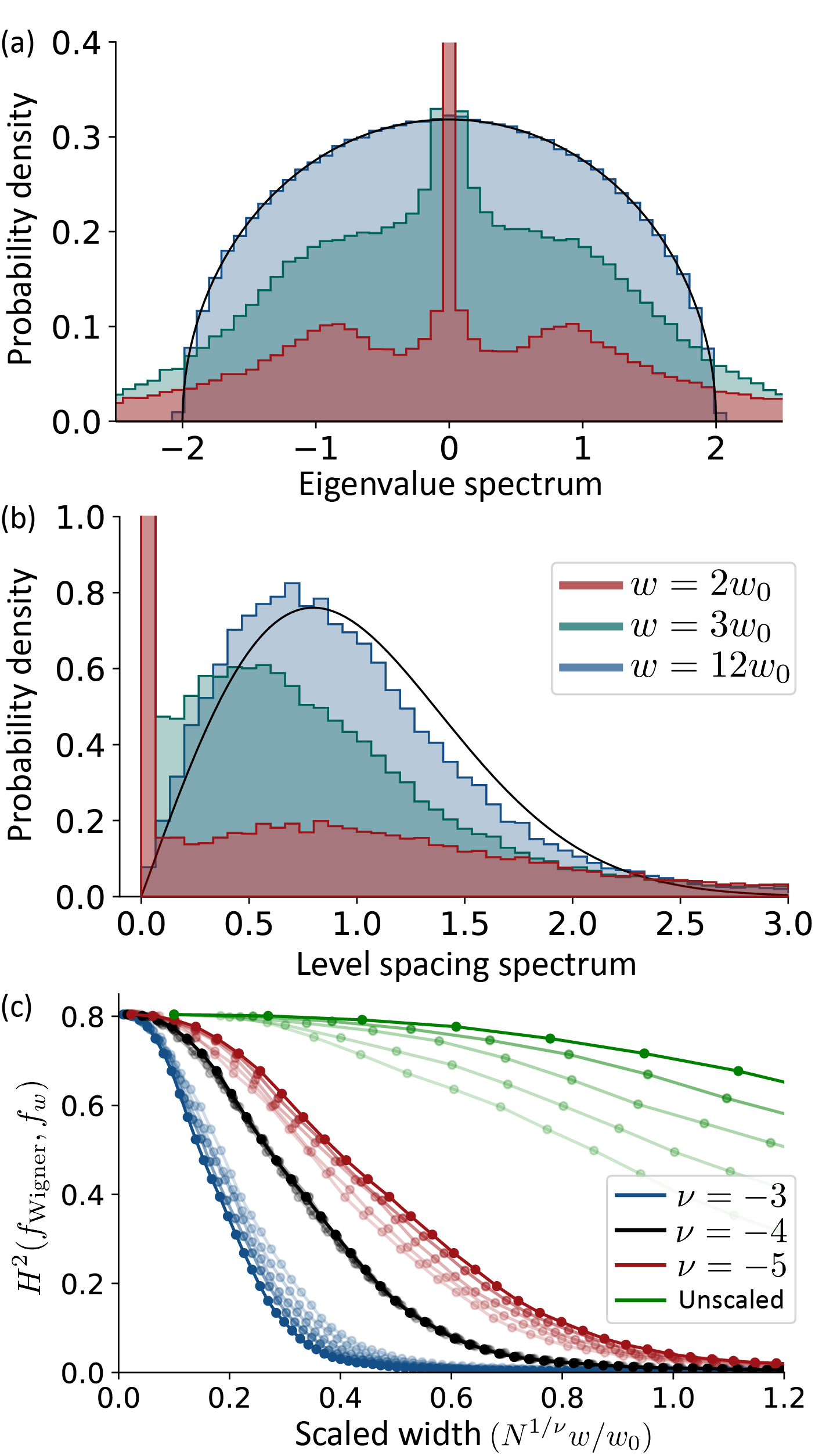}
    \caption{Eigenvalue spectrum of the CCQED connectivity, demonstrating random matrix statistics.  (a) The eigenvalue distribution $f_w$ and (b) level-spacing spectrum for the nonlocal interaction matrix $J_{ij}=\cos(2\mathbf{r}_i\cdot\mathbf{r}_j/w_0^2)$. $N=1000$ spins are averaged over 50 realizations of $J_{ij}$ matrices. The distributions are shown for $w=2w_0$ (red), $w=3w_0$ (green), and $w=12w_0$ (blue). Each distribution is normalized to have unit variance. The solid black lines show the Wigner semicircle eigenvalue distribution (panel a) and the Wigner surmise for the level spacing distribution of Gaussian orthogonal ensemble random matrices (panel b). Panel (c) shows the Hellinger distance between the averaged CCQED connectivity eigenvalue distribution $f_w$ (shown in panel a), and the Wigner semicircle distribution $f_{\mathrm{Wigner}}$.  The distance is plotted as a function of the rescaled width $N^{1/\nu}w/w_0 $ for different power laws $\nu=-3,-4,-5$ corresponding to different colors, as indicated.  Lines of varying intensity correspond to different $N$ ranging from $N=100$ (faintest) to $N=1000$ (darkest). The traces overlap well for the $\nu=-4$ scaling. } 
    \label{fig:Jij-random}
\end{figure}

However, a key difference between the CCQED connectivity and the SK connectivity is the presence of correlations between different matrix elements.  The former's correlation strength decreases with width, see Eq.~\eqref{eqn:Corr_Jij} and Fig.~\ref{fig:Jij-regimes}(d). We can obtain insights into how large the width must be in order to suppress these correlations, thereby crossing into an SK-like spin glass phase, by comparing the statistical structure of the CCQED connectivity eigenvalue distribution to that of the SK model connectivity.  In particular,  the eigenvalue distribution obeys the same Wigner's semicircular law $f_{\mathrm{Wigner}}(x)=\sqrt{4-x^2}/2\pi$ as does the SK connectivity with zero mean i.i.d.~Gaussian elements of variance ${1}/{N}$~\cite{Wigner55:char,Wigner58:roots}.  Moreover, the distribution of spacings $s$ between adjacent eigenvalues (normalized by the mean distance) in both obey Wigner's surmise $p_{\mathrm{Wigner}}(s) = \frac{\pi s}{2} e^{-\pi s^2 / 4}$, reflecting repulsion between adjacent eigenvalues~\cite{Wigner56:conf}.  Figure~\ref{fig:Jij-random}(a,b) plots the  eigenvalue distribution $f_w(x)$ and level-spacing distribution $p_w(s)$ for several widths $w$ for the CCQED connectivity.  Both distributions approach those of the SK model for widths beyond a few $w_0$.  As we discuss next, the required ratio $w/w_0$ to reach the SK regime depends on the system size $N$. 

Figure~\ref{fig:Jij-random}(c) plots the difference between the confocal eigenvalue distribution, denoted by $f_w$, and the Wigner semicircular law $f_{\mathrm{Wigner}}$ in terms of the Hellinger distance, which is  defined for arbitrary probability distributions $p(x)$ and $q(x)$ as $H^2(p,q) = \int dx (\sqrt{p(x)}-\sqrt{q(x)})^2/2$. The distance metric equals 1 for completely non-overlapping distributions and 0 for identical distributions. We find that the Hellinger distance follows a universal ($N$ independent) curve as a function of $\tilde{w} N^{-1/4}$. The structure of this curve demonstrates that as long as $w>w_{SK}= \alpha N^{1/4}w_0$, where $\alpha$ is a constant of order unity, then the spectrum of the confocal cavity connectivity assumes Wigner's semicircular law, just like that of the SK model connectivity. At this large value of $w$, the strength of correlations between different matrix elements in the confocal cavity, given by Eq.~\eqref{eqn:Corr_Jij}, is $O({1}/{N})$.  Such weak correlations, combined with an $O(1)$ variance and a negligible mean of the matrix elements, endow the CCQED connectivity with similar spectral properties to that of the SK connectivity. Given these similarities, we thus expect that the CCQED model possesses an SK-like spin glass phase at widths $w>w_{SK}$.  While the Hellinger distance in Fig.~\ref{fig:Jij-random}(c) falls to zero, it does so smoothly as $N^{1/4}$; whether a phase transition or crossover occurs between the associative memory and spin glass behaviors is unclear and warrants future investigation.

\begin{figure}[t!]
    \centering
    \includegraphics[width=\columnwidth]{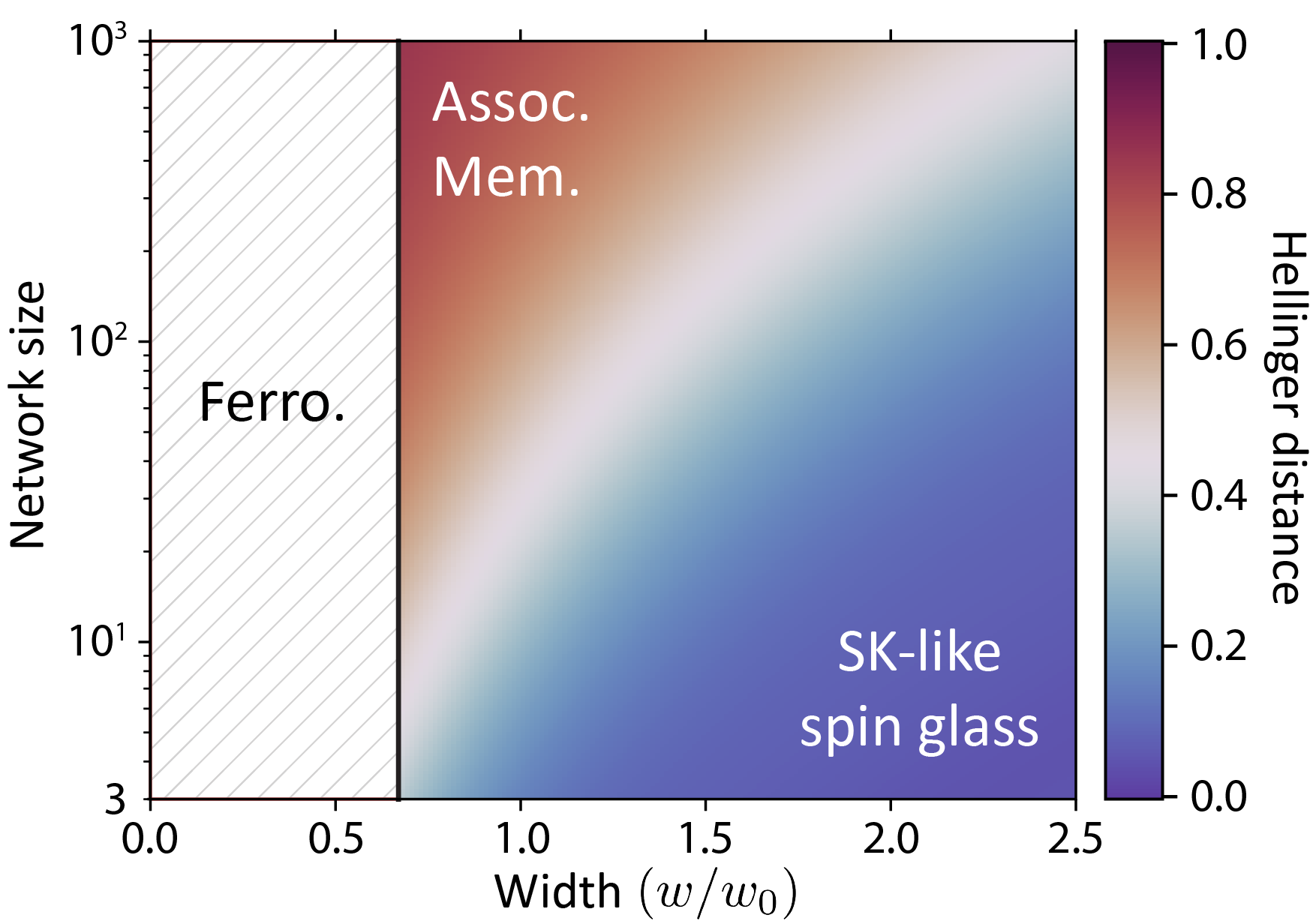}
    \caption{Regimes of CCQED connectivity behavior versus
    width $w$ and system size $N$.  The three shaded regions correspond to ferromagnetic (F), associative memory, and SK-like spin glass (SG) regimes, as discussed in the text.  Note that associative memories can still be encoded even in the spin glass regime due to SD dynamics; see Secs.~\ref{sec:SD} and~\ref{sec:learning}.}
    \label{fig:regimes}
\end{figure}

Figure~\ref{fig:regimes} summarizes the three regimes of the CCQED connectivity matrix.  The evolution from  ferromagnetic, to associative memory, to spin glass versus $w$ is analogous to the behavior exhibited by the Hopfield model as the ratio $P/N$ of the number of memory patterns to neurons increases in the Hebbian connectivity.  Having demonstrated that a significant number of metastable states exist in the cavity QED system, Sec.~\ref{sec:SD} will now discuss the natural dynamics of the cavity.  We will show that these differ from standard 0TMH or Glauber dynamics.  As mentioned, changing the dynamics changes the basins of attraction associated with each metastable state.  Remarkably, as we will show in Sec.~\ref{sec:SDHN}, the CCQED dynamics leads to a dramatic increase in the robustness of the associative memory by ensuring that many metastable states with sufficiently large basins for robust pattern completion exist, even at large $\tilde{w}$.


\section{Spin dynamics of the superradiant cavity QED system}
\label{sec:SD}

We now discuss the spin dynamics arising intrinsically for atoms pumped within an optical cavity.  To do this, we start from a full model of coupled photons and spins, and show how both the connectivity matrix discussed above and the natural dynamics emerge.   Aspects of these have been presented in earlier work; e.g., the idea of deriving an associative memory model from coupled spins and photons was discussed in Refs.~\cite{Gopalakrishnan:2011jx,Gopalakrishnan:2012cf}, with spin dynamics discussed in Ref.~\cite{Fiorelli:2020kd}.  Because---as we discuss in Sec.~\ref{sec:SDHN}---the precise form of the open-system dynamics is crucial to the possibility of memory recovery, we include a full discussion of those dynamics here to make this paper self contained.

Our discussion of the spin dynamics proceeds in several steps.  Here, we start from a model of atomic spins coupled to cavity photon modes; this is derived in Appendix~\ref{app:RamanScheme}.  We then discuss how to adiabatically eliminate the cavity modes in the regime of strong pumping and in the presence of dephasing, leading to a master equation for only the atomic spins.  This equation describes rates of processes in which a single atomic spin flips, and these rates include a superradiant enhancement, dependent on the state of other spins in the same ensemble.  Such dynamics can be simulated stochastically.  Finally, we show how, for large enough ensembles, a deterministic equation for the average magnetization of the ensemble can be derived. This equation is shown to describe a discrete form of steepest descent dynamics.

The system dynamics is given by the master equation:
\begin{equation}
    \dot \rho = -i[H,\rho]+\kappa\sum_m \mathcal{L}[a_m],
\end{equation}
where $a_m$ is the annihilation operator for the $m$th cavity mode and the Lindblad superoperator is $\mathcal{L}[X]=2X\rho X^\dag - \{X^\dag X,\rho \}$.  The dissipative terms describe cavity loss at a rate $\kappa$~\footnote{We assume that all modes considered decay at the same rate $\kappa$, which seems consistent with experimental observations up to at least mode indices of 50~\cite{Kollar:2014us,Vaidya2018}.}.  This is the only dissipative process we consider, as we assume the pump laser  is  sufficiently far detuned from the atomic excited state that we may ignore  spontaneous emission.  As noted earlier, spontaneous emission would  lead to heating, which sets a maximum duration of the experiment. As derived in Appendix~\ref{app:RamanScheme}, the Hamiltonian takes the form of a multimode generalization of the Hepp--Lieb--Dicke model~\cite{dicke54,hepp73,Garraway2011,kirton2019:Review}:
\begin{multline}\label{eqn:multimodeDicke}
    H =  \sum_m \big(-\Delta_m a^\dag_m a_m+\longfield_m(a^\dag_m+a_m)\big) +\transfield\sum_{i=1}^N S_i^z  \\
    + \sum_{i=1}^N\sum_{m}g_{im}S_i^x(a^\dag_m+a_m) + \noise(t)\sum_{i=1}^N S_i^x.
\end{multline}
The first two terms describe the cavity alone.  We work in the rotating frame of the transverse pump: $\Delta_m$ denotes the detuning of the transverse pump from the $m$th cavity mode.  We will assume that the transverse pump is red detuned, and so $\Delta_m <0$.  We also include a longitudinal pumping term, written in the transverse mode basis as $\longfield_m$. Such a pump allows one to input memory patterns (possibly corrupted) into the cavity QED system.

To describe the atomic ensembles, we introduce collective spin operators $S_i^\alpha=1/2\sum_{j=1}^{M}\sigma_j^\alpha$, where $\sigma_j^\alpha$ are the Pauli operators for the individual spins within the $i$th localized ensemble, which contains $M$ spins.  The term $\transfield S_i^z$ describes the bare level splitting between the atomic spin states.  The coupling between photons and spins is denoted $g_{im}=\Xi_{m}(\mathbf{r}_i)\Omega g_0 \cos(\phi_m)/\Delta_A$. This expression involves the transverse profile $\Xi_m(\mathbf{r})$ of the $m$th cavity mode and the effect from the Gouy phase $\cos(\phi_m)$; see Appendix~\ref{app:RamanScheme}.  For a confocal cavity, these are Hermite-Gaussian modes; their properties are extensively discussed elsewhere~\cite{siegman,Vaidya2018,Guo2019Sign,Guo2019Emergent}.

The final term in Eq.~\eqref{eqn:multimodeDicke} introduces a classical noise source $\noise(t)$.  In Appendix~\ref{app:classicalbath}, we show this coupling generates dephasing in the $S^x$ subspace so as to restrict the dynamics to classical states, simplifying the dynamics.  Such noise may arise naturally from noise in the Raman lasers. In addition, such a term can also be deliberately enhanced either through increasing such noise, or by introducing a microwave noise source oscillating around $\transfield$.  We choose to consider the dynamics with such a noise term to enable us to draw comparisons to other classical associative memories.  Without such a term, understanding the spin-flip dynamics would be far more complicated, as it would require a much larger state space, with arbitrary quantum states of the spins.  Exploring this quantum dynamics, when this noise is suppressed, is a topic for future work, as we discuss in Sec.~\ref{sec:conclusion}.

This multimode, multi-ensemble generalization of the open Dicke model exhibits a normal--to--superradiant phase transition, similar to that known for the single-mode Dicke model. The effects of multiple cavity modes have been considered for a smooth distribution of atoms~\cite{Gopalakrishnan09,Gopalakrishnan10}, where it was shown that beyond-mean-field physics can change the nature of the transition. In contrast, in Eq.~\eqref{eqn:multimodeDicke} we consider ensembles that are small compared to the length scale associated with the cavity field resolving power~\footnote{In other words, small with respect to the local interaction length scale, around a micron~\cite{Vaidya2018}.}, so all atoms in an ensemble act identically.
As such, in the absence of inter-ensemble coupling, the normal--to--superradiant phase transition occurs independently for each ensemble of $M$ atoms at the mean-field point $g_{i,\mathrm{eff}}=g_c$, where $4Mg_c^2=\transfield(\detune^2+\kappa^2)/\detune$, and $g_{i,\mathrm{eff}}$ is the effective coupling for ensemble $i$ to a cavity supermode (a superposition of modes coupled by the dielectric response of the localized atomic ensemble)~\cite{Kollar2017sm}. The normal phase is characterized by $\langle S^x \rangle=0$ and a rate $\propto{M}$ of coherent scattering  of pump photons into the cavity modes. In the superradiant phase, $\langle S^x \rangle\neq 0$, and the atoms coherently scatter the pump field at an enhanced rate $\propto M^2$. This is experimentally observable  via a macroscopic emission of photons from the cavity and a $\mathbb{Z}_2$ symmetry breaking reflected in the phase of the light (0 or $\pi$)~\cite{Baumann11,Kollar:2014us,Kroeze:2019ex}.

In the absence of coupling, each ensemble would independently choose how to break the $\mathbb{Z}_2$ symmetry, i.e., whether to point up or down. Photon exchange among the ensembles couples their relative spin orientation and modifies the threshold.
When all $N$ ensembles are phase-locked in this way, the coherent photon scattering into the cavity in the superradiant phase becomes $\propto (NM)^2$, in contrast to $\propto NM$ in the normal phase.  Throughout this paper, our focus will be on understanding the effects of photon exchange, when the system is pumped with sufficient strength that all the ensembles are already deep into the superradiant regime.  The behavior near threshold, and shifts to the threshold due to inter-ensemble interactions, are discussed again in Sec.~\ref{sec:conclusion}. We now describe how to consider the collective spin ensemble dynamics deep in the superradiant regime.

To obtain an atom-only description deep in the superradiant regime, it is useful to displace the photon operators by their mean-field expectations for a given spin state: $a_m \to a_m+(\longfield_m+\sum_i g_{im}S_i^x)/(\Delta_m+i\kappa)$.  This can be done via a Lang--Firsov polaron transformation~\cite{Lang1963a,Lang1963b},  as defined by the unitary operator:
\begin{equation}
    U = \exp\bigg[\sum_{i=1}^N\sum_m (\varphi_m+g_{im}S_i^x)\left(\frac{a_m^\dag}{\Delta_m+i\kappa}-\mathrm{H.c.}\right) \bigg].
\end{equation}
Note that this remains a unitary transformation even with the inclusion of cavity loss $\kappa$. The polaron transform changes both the Hamiltonian and the Lindblad parts of the master equation, redistributing terms between them. 
The transformed dissipation term remains $\kappa \sum_m \mathcal{L}[a_m]$, while the transformed Hamiltonian is:
\begin{multline}
    \label{eqn:PolaronTransformedH}
    \tilde H = -\sum_m \Delta_m a^\dag_m a_m +H_{\mathrm{Hopfield}}\\
    +\frac{\transfield}{2}\sum_{i=1}^N \big(i D_{F,i}S_i^-+\mathrm{H.c.}\big)+\noise(t)\sum_{i=1}^N S_i^x.
\end{multline}
We define polaron displacement operators as
\begin{equation}
  D_{F,i}
  =
  \exp\left[
     \sum_m g_{im} 
    \left(
      \frac{a^\dagger_m }{\Delta_m +  i\kappa} - \mathrm{H.c.}
    \right)
  \right],
\end{equation}
and $S_i^\pm=S_i^y\pm iS_i^z$ are the (ensemble) raising and lowering operators in the $S^x$ basis. The Hopfield Hamiltonian emerges naturally from this transform: 
\begin{equation}
    \label{eqn:HopfieldFromCavity}
    H_{\mathrm{Hopfield}} = -\sum_{i,j=1}^NJ_{ij}S_i^x S_j^x - \sum_{i=1}^N \spinfield_i S_i^x,
\end{equation}
with the connectivity and longitudinal field given by:
\begin{equation}
    J_{ij}=-\sum_{m}\frac{\Delta_m g_{im} g_{jm}}{\Delta_m^2+\kappa^2},\quad \spinfield_i = -2\sum_m \frac{\longfield_m\Delta_m g_{im}}{\Delta_m^2+\kappa^2}.
\end{equation}

With the Hamiltonian in the form of Eq.~\eqref{eqn:PolaronTransformedH}, we can now adiabatically eliminate the cavity modes by treating the term proportional to $\transfield$ perturbatively via the Bloch-Redfield procedure.  As derived in Appendix~\ref{app:classicalbath}, this yields the atomic spin-only master equation
    \begin{equation} \label{eqn:masterEqn}
        \dot \rho = -i[H_{\mathrm{eff}},\rho]
        +\sum_{i=1}^N \bigg( K_i(\dEi^+) \mathcal{L}[S_i^+] + K_i(\dEi^-) \mathcal{L}[S_i^-] \bigg).
    \end{equation}
In the above expression, $H_{\mathrm{eff}}$ is an effective Hamiltonian including both $H_{\mathrm{Hopfield}}$ and a Lamb shift contribution~\footnote{By which we mean, the energy shifts due to the coupling of the spins to the continuum of modes in the bath~\cite{Breuer2002}.}, $K_i(\dE)$ is a rate function discussed below, and $\dEi^+$  ($\dEi^-$) are the changes in energy of $\ham$ after increasing (decreasing) the $x$ component of collective spin in the $i$th ensemble.  Note that although this expression involves ensemble raising and lowering operations, the master equation describes processes where the spin increases or decreases by one unit at a time.  For brevity, we refer to these processes as spin flips below. Because these are collective spin operators, there will be a ``superradiant enhancement'' of these rates, as we discuss below in Sec.~\ref{sec:unravelling}.

As mentioned above, classical noise dephases the quantum state into the $S^x$ subspace in which each ensemble exists in an $S_i^x$ eigenstate. By doing so, the  state may be described by the vector $(s^x_1,s^x_2,\cdots,s^x_3)$, where $s^x_i$ is the $S_i^x$ eigenvalue of the $i$th spin ensemble.  Since $H_{\mathrm{eff}}$ commutes with $S^x$,  it generates no dynamics in this subspace. The dynamics thus arises solely through the dissipative Lindblad terms, corresponding to incoherent $S^x$ spin-flip events.

The energy difference, upon changing the spin of the $i$th ensemble by one unit $\pm 1$, can be explicitly written as
\begin{equation}
    \label{eqn:flipcost}
    \dEi^\pm = -J_{ii}\mp 2 \sum_{j=1}^N J_{ij}s_j^x.
\end{equation}
For the terms with $j\neq i$, these represent the usual spin-flip energy in a Hopfield model. An additional self-interaction term $J_{ii}(\mp 2 s_i^x +1)$ arises from the energy cost of changing the overall spin of the $i$th ensemble.  The self-interaction $J_{ii}$ thus provides a cost for the $i$th spin to deviate from $s_i^x=\pm S_i$, where $S_i=M/2$ is the modulus of spin of ensemble $i$. As written in Eq.~\eqref{eqn:JijForm}, $J_{ii}$ is enhanced by a term $\beta$, dependent on the size of the atomic ensembles. If sufficiently large, this could freeze all ensembles in place, by making any configuration with all $s_i^x=\pm S_i$ a local minimum.  However, the strength of $J_{ii}$ can be reduced by tuning slightly away from confocality, which smears-out the local interaction~\cite{Vaidya2018}.  We will see below that for the realistic parameters employed in Sec.~\ref{sec:ccQED}, no such freezing is observed.  It is important to note that the largest self-interaction energy cost occurs at the first spin-flip of a given ensemble---i.e., subsequent spin flips become easier not harder. One may also note that as the size of ensemble $M$ increases, the interaction strength ratio of the self-interaction to the interaction from other ensembles is not affected; Eq.~\eqref{eqn:flipcost} shows all terms increase linearly with ensemble size.

As derived in Appendix~\ref{app:classicalbath}, the functions that then determine the rates of spin transitions take the form:
\begin{multline}\label{eqn:spinFlipRates}
  K_i(\dE)
  = 
  \frac{\transfield^2}{8}\, \Re 
  \int_0^\infty 
  d \tau \exp\Big[ -i \dE\tau-C(\tau) \\
  - \sum_m \frac{ g_{im}^2}{(\Delta_m^2+\kappa^2)}\big(1- e^{-(\kappa-i\Delta_m)\tau}\big)\Big],
\end{multline}
where the function $C(\tau)$ depends on the correlations of the classical noise source $\noise(t)$.  It does so via
$J_c(\omega)$, the Fourier transform of $\langle \noise(t) \noise(0)\rangle$, using
\begin{equation}
C(\tau) = \int_0^\infty \frac{d\omega}{2\pi} \frac{J_c(\omega)}{\omega^2}\sin^2\left(\frac{\omega \tau}{2}\right).
\end{equation}
Details of the derivation of this expression are given in Appendix~\ref{app:classicalbath}, along with explicit calculations for an ohmic noise source. 

\subsection{Spin-flip rates in a far-detuned confocal cavity}
\label{sec:confocal_dynamics}

We now show how the expressions for the spin-flip rate $K(\dE)$ simplify for a degenerate, far-detuned confocal cavity. In an ideal confocal cavity, degenerate families of modes exist with all modes in a given family having the same parity.  Considering a pump near to resonance with one such family, we may restrict the mode summation to that family, and take  $\Delta_m=\detune$ for all modes $m$. In this case, the sum over modes in Eq.~\eqref{eqn:spinFlipRates} simplifies, and becomes proportional to $J_{ii}$.  If we further assume $\detune\gg J_{ii}$---the far-detuned regime---we can Taylor expand the exponential in Eq.~\eqref{eqn:spinFlipRates} to obtain the simpler spin-flip rate function
\begin{equation}\label{eqn:confocalRate}
        K_i(\dE) =  h(\dE)
        +\frac{ e^{-J_{ii}/|\detune|}J_{ii}\transfield^2\kappa  }{8|\detune|[(\dE-\detune)^2+\kappa^2]},
\end{equation}
where $h(\dE)$ is a sharply peaked function centered on $\dE=0$.  Its precise form depends on the spectral density of the noise source and is given explicitly in Appendix~\ref{app:classicalbath}.  Classical noise broadens $h(\dE)$ into a finite-width peak. Considering experimentally realistic parameters, this width is at least an order--of--magnitude narrower than the range of typical spin-flip energies.  As a result, its presence does not significantly affect the dynamics. The main contribution to the spin-flip rate thus comes from the second term, a Lorentzian centered at $\dE=\detune$. 

\begin{figure}
    \centering
    \includegraphics[width=\columnwidth]{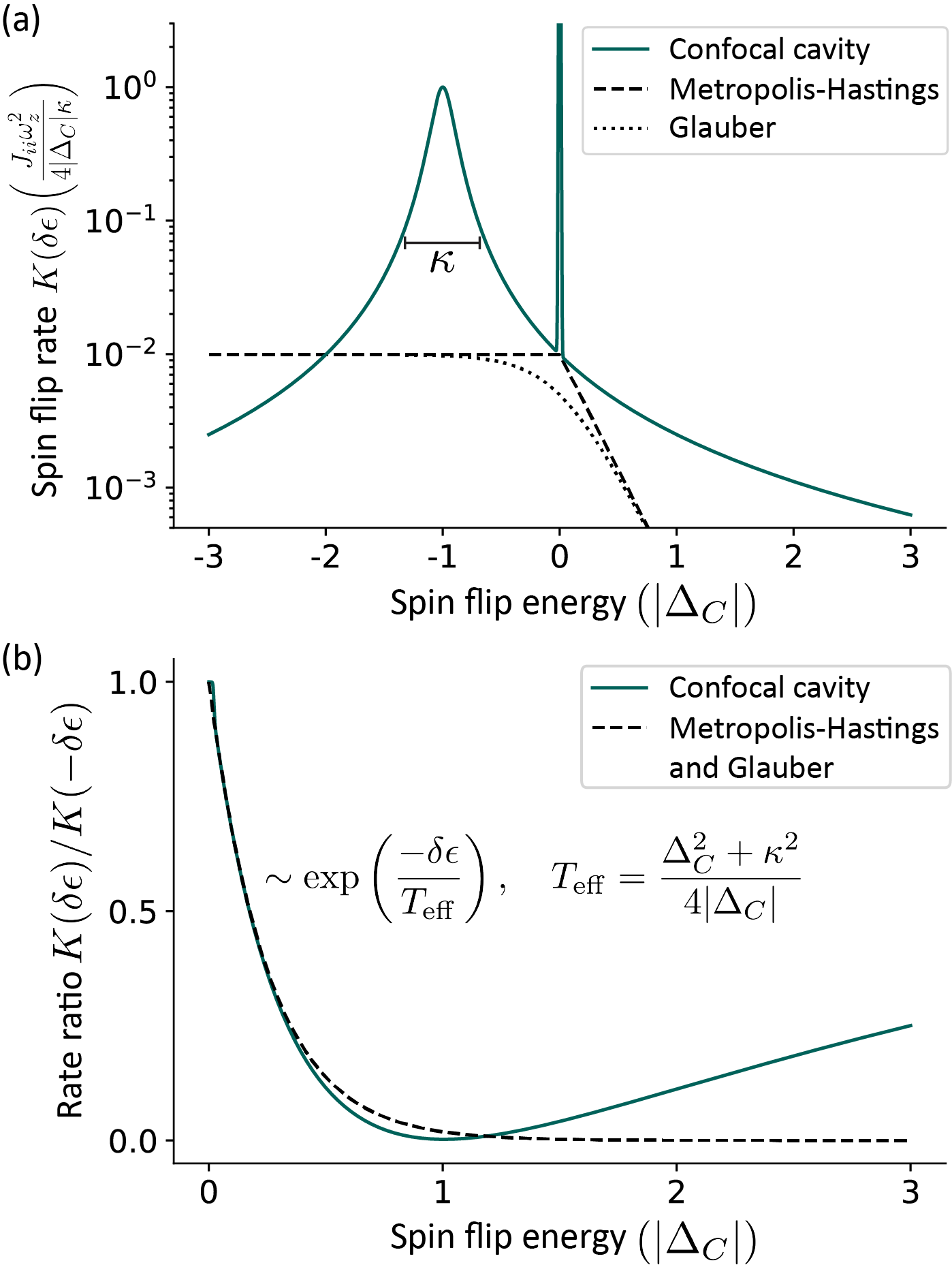}
    \caption{(a) The spin-flip rate function $K(\dE)$ in the confocal limit (solid line) versus spin-flip energy $\delta\epsilon$. Units are in terms of the pump--cavity detuning $|\detune|$. Rates for Metropolis-Hastings (dashed line) and Glauber (dotted line) dynamics are also shown. (b) Ratio of energy-decreasing to energy-increasing spin-flip rates $K(\pm\dE)$ versus spin-flip energy (solid line). Comparing  finite-temperature Metropolis Hastings dynamics (dashed line) to the actual cavity dynamics allows one to identify a low-energy effective temperature $T_{\mathrm{eff}}= (\detune^2+\kappa^2)/4|\detune|$~\cite{Torre13,Maghrebi2016,kirton2019:Review}. }
    \label{fig:rateFunc}
\end{figure}

Figure~\ref{fig:rateFunc}(a) plots the spin-flip rate of Eq.~\eqref{eqn:confocalRate}. By choosing negative $\detune$---i.e., red detuning---the negative offset of the Lorentzian peak from $\dE=0$ ensures that energy-lowering spin-flips occur at a higher rate than energy-raising spin-flips, thereby generating cooling dynamics. We can further define an effective temperature for the dynamics for spin-flip energies sufficiently small in magnitude.  To show this, we inspect the ratio of energy-lowering to energy-raising spin-flip rates $K(\dE)/K(-\dE)$. To obey detailed balance---as would occur if coupling to a thermal bath---this ratio must be of the form $\exp(-\dE/T)$, where $T$ is the temperature of the bath.  For small $|\dE|$, this means one should compare the rate ratio to a linear fit:
\begin{equation}\label{eqn:teff}
        \frac{K_i(\dE)}{K_i(-\dE)}\approx 1 -\frac{\dE}{T_{\mathrm{eff}}},
        \qquad T_{\mathrm{eff}} = \frac{\detune^2 + \kappa^2}{4|\detune|}.
\end{equation}
To determine whether this temperature is large or small, it should be compared to a typical spin-flip energy $\dE$: the system is hot (cold) when the ratio $T_{\mathrm{eff}}/\dE$ is much greater (less) than unity.  The ratio of spin flip rates is shown in Fig.~\ref{fig:rateFunc}(b), along with the Boltzmann factor using $T_{\mathrm{eff}}$ defined in Eq.~\eqref{eqn:teff}.  The rate ratio shows a small kink close to $|\dE|=0$ arising from the noise term, but otherwise closely matches the exponential decay up until $|\dE|\simeq|\detune|$.  We may ensure the spin-flip energies are $\leq|\detune|$ by choosing a pump strength $\Omega \propto \detune/\sqrt{M}$.  The spin dynamics drive the system toward a thermal-like state in this regime.

Despite the presence of an effective thermal bath, the dynamics arising from the confocal cavity are rather different from those of the standard finite-temperature Glauber or Metropolis-Hastings dynamics.  This can be seen by comparing the functions $K(\dE)$ that would correspond to these dynamics.
Glauber dynamics implements a rate function of the form
\begin{equation}
    K_{\mathrm{Glauber}}(\dE) \propto \frac{\exp(-\dE/T)}{1+\exp(-\dE/T)}, 
\end{equation}
while for Metropolis-Hastings,
\begin{equation}
    K_{\mathrm{MH}}(\dE) \propto \begin{cases}
    1&\dE\leq 0 \\
    \exp(-\dE/T)&\dE>0
    \end{cases}.
\end{equation}
The zero temperature limits of both these functions are step functions; e.g.,  $K_{\mathrm{0TMH}}\propto\Theta(-\dE)$. These functions plotted  in Fig.~\ref{fig:rateFunc}(a), taking $T=T_{\mathrm{eff}}$ with an arbitrary overall rescaling to match the low-energy rate of the cavity QED dynamics. The cavity QED dynamics exhibits an enhancement in the energy-lowering spin-flip rates peaked at $\detune$. By contrast, the rate function for Metropolis-Hastings dynamics is constant for all energy lowering spin-flips.  It is  nearly constant at low temperatures for Glauber dynamics. In comparison, the cavity QED dynamics specifically favors those spin flips that dissipate more energy---the cavity allows these spins to flip at a higher rate.  As we will see in Sec.~\ref{sec:SDHN}, this ``greedy'' approach to steady state significantly changes the basins of attraction of the fixed points. 

The magnitude of $T_{\mathrm{eff}}$ is a potential  problem if we were to consider single atoms rather than ensembles of atoms.  In such a case,  $J_{ij}/T_{\mathrm{eff}} \simeq g^2 \detune^2 / (\detune^2 + \kappa^2)^2$.  Since typical parameters satisfy $g \ll \detune, \kappa$~\cite{Kollar:2014us}, this ratio implies that the system lies within a high-temperature regime.   However, this need not be the case because we can employ ensembles of identical spins as our nodal element in the network. As seen from Eq.~\eqref{eqn:flipcost}, if all other ensembles are assumed to have $s_i^x=\pm M/2$, then the relevant energy scale for a spin flip $\dEi$ is enhanced by the number of spins in the ensemble. This allows one to reach  $\dEi/T_{\mathrm{eff}} \propto M \gg 1$---i.e., a low-temperature regime by increasing $M$, the number of atoms per ensemble.  The assumption that all ensembles are fully polarized is well founded:  As noted earlier, the on-site interactions $J_{ii}$ drive the spins within an ensemble to align as if the system were composed of rigid (easy-axis) Ising spins. Moreover, the spin ensemble spin-flip rates are superradiantly enhanced, meaning that the time duration of any ensemble flip is short and $\propto 1/M$.  As described below, the superradiant enhancement also reduces the timescale required to reach equilibrium. As we discuss in Sec.~\ref{sec:conclusion}, to reach the quantum regime would require us to consider $M \simeq 1$, which in turn requires enhancements of the single-atom cooperativity so that the low-temperature regime may be reached at the level of single atoms per node.

\subsection{Stochastic unravelling of the master equation}\label{sec:unravelling}

To directly simulate Eq.~\eqref{eqn:masterEqn}, we make use of a standard method for studying the time evolution of a master equation: stochastic unraveling into quantum trajectories~\cite{gardiner2004quantum,Daley2014}. In this method, the unitary Hamiltonian dynamics are punctuated by stochastic jumps due to the Lindblad operators. For our time-local master equation, the dynamics realizes a Markov chain. That is, the evolution of the state depends solely on the current spin configuration and the transition probabilities described by the spin-flip rates. Moreover, the transition probabilities are the same for all spins within a given ensemble.

A stochastic unraveling proceeds as follows. An initial state is provided as a vector $\mathbf{s}=(s_1^x,s_2^x,\cdots,s_N^x)$. The Lindblad terms in the master equation Eq.~\eqref{eqn:masterEqn} describe the total rates at which spin flips occur. However, the total spin-flip rates additionally experience a superradiant enhancement from the matrix elements of $S_i^\pm$. The collective spin-flip rates within an ensemble are thus
\begin{equation}
    [S_i(S_i+1)-s^x_i(s^x_i\pm1)]\,K_i(\dEi^\pm),
\end{equation}
where the upper (lower) sign indicates the up (down) flip rate.  As such, the rate of transitions within a given ensemble increases as that ensemble begins to flip, reaching a maximum when $s^x_i=0$ and then decreases as the ensemble completes its orientation switch. The ensemble spin-flip rates are computed for each ensemble at every time step. To determine which spin is flipped, waiting times are sampled from an exponential distribution using the total spin-flip up and down rate for each ensemble. The ensemble with the shortest sampled waiting time is chosen to undergo a single spin flip. The time in the simulation then advances by the waiting time for that spin flip. The process then repeats.  This requires the total rates to be recomputed at every time step for the duration of the simulation. Results of such an approach are shown in Fig.~\ref{fig:ensemble_dynamics}(a).

\subsection{Deterministic dynamics for large spin ensembles}\label{sec:ensembles}

A full microscopic description of the atomic spin states becomes unwieldy when considering large numbers of atoms per ensemble. Fortunately, the large number of atoms also means the full microscopic dynamics becomes unnecessary for describing the experimentally observable quantities. The relevant quantity describing an ensemble is not the multitude of microscopic spin states for each atom, but the net macroscopic spin state of the ensemble. We can therefore build upon the above treatment to produce a deterministic macroscopic description of the dynamics: Our approach is to construct a mean-field description that is individually applied to each ensemble. This description becomes exact in the thermodynamic limit $S_i\to\infty$ and faithfully captures the physics of ensembles with $\geq{10^3}$ atoms under realistic experimental parameters. 

We begin by defining the macroscopic variables we use to describe the system. These are the normalized magnetizations of the spin ensembles $m_i\equiv \langle S_i^x \rangle / S_i\in [-1,1]$ that depend on the constituent atoms via only their sum. The $m_i$ remain of $O(1)$ independent of $S_i$, while fluctuations due to the random flipping of spins within the ensemble scale $\propto 1/\sqrt{S_i}$. The fluctuations about the mean-field value thus become negligible at large $S_i$, and the $m_i$ follow deterministic equations of motion. Following the derivation in Appendix~\ref{app:mean-field} (see also~\cite{Fiorelli:2020kd}), we find that the $m_i$ follow the coupled differential equations   
\begin{multline}
\label{eqn:MFODE}
        \frac{d}{dt}m_i = \mathrm{sgn}(\localfield_i) S_i \big|K_i(\dEi^+)-K_i(\dEi^-)\big|(1-m_i^2) \\
        + K_i(\dEi^+)(1-m_i) -K_i(\dEi^-)(1+m_i),
\end{multline}
where $\localfield_i=\sum_j J_{ij} S_j m_j$ is the local field experienced by the $i$th ensemble, given the configuration of the other ensembles. The equations are coupled because, as defined above, each $\dEi^\pm=-J_{ii}\mp2\sum_j J_{ij} S_j m_j$ depends on all the other $m_j$. Note that the self-interaction cost to flipping an atomic spin is largest for the first spin that flips within a given ensemble.  

The term proportional to $S_i$ in Eq.~\eqref{eqn:MFODE} describes the superradiant enhancement in the spin-flip rate. This term would be zero if $S_i=1/2$, since $m_i =s_i^x/S_i = \pm 1$ in that case. For large ensembles this term acts to rapidly align the ensemble with the local field when $|m_i|<1$. The other terms act to provide the initial kick away from the magnetized $|m_i|=1$ states, but receive no superradiant enhancement in the spin-flip rate. A separation of time scales emerges between the rapid rate at which an ensemble flips itself to align with the local field, described by the superradiant term, and the slower rate at which an ensemble can initiate a flip driven by the other terms. The dynamics that emerge correspond to periods of nearly constant magnetizations $|m_i|=1$, punctuated by rapid ensemble-flipping events $m_i\to -m_i$.

To gain analytical insight into the equations of motion, we make use of the separation of timescales that emerges in the large $S_i$ limit. When the ensembles are not undergoing a spin-flip event, they are in nearly magnetized states $|m_i|\approx 1$ and can be approximated as constants. The spin-flip energies $\dEi^\pm$ and rates $K_i(\dE)$ then become constant as well.  This enables one to decouple the equations of motion Eqs.~\eqref{eqn:MFODE} to consider the flip process of a single ensemble.  This in turn provides a simple analytical solution for the time evolution of the flipping magnetization,
\begin{equation}
    m_i(t) = \mathrm{sgn}(\localfield_i)\tanh\left[S_i\left|K_i(\dEi^+)-K_i(\dEi^-)\right|\left(t-t_0^i\right) \right],
\end{equation}
where the constants $t_0^i$ are determined from the initial conditions to be
\begin{equation}
    \label{eqn:fliptime}
    t_0^i = -\frac{\mathrm{sgn}(\localfield_i m_i)\log(8S_i)}{8S_i\left|K_i(\dEi^+)-K_i(\dEi^-)\right|}
\end{equation}

\begin{figure}[t!]
    \centering
    \includegraphics[width=\columnwidth]{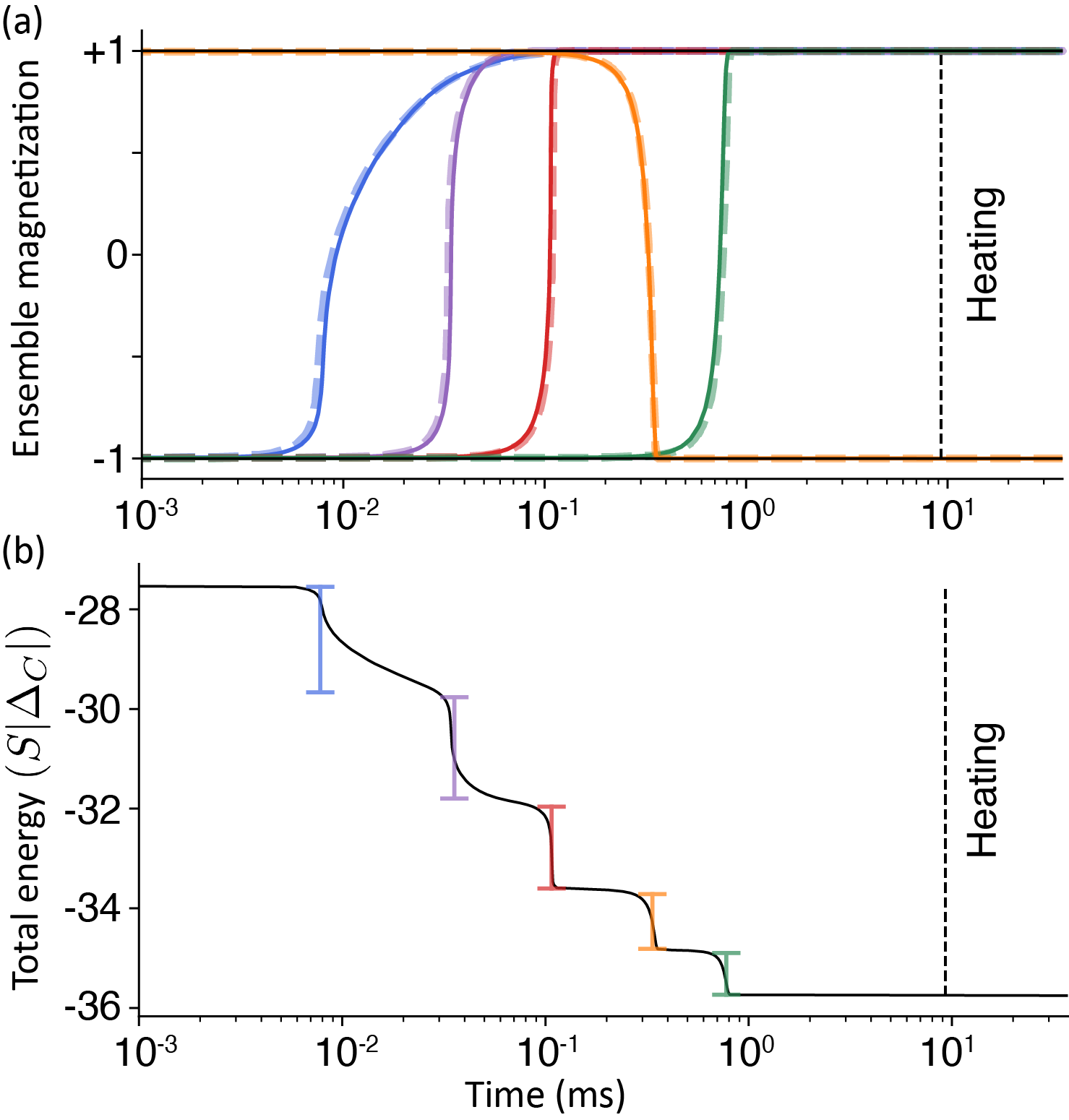}
    \caption{(a) Simulation of a system of $N=\text{100}$ ensembles.  All are initialized in a local minimum state except five ensembles that are flipped in the wrong orientation. Each ensemble is composed of $M=2S=10^5$ spins. A comparison is made between dynamics derived from the equations of motion in the thermodynamic limit (dashed lines) and a full stochastic unraveling of the master equation (solid lines). The five traces show relaxation of those five spin ensembles to the local minimum state. The other $N-3$ spin ensembles remain magnetized at $\pm1$ and are plotted in black. Data from the stochastic master equation simulation lie close to that of the equation of motion. The timescale associated with spontaneous decay (and thus heating) is marked by a dotted vertical line; the system reaches steady state beforehand. All parameters are as listed in Sec.~\ref{sec:ccQED} and $\omega_z=1$~MHz.  (b) The total energy versus time, in units of $S|\detune|$. As expected of discrete steepest descent dynamics, steps are clearly visible and they occur in order of the energy removed per spin flip.}
    \label{fig:ensemble_dynamics}
\end{figure}

These approximate solutions to the equations of motion accurately predict that ensembles will flip to align with their local field, and the ordering of the $t_0^i$ predicts the order in which the ensembles will flip. Ensembles that are already aligned with the local field have a $t_0^i<0$ and will not flip in the future unless the local field changes. On the other hand, ensembles not aligned with the local field have  a $t_0^i>0$ and will flip. The ensemble with the smallest $t_0^i>0$ will flip first. This leads to a discrete version of \emph{steepest descent} dynamics, since Eq.~\eqref{eqn:fliptime} implies that the smallest $t_0^i$ corresponds to the largest energy change $|\dEi|$. One may approximate all $m_i$ as constant up until the vicinity of the first $t_0^i$, at which point the $i$th spin flips and alters the local fields. The approximation then holds again for the new local fields until the next spin-flip event occurs. This process continues until convergence is achieved once all ensembles align with their local field. Overall, the mean-field dynamics takes a simple form: at every step, the system determines which ensemble would lower the energy the most by realigning itself, then realigns that ensemble in a collective spin-flip, continuing until all $\dEi>0$. The final configuration is a (meta)stable state corresponding to the minimum (or local minimum) of the energy landscape described by $\ham$.

Figure~\ref{fig:ensemble_dynamics}(a) shows a typical instance of the dynamics described by the equations of motion in Eqs.~\eqref{eqn:MFODE} and compares it to a stochastic unraveling of the master equation in Eq.~\eqref{eqn:masterEqn}. The $10^7$ spins are divided into 100 ensembles, each representing an $S=10^5/2$ collective spin. We use a $J_{ij}$ matrix constructed from the CCQED connectivity with spin distribution width $w=1.5w_0$. The ensemble magnetizations are initialized close to a local minimum configuration. Specifically, five ensembles are chosen at random and misaligned with their local field while the rest are aligned with their local field to specify the initial condition. We see that the five initially misaligned ensembles realign themselves to their local fields.  Their convergence to the local minimum state described by the $J_{ij}$ matrix occurs before the timescale set by spontaneous emission. We also see that the mean-field equations of motion closely match the stochastic unraveling. 

We confirm that these dynamics are consistent with steepest descent (SD) by considering the evolution of energy, as shown in  Fig.~\ref{fig:ensemble_dynamics}(b). The total energy of the spin ensembles monotonically decreases, with clearly visible steps corresponding to spin-flip events. These steps occur in order of largest decrease in energy. The effect of SD dynamics on the robustness of stored memories, and more generally on the nature of basins of attraction in associative memory networks, has remained unexplored to date.  We address this question in the next section.


\section{Implications of steepest decent dynamics for associative memory}\label{sec:SDHN}

In the previous section, we found that for large-spin ensembles, the cavity-induced dynamics is of a discrete ``steepest descent'' form; i.e., at each time-step the next spin-flip event of an ensemble is that which lowers the energy the most. (This is in contrast to 0TMH dynamics where spins are flipped randomly provided that the spin-flip lowers the energy.) We now explore how changing from 0TMH to SD dynamics affects associative memories. To understand the specific effects of the dynamics, in this section we consider both ``standard" Hopfield neural networks with connectivity matrices $J_{ij}$ drawn from Hebbian and pseudoinverse learning rules and Sherrington--Kirkpatrick (SK) spin glasses. We find that SD not only improves the robustness and memory capacity of Hopfield networks, but also gives rise to extensive basins of attraction even in the spin glass regime. An introduction to Hopfield neural networks was given in Section~\ref{sec:HModel}. 

We will compare 0TMH dynamics, where any spin flip that lowers the energy of the current spin configuration is equally probable versus SD dynamics, where the spin flip that lowers the energy the most always occurs. Steepest descent is a deterministic form of dynamics, while 0TMH is probabilistic, and so fixed points can be expected to be more robust under SD.  We also note that natural SD dynamics, such as that exhibited by the pumped cavity QED system, is more efficient than simulated SD dynamics (given an equal effective time step duration).  This is because numerically checking all possible spin flips to determine which provides the greatest descent is an $\mathcal{O}(N^2)$ operation.  Thus, interestingly, the natural steepest descent dynamics effectively yields an $\mathcal{O}(N^2)$ speed-up over simulations, by effectively computing a maximum operation over $N$ local fields (each of which is computed via a sum over $N$ spins) in $\mathcal{O}(1)$ time. 

\subsection{Enhancing the robustness of classical associative memories through steepest descent}

While the locations in configuration space of the local energy minima of Eq.~\eqref{eqn:H_Hopfield}, or equivalently the fixed point attractor states of Eq.~\eqref{update}, are entirely determined by the connectivity $J_{ij}$, the basins of attraction that flow to these minima depend on the specific form of the energy minimizing dynamics. To quantify the size of these basins, we require a measure of distance between states. A natural distance measure is the Hamming distance, defined as follows. Consider two spin states $\mathbf{s}$ and $\mathbf{s^\prime}$ corresponding to $N$-dimensional binary vectors with elements $s_i=\pm1$. The Hamming distance between $\mathbf{s}$ and $\mathbf{s^\prime}$ is defined as $d = \sum_{i=1}^N |s_i - s^\prime_i|/2$. Thus, the Hamming distance between an attractor memory state and another initial state simply counts the number of spin-flip errors in the initial state relative to the attractor. This notion of Hamming distance enables us to determine a memory recall curve for any given attractor state under any particular energy minimizing dynamics.  We first pick a random initial state at a given Hamming distance $d$ from the attractor state, by randomly flipping $d$ spins. We then check whether it flows back to the original attractor state under the given dynamical scheme. If so, a successful pattern completion, or memory recall event, has occurred.  We compute the probability of recall by computing the fraction of times we recover the original attractor state over random choices of $d$ spin flips from the initial state.  This yields a memory recall probability curve as a function of $d$.  We define the size of the basin of attraction under the dynamics to be the maximal $d$ at which this recall curve remains above $0.95$.

\begin{figure}[t!]
    \centering
    \includegraphics[width=\columnwidth]{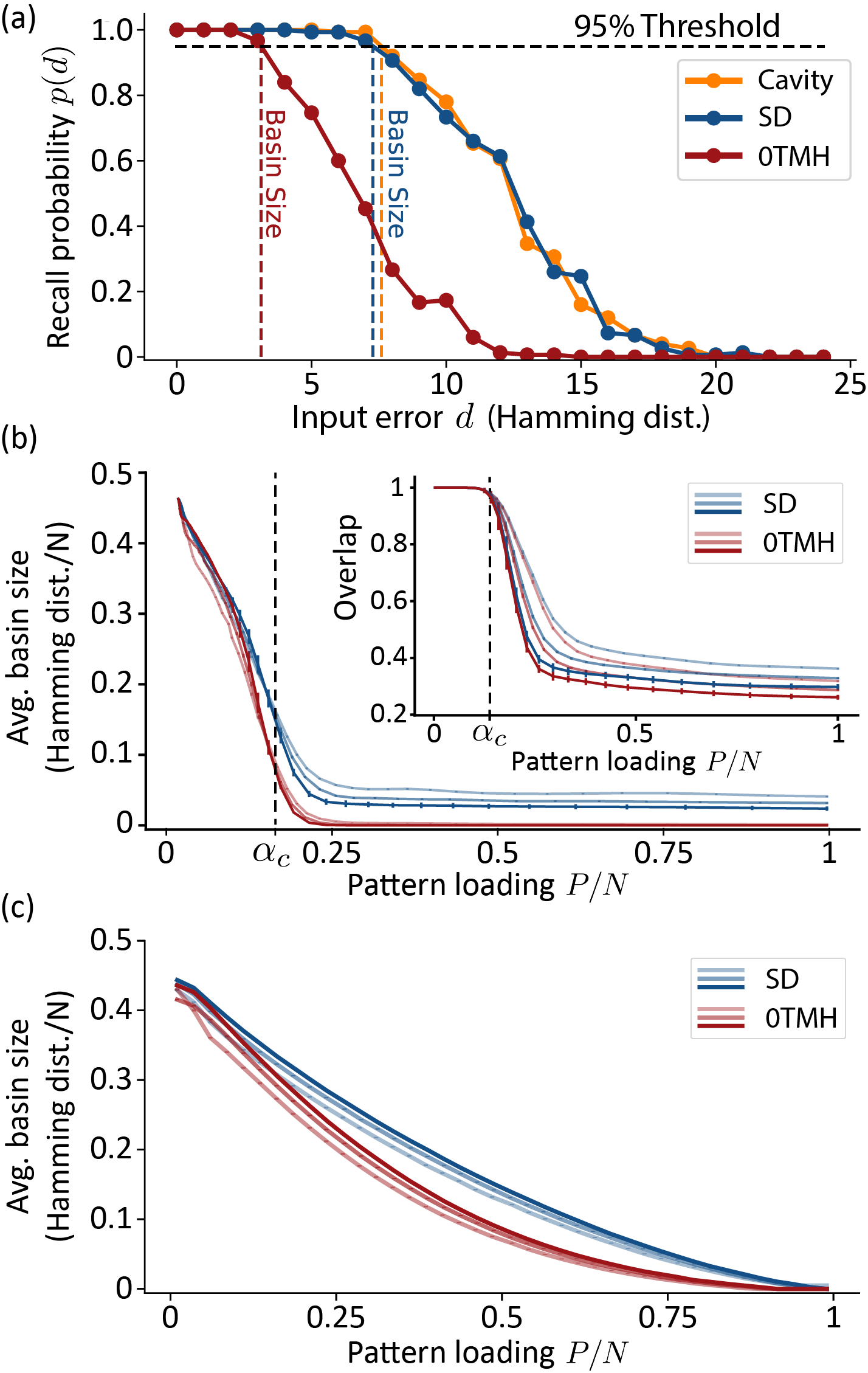}
    \caption{(a) Recall probability as a function of the input error applied, quantified by Hamming distance between the input and pattern state.   This is plotted for three forms of dynamics: 0TMH (red), SD (blue), and CCQED (orange).  The latter is given by Eq.~\eqref{eqn:MFODE}.  This illustrates how the dynamics affects recall probability, and thus basin size (defined by the point where the recall curve drops below 95\%). We use a connectivity matrix $J_{ij}$ corresponding to a pseudoinverse Hopfield model with $N=100$ spins and $P=0.6N$ patterns stored.  We choose this learning rule because it illustrates a large difference between the two dynamics.  (b) Average basin size of attractors for connectivity of a Hebbian learning rule.  The intensity of the trace increases with system size, $N=200, 400, 800$, while color indicates the form of dynamics. The critical ratio $\alpha_c\approx0.138$ marks the known pattern loading threshold where Hebbian learning fails in the thermodynamic limit~\cite{Amit1985}. Inset:  overlap between the attractor of the dynamics $\mathbf{s}^\mu$ and the bare memory $\boldsymbol{\xi}^\mu$, as discussed in the text.  (c) Average basin size for the pseudoinverse learning rule versus the ratio of patterns stored to number of nodes, $P/N$. Steepest descent dynamics outperforms 0TMH in all cases, yielding larger basin sizes for $P/N\agt0.1$, and thus, greater memory capacity.}
    \label{fig:SDHopfieldNets}
\end{figure}

Figure~\ref{fig:SDHopfieldNets}(a) shows memory recall curves, basin sizes, and their dependence on the form of the energy minimizing dynamics for a Hopfield network trained using the pseudoinverse learning rule; see Eq.~\eqref{pseudoheb}. Three different forms of dynamics are shown: 0TMH, SD, and the CCQED dynamics of Eqs.~\eqref{eqn:MFODE}.  As expected, the CCQED dynamics closely match the SD dynamics. The basin size depends on dynamics: SD and CCQED dynamics lead to a larger basin of attraction than 0TMH.  Indeed, the entire memory recall curve is higher for SD than 0TMH.

This increase in the robustness of the memory, at least for small $d$, leads directly to an increase in basin size. To understand this, consider the following argument.  Imagine the Hamming surface of $2^{\binom{N}{d}}$ configurations a distance $d$ from a fixed point at the center of the surface. The recall curve $p(d)$ is the probability the dynamics returns to the fixed point at the center of the surface when starting from a random point on the surface. Under 0TMH, due to the stochasticity of the random choice of spin flip that the lowers energy, many individual configurations on the Hamming ball could flow to multiple different fixed points, thereby lowering the probability $p(d)$ for returning to the specific fixed point at the center of the Hamming surface. However, under the deterministic SD dynamics, each point on the Hamming surface must flow to one and only one fixed point. Of course, many points on the Hamming surface could, in principle, flow under SD to a different fixed point other than the fixed point at the center.  But, as verified in simulations (not shown), for small enough $d$, more points on the Hamming surface flow under SD to the central fixed point than under 0TMH.  As such, recall probability increases for small $d$, as indeed shown in Fig.~\ref{fig:SDHopfieldNets}(a). This deterministic capture of many of the states on the Hamming surface by the central fixed point effectively enhances the robustness of the memory.
 
We can study how the SD and 0TMH dynamics affect the dependence of basin size on the number of patterns stored, and hence the memory capacity.  We separately consider the two traditional learning rules,  Hebbian  and pseudoinverse. In both cases, we expect a trade-off between the number of patterns $P$ stored and the basin size, with the latter shrinking as the former increases. Figures~\ref{fig:SDHopfieldNets}(b,c) demonstrate this trade-off for both learning rules.  However, in both cases, switching from 0TMH to SD ameliorates this trade-off; at any level of memory load, the average basin size increases under SD versus 0TMH. 

To calculate each point, we generate $P$ random desired memory patterns $\boldsymbol{\xi}^\mu$ in order to construct the $J_{ij}$ matrix according to the learning rule under consideration. For each pattern, we determine its associated attractor state $\mathbf{s}^\mu$. In the case of the pseudoinverse learning rule, the attractor state associated with a desired memory $\boldsymbol{\xi}^\mu$ is simply identical to the desired memory. However, in the Hebbian rule, the associated attractor state $\mathbf{s}^\mu$ is merely close to the desired memory $\boldsymbol{\xi}^\mu$.  We find this attractor state $\mathbf{s}^\mu$ by initializing the network at the desired memory $\boldsymbol{\xi}^\mu$ and flowing to the first fixed point under 0TMH. The inset in Fig.~\ref{fig:SDHopfieldNets}(b) plots the overlap $N^{-1} \sum_i \xi^\mu_i s^\mu_i$ as a function of the pattern loading $\alpha=P/N$ of the Hebbian model. This overlap is close to $1$ for $\alpha<\alpha_c \approx 0.138$ and drops beyond that, indicating that beyond capacity, the Hebbian rule cannot program fixed points close to the desired memories. 

We focus specifically on the fixed points $\mathbf{s}^\mu$ to dissociate the issue of programming the locations of fixed points $\mathbf{s}^\mu$ close to the desired memories $\boldsymbol{\xi}^\mu$ from those of examining the basin size of existing fixed points and the dependence of this basin size on the dynamics.   To determine the basin size for these fixed points, we compute the maximal input error for which the recall probability remains above $95\%$.  This is performed for both learning rules under both 0TMH and SD dynamics.  We then average the recovered basin size over the $P$ patterns.

Figure~\ref{fig:SDHopfieldNets}(b) presents the results for Hebbian learning, and SD dynamics appears to increase the basin size for pattern-loading ratios $P/N>0.1$. Note that the basin size under 0TMH is not extensive in the system size $N$ beyond the capacity limit $P/N=\alpha_c\approx 0.138N$. However, remarkably, under SD the basin size of the fixed points $\mathbf{s}^\mu$ \textit{are} extensive in $N$, despite the fact that the Hopfield model is in a spin glass phase at this point. Thus, the SD dynamics can dramatically enlarge basin sizes compared to 0TMH, even in a glassy phase. However, this enlargement of basin size does not by itself constitute a solution to the problem of limited associative memory capacity, because it does not address the programmability issue; above capacity, the fixed points $\mathbf{s}^\mu$ are not close to the desired memories $\boldsymbol{\xi}^\mu$; see Fig.~\ref{fig:SDHopfieldNets}(b) inset. Steepest descent can only enlarge basins, not change their locations. As such, for Hebbian learning, the memory capacity $C\approx0.138N$ cannot be enhanced under any choice of energy minimizing dynamics unless an additional programming step is implemented. 

In contrast to Hebbian learning, the pseudoinverse learning rule does not have a  programmability problem by construction. The connectivity possesses a fixed point $\mathbf{s}^\mu$ identical to each desired memory $\boldsymbol{\xi}^\mu$.  Figure~\ref{fig:SDHopfieldNets}(c) shows that SD confers a significant increase in basin size for pseudoinverse learning as well. At large $P/N$, the basin sizes under SD are more than twice as large as under 0TMH dynamics. 

\subsection{Endowing conventional spin glasses with associative memory-like properties}

Basin sizes do not typically scale extensively with $N$ under 0TMH dynamics. This is because the number of metastable states grows exponentially (in the system size $N$)   in both the Hopfield model (with Hebbian connectivity for $P/N \gg 0.138$) and in the SK spin-glass model~\footnote{For example, the total number of local minima in the SK model is estimated to be $\mathcal{O}(e^{\alpha N})$, where $\alpha=0.201(2)$~\cite{nemoto1988metastable}.}.  However, we now present numerical evidence that SD dynamics endows these same energy minima with basin sizes that exhibit extensive scaling with $N$, even in a pure SK spin glass model. Thus, we find that the SK spin glass, under the SD dynamics, behaves like an associative memory with an exponentially large number of memory states with extensive basins. (A programming step would be required; see Sec.~\ref{programming}.)

\begin{figure}[t!]
    \centering
    \includegraphics[width=\columnwidth]{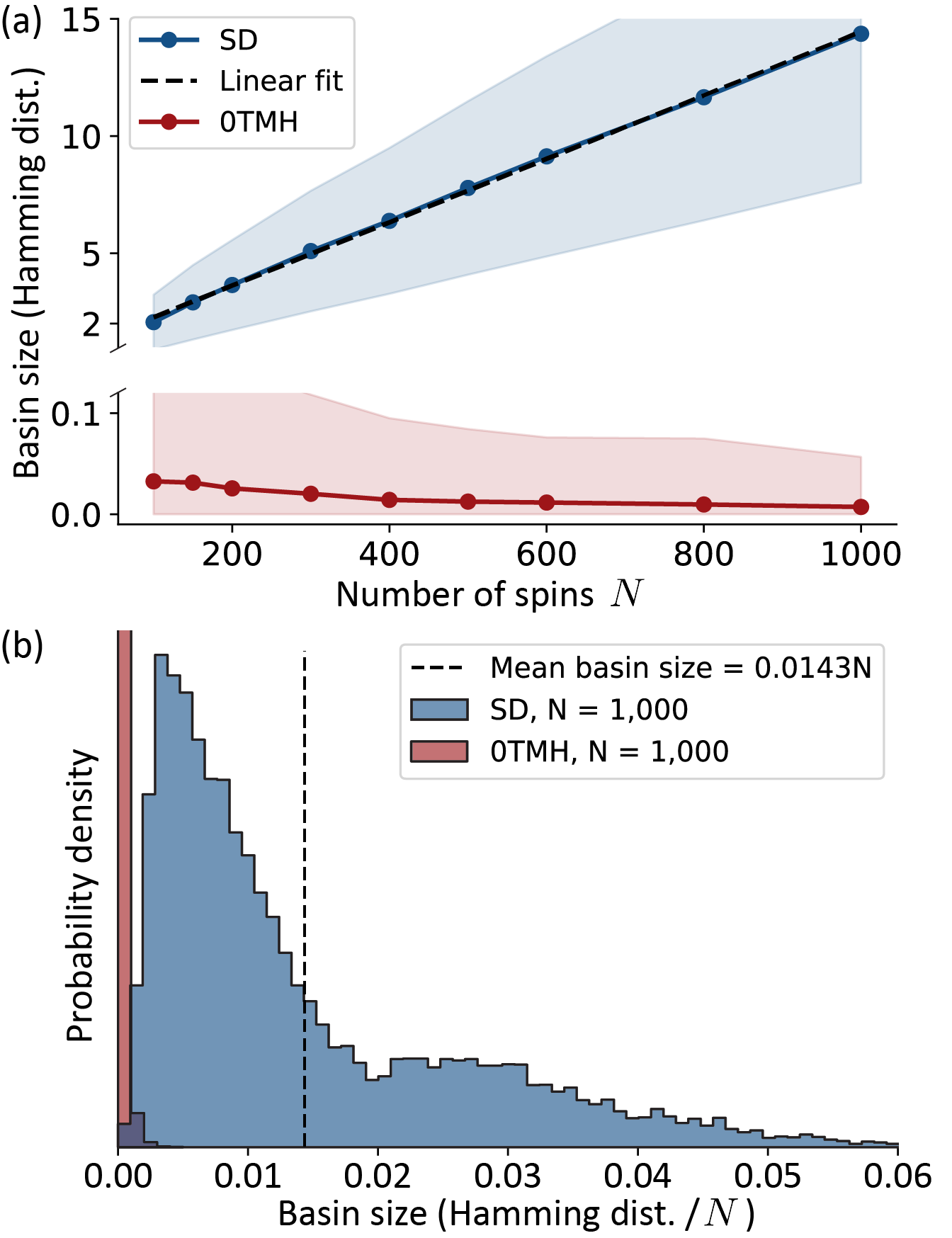}
    \caption{(a) Average basin size of randomly found local minima for the SK model using 0TMH (red) or SD (blue). The shaded region corresponds to $\pm 1$ standard deviation in the basin sizes. For each $N$, the  basin size is averaged over 50 realizations of $J_{ij}$ matrices, with 500 random initial states per $J_{ij}$ realization.  (b) Histogram of basin sizes using 0TMH (red) and SD (blue) for the $N=1000$ data plotted in (a). The mean basin size for SD is marked by the vertical dashed line.}
    \label{fig:SK_basins}
\end{figure}

More quantitatively, we numerically compute the size of basins present in an SK spin glass using both kinds of dynamics; see Fig.~\ref{fig:SK_basins}. Connectivity matrices $J_{ij}$ are initialized with each element drawn, \textit{i.i.d.}, as Gaussian random variables of mean zero and variance 1; normalization of the variance is arbitrary at zero temperature. Rather than compute a uniform average over all metastable states of the $J_{ij}$ matrix---a task that is numerically intractable for large system sizes---we instead sample metastable states by initializing random initial states and letting those states evolve under 0TMH dynamics. Once a metastable state is reached, the basin of attraction is measured under both 0TMH and SD dynamics as discussed above. 

Figure~\ref{fig:SK_basins}(a) plots the average basin size as a function of the system size $N$, averaging both over realizations of $J_{ij}$ matrices and over the random initial states. Under 0TMH dynamics, the mean basin size decreases with increasing $N$, approaching zero at large $N$.  This shows that for 0TMH dynamics, metastable states of the spin glass become very fragile, as expected from a traditional understanding of metastable states in a spin glass. In contrast, for SD dynamics, the average basin size scales extensively with system size in a roughly linear fashion. A phenomenological fit results in scaling of $0.0135(1)N+0.91(7)$.

These results help reveal the structure of the energy landscape. Vanishing basin size under 0TMH implies that a single spin-flip perturbation is sufficient to open new pathways of energy descent to different local minima. As soon as such pathways exist, 0TMH will find them, leading to vanishing basin size. For SD, we however see that the \textit{steepest} path of energy descent is almost always the one leading back to the original local minimum that was perturbed, as long as the number of spin flips is less than about $0.0135 N$. This picture explains how metastable states are stable against small perturbations under SD, but unstable under 0TMH.

The extensive scaling of basin sizes found for SD dynamics not only means that metastable states with a finite basin of attraction exist, but in fact suggest the number of such states is exponential in system size, far greater than the capacity of any typical associative memory. While sampling metastable states via relaxation of random initial states under the standard 0TMH dynamics does not yield a uniform sampling, Fig.~\ref{fig:SK_basins}(b) shows that nearly all such discovered metastable states have a finite basin of attraction under SD dynamics. The distribution is peaked away from zero basin size, whereas with 0TMH the distribution is strongly peaked around zero. This suggests that extensive basin sizes under SD dynamics may be a property of almost all metastable states in the SK spin-glass. 

We conclude that the SK spin glass enhanced with SD dynamics operates, remarkably, like an associative memory, with exponentially many spurious memories. These memories are random and non-programmable via any existing learning rule, but have extensive basins of attraction, tolerating up to $0.0135 N$ input errors on average while maintaining a $95\%$ probability of recall. While this basin size is small compared to that of stored patterns in standard associative memories---cf.~Fig.~\ref{fig:SDHopfieldNets}---it is nevertheless extensive and there are exponentially more such memories than in standard Hopfield networks.

\section{Associative memory in pumped confocal cavities}\label{sec:learning}

We have thus far shown in Sec.~\ref{sec:connectivity} that CCQED supports a large number of metastable states,  and in Sec.~\ref{sec:SD} that the natural cavity QED dynamics produces a discrete steepest descent (SD) dynamics in energy.  In Sec.~\ref{sec:SDHN}, we showed that SD dynamics can enhance the robustness of memory when applied to traditional Hopfield models, and moreover, can endow the SK spin-glass with associative memory-like properties.  

In this Section, we pull these key elements together to demonstrate how robust associative memory can be created in a transversely pumped confocal cavity. 
We begin by showing that metastable states in such systems indeed possess large basins of attraction. We then develop a method to solve the addressing, or programmability problem, which involves converting the patterns we wish to store into the patterns that naturally arise as local energy minima with large basins of attraction.  Our pattern storage scheme is applicable to any connectivity matrix $J_{ij}$.  We verify its performance for the CCQED connectivity matrix both in the associative memory regime and the in SK spin glass regime. 

\subsection{Basins of attraction with confocal cavity QED connectivity}

We now examine the basin size of the metastable states found in Sec.~\ref{sec:connectivity} by tuning the width $w$ of the distribution of ensemble positions. We measure the basin size of metastable states just as we did in Fig.~\ref{fig:SDHopfieldNets} for Hopfield models and Fig.~\ref{fig:SK_basins} for the SK model---by the relaxation of random perturbations. In particular, a random CCQED connectivity matrix $\mathbf{J}$ is realized by sampling the spin ensemble positions from a 2D Gaussian distribution of fixed width $w$. A large number of random initial states are then prepared and allowed to relax via the native SD dynamics to a metastable state. The basin size of that metastable state is then found by initializing nearby random states and estimating the probability that these perturbed states evolve back to the metastable state as a function of the Hamming distance of the perturbation. 

\begin{figure}[t!]
    \centering
    \includegraphics[width=\columnwidth]{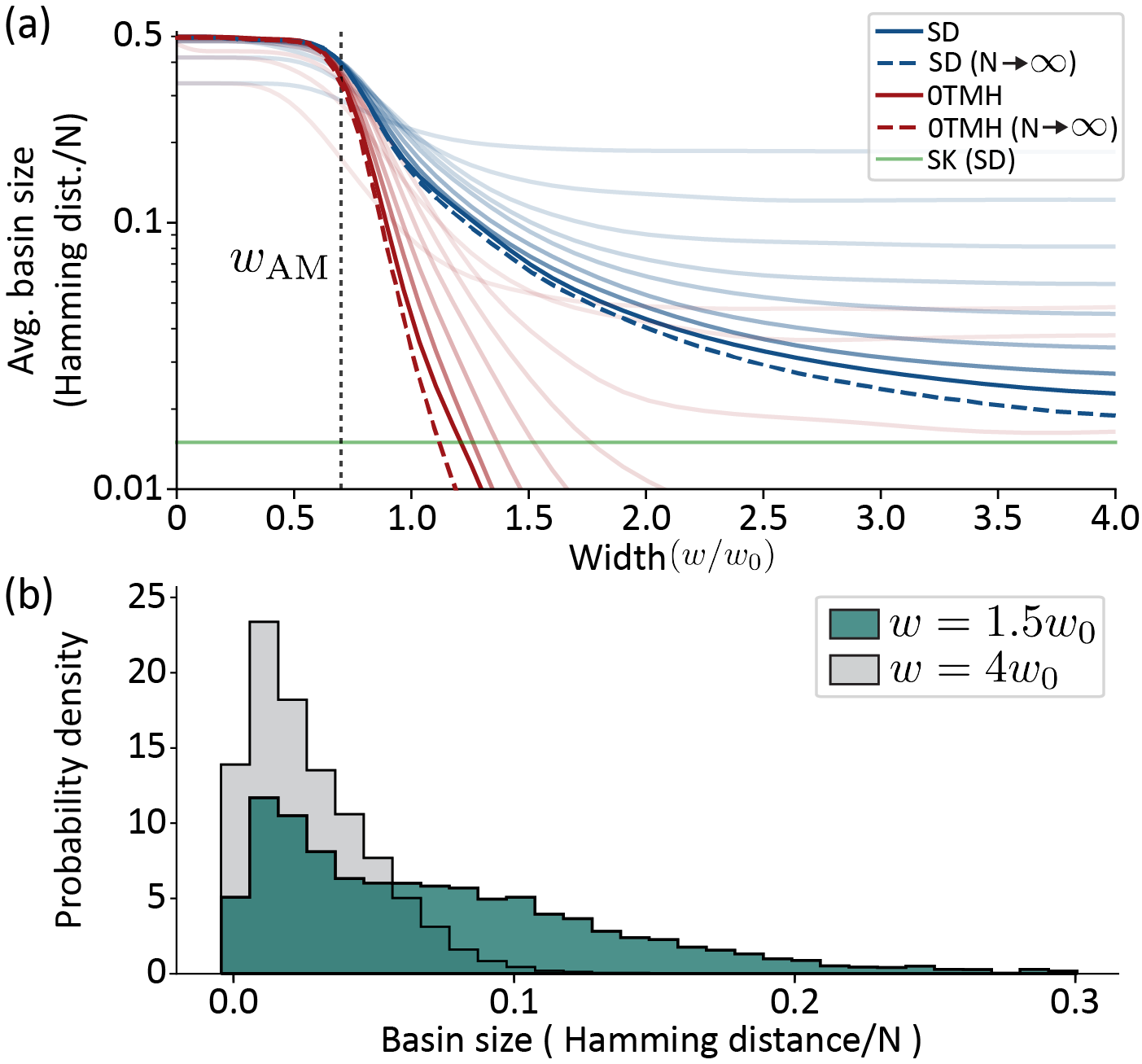}
    \caption{(a) The average basin size in the CCQED system versus width of the spin ensemble distribution $w$.  Metastable states are found by relaxing random initial seeds. Basin sizes for both 0TMH dynamics (red) and the (native) SD dynamics (blue) are shown. System sizes ranging from $N=3$ to $N=800$ are considered, with darker traces corresponding to larger sizes.  We extrapolate to the limit $N\to \infty$ by fitting a linear function of $1/N$ to the finite $N$ data.  The intercept at $1/N=0$ yields the thermodynamic limit (dashed lines). The ferromagnetic--to--associative memory phase transition, indicating the onset of multiple metastable states, is marked by a vertical black dashed line at $w_{\mathrm{AM}}$. The green line shows the typical basin size under SD dynamics for the SK connectivity, $0.0135(1) N$, as extracted from Fig.~\ref{fig:SK_basins}. (b) Histogram of the basin sizes found for widths $w=1.5w_0$ and $w=4w_0$.}
    \label{fig:cosBasinSizes}
\end{figure}

Figure~\ref{fig:cosBasinSizes} shows how the basin size evolves with $w$ for the CCQED connectivity. For each width $w$, the measured basin sizes are averaged over many realizations of the CCQED matrices $\mathbf{J}$, with 200 random initial states per matrix.  The number of realizations of $\mathbf{J}$ required for convergence varies with $N$, ranging from 1,600 realizations for $N=3$ spins to 32 realizations for $N=800$ spins.  This is due to self-averaging at large system size. Figure~\ref{fig:cosBasinSizes}(a) presents the average basin size as a function of $w$, using both the native SD dynamics and 0TMH for comparison. A sharp decrease in basin size occurs near the ferromagnetic--to--associative memory transition at $w_{\mathrm{AM}}\approx 0.67 w_0$. For 0TMH dynamics, the mean basin size per spin quickly falls to zero after crossing $w_{\mathrm{AM}}$. Remarkably, however, steepest descent dynamics yields larger  basins beyond $w_{\mathrm{AM}}$.  These are extensive in size.  Thus, just as in the case of the Hopfield model beyond capacity and in the SK spin-glass, switching from 0TMH to SD dramatically expands the size of basins from nonextensive to extensive. Moreover, there exist $\mathcal{O}(e^{N^{0.4}})$  metastable states in this regime, which is less than the $\mathcal{O}(e^{N})$ expected of a spin glass; see Fig.~\ref{fig:Jij-transitions}.  Overall, this demonstrates that a confocal cavity in this intermediate width regime functions as a high-capacity, robust associative memory with many metastable states with extensive basin sizes, albeit with random non-programmable memories. For larger widths, the average basin size asymptotically reduces to that exhibited in the SK model. This observation, when combined with the analysis of Sec.~\ref{sec:connectivity}, provides further evidence that the confocal cavity connectivity yields an SK spin glass at large width. 

Figure~\ref{fig:cosBasinSizes}(b) shows examples of the distribution of basin sizes under SD dynamics for widths $1.5w_0$ and $4w_0$.  For $w=1.5w_0$, the variance in basin sizes is large despite a small mean size of ${\sim}0.05N$. Pattern completion is robust in this regime for a large percentage of basins since many exceed $0.1N$ in size. For $w=4w_0$, the basin size distribution is narrower, but still peaked away from zero.  However, nearly all basins sizes are less than $0.1N$, as expected in the spin glass regime; see Fig.~\ref{fig:Jij-regimes}(c).

\subsection{Solving the programmability problem: a scheme for memory storage}\label{programming}

We have demonstrated the presence of many metastable states with large basins of attraction (of extensive size) in the transversely pumped CCQED system. Such a system evidently manifests a robust recall phase with a large number of random memories. However, these memories are not directly programmable, in the sense that we cannot place the local energy minima wherever we wish in configuration space by tuning the connectivity.  In contrast, for the Hopfield model with the Hebbian connectivity of Eq.~\eqref{heb}, we can place energy minima close to up to $0.138 N$ desired points $\boldsymbol{\xi}^\mu$ in configuration space, as long as these points are random and uncorrelated. Also, the more complex pseudoinverse rule of Eq.~\eqref{pseudoheb} enables us to place local energy minima at $N$ arbitrary points (though there is no guarantee of any minimal basin size if some points are close to each other).  

We now show how to effectively program the CCQED system to store any desired patterns using an encoder that translates the desired patterns to be stored into the non-programmable patterns that are naturally stabilized by the intrinsic cavity dynamics. While the pattern storage scheme we present here is designed with the physics of CCQED in mind, we note that the method is also applicable to any Hopfield or spin-glass like network connectivity with non-programmable energy minima with extensive basin sizes.

The basic idea behind the encoder is to find a linear transformation that maps any set of desired patterns, as represented by light-fields,  into metastable spin configurations of the cavity.  Once found, this transformation is then applied to any input state (i.e., light field) before allowing the confocal cavity network to dynamically evolve to a metastable state.  The (optical) output can then undergo an inverse transform, back into the original basis. In this way, we harness the dynamics of the CCQED spin network to store general patterns, and recall them through pattern completion,  without ever needing to change the CCQED connectivity matrix. 

To formally define the properties required of the transformation, $\enc$, we consider a set of patterns that we wish to store, $\{\patt^p\}$ for $p=1$ to $P$. We require that there are at least $P$ metastable states of the Hopfield energy landscape $\{\cavmin^p\}$ that need to be found and cataloged.  These may be found by, e.g., allowing random initial states to relax---note that there may be more than $P$ such states. For an arbitrary input state $\mathbf{x}$, the network operates as follows:
\begin{enumerate}
    \item The input state $\mathbf{x}$ is transformed to an encoded state $\tilde{\mathbf{x}}=\enc (\mathbf{x})$. The transformation should smoothly map the desired pattern states onto metastable states of the Hopfield network: we thus require $\enc(\patt^p)=\cavmin^p$ for all $p$. 
    \item The Hopfield network is initialized in the encoded state $\tilde{\mathbf{x}}$ and allowed to evolve to a metastable state $\tilde{\mathbf{y}}$.
    \item The evolved state $\tilde{\mathbf{y}}$ is then subject to an inverse transformation $\dec$. The final output is  $\mathbf{y}=\dec(\tilde{\mathbf{y}})$. The transformation $\dec$ should map the metastable states of the energy landscape back onto the pattern states so that $\dec(\cavmin^p)=\patt^p$. Note that $\dec(\enc(\patt^p))=\patt^p$, so that $\enc$ and $\dec$ act as inverse transformations when acting on pattern states. 
\end{enumerate}

We now discuss how to find the smooth transformations $\enc$ and $\dec$ obeying the properties above. The simplest option is a linear transformation, for which we should define $\enc(\mathbf{x})=\encmat\cdot\mathbf{x}$, where $\encmat$ is the matrix that minimizes the error function
\begin{equation}
    E = \sum_{p=1}^P\left| \encmat \cdot \patt^p-\cavmin^p \right|^2+\lambda||\encmat||^2.
\end{equation}
Here, $||\cdot||$ is the Hilbert-Schmidt norm. The terms in the sum ensure that the patterns are mapped to the metastable states of the cavity QED system. However, this requirement under-constrains $\encmat$ for $P<N$. To remedy this, we add the term $\lambda||\encmat||^2$ to penalize large elements of $\encmat$;  $\lambda$ is a regularization hyperparameter. 
This least-squares problem can be solved analytically, yielding:
\begin{equation}
\label{eqn:encoder}
    \encmat = \Big(\sum_{p=1}^P \cavmin^p\patt^{pT} \Big)\Big(\lambda\mathbb{1} + \sum_{p=1}^P\patt^p\patt^{pT} \Big)^{-1}.
\end{equation}
While this solves the linear transform problem, the linear encoding itself has a problem that we must address. Spin configurations $\tilde{\mathbf{x}}$ should correspond to vectors whose elements are $\pm1$, corresponding to the magnetization of the $S^x$ components of the spin ensembles. We require that the output of the transformation takes this form.  Thus, we must modify the encoding so that the full transformation for an arbitrary input state is  $\enc(\mathbf{x}) = \mathrm{sgn}(\encmat\cdot\mathbf{x})$, where the sign function is understood to be applied element-wise.

The encoder matrix defined in Eq.~\eqref{eqn:encoder} will not be invertible, in general. However, by simply reversing the roles of the patterns and cavity minima in constructing the encoder matrix, we can construct a corresponding decoder matrix that maps the metastable states back to the desired patterns:
\begin{equation} \label{eqn:decoder}
    \decmat = \Big(\sum_{p=1}^P \patt^{p}\cavmin^{pT} \Big)\Big(\lambda\mathbb{1} + \sum_{p=1}^P\cavmin^p\cavmin^{pT} \Big)^{-1}.
\end{equation}

The memory capacity of an associative memory is typically defined as the maximum number of patterns that can be stored as metastable states of the energy landscape. Under this standard definition, at most $N$ total patterns can be mapped exactly onto attractor states of the CCQED connectivity because the transformation map is linear.  This means that the maximum memory capacity of the encoded confocal cavity is $N$, which is the same as that of the pseudoinverse learning method, but significantly outperforms  Hebbian learning with its capacity of ${\sim}0.138N$. However, a more practical definition of capacity would depend also on the size of the basins of attraction, and hence the robustness of the stored patterns. We now verify the memory capacity of $N$ accounting for basin size.

We numerically benchmark the encoded pattern storage method with the CCQED connectivity and SD dynamics that are present in the cavity. Fixing the system size $N$, we generate $P$ random patterns to store in the cavity. Because the CCQED connectivity hosts a large number of metastable states with large basins in the regime $0.7w_0\leq w\leq 2w_0$, we set the width $w$ to $1.5w_0$ and construct a connectivity matrix. Metastable states are cataloged by relaxing $P$ random initial seeds using SD.  (Note that this procedure is biased toward selecting metastable states with the largest basins of attraction; these are preferable for our scheme.)  This is then followed by construction of the encoder and decoder using Eqs.~\eqref{eqn:encoder} and~\eqref{eqn:decoder}. The basin sizes of the pattern states are determined by applying increasing amounts of input error to the patterns until the probability of recall drops below $95\%$.

\begin{figure}
    \centering
    \includegraphics[width=\columnwidth]{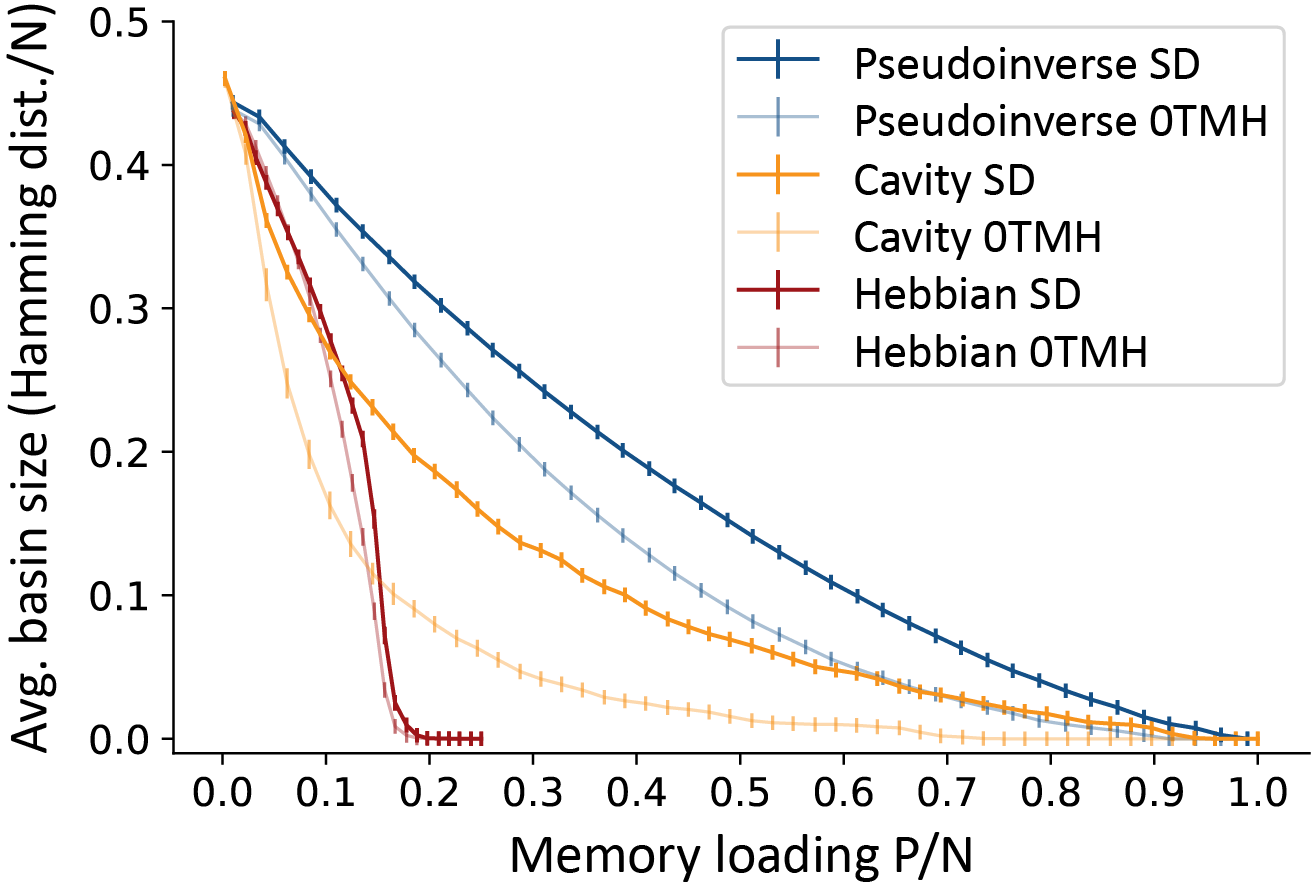}
    \caption{Average basin size of patterns stored in a CCQED system as a function of $P/N$ using the linear transformation encoding scheme (yellow).
    Results are shown for $N=800$, however similar results are found for system sizes of $N=50$--$800$. For comparison, the average basin sizes are plotted for the  Hebbian (red) and pseudoinverse (blue) learning methods. In each case, we show the results of both SD (solid) and 0TMH (dashed) dynamics; $w = 1.5w_0$.}
    \label{fig:confocalMem}
\end{figure}

Figure~\ref{fig:confocalMem} compares the performance of encoding memories onto the metastable states of the CCQED connectivity against the results of Hebbian and pseudoinverse learning. The average basin size of our CCQED scheme is larger than Hebbian learning. The results do not depend strongly on system size.  For the CCQED scheme, extensive basin sizes are indeed observed up to the maximum $P=N$ stored patterns. This memory capacity represents roughly an order--of--magnitude improvement over the standard Hebbian learning model ($0.138N$) and coincides with the pseudoinverse learning rule. The basin sizes in CCQED are not as large as pseudoinverse learning, though they are more physically realizable.  Note that for the Hebbian learning, we measure recall as the probability of recovering the desired memory.  As such, the drift of local minima from the desired patterns contributes to the collapse of basin size at $P/N=0.138$.    

Remarkably, we find that for memory loading ratios $P/N \leq 0.4$, the average basin size with the encoding and decoding transformation significantly exceeds $0.1N$. This is notable because earlier we found that the average basin sizes are less than $0.1N$ at this width, $w=1.5w_0$; see Fig.~\ref{fig:cosBasinSizes}. It seems that the encoding and decoding transformations significantly enhance the robustness of the associative memory. Inspection of the form of the linear transformation reveals that it focuses the $P<N$ input states onto a subset of the metastable states and vice-versa for the decoder matrix. This effectively gives it a head start by reducing the Hamming distance between perturbed input states and the nearest metastable state. 

The pattern storage scheme also works in the SK spin glass regime, either as applied to the CCQED connectivity at large $w$ or to the actual SK model. In both cases, $N$ memories may be stored. The linear transformation scheme is thus quite flexible:  it is applicable to any model with extensively scaling basins of attraction. An interesting challenge for future work would be to see if it is possible to extend the current scheme to \textit{nonlinear} mappings between pattern states and attractors in order to achieve a superextensive scaling of the memory capacity with system size $N$. This would provide a physically realizable fully programmable associative memory with capacity well beyond the $P=N$ pattern limit. In this case, the fact the SK spin glass possesses an exponential number of metastable states with extensive basin sizes under SD dynamics could become particularly relevant for associative memory applications. 

\subsection{Weight chaos and the robustness of memory patterns}

Considering the practical implementation of the associative memory, an important question is that of `weight-chaos' ~\cite{Bray:1987ip,Zhu:2016iea,Albash:2019db,Pearson:2019ig}. This is the problem that, for standard Hopfield learning rules, flipping the sign of a few (or even one) $J_{ij}$ matrix element can result in a model with a completely different set of metastable states.  This arises due to the chaotic nature of glassy systems.  In our case, weight chaos can arise from two sources. Imprecision in the placement of spin ensembles leads to fluctuations in the positions $\mathbf{r}_i$ and thus in the nonlocal interaction $J_{ij}\sim \cos(2\mathbf{r}_i\cdot\mathbf{r}_j/w_0^2)$.  Additionally, atom number fluctuations per spin ensemble can occur between experimental realizations, leading to fluctuations in $J_{ij}$ proportional to $1/\sqrt{M}$.  Neither of these can change the sign of large $J_{ij}$ elements, thereby avoiding the most severe, randomizing repercussions of weight chaos.

\begin{figure}
    \centering
    \includegraphics[width=\columnwidth]{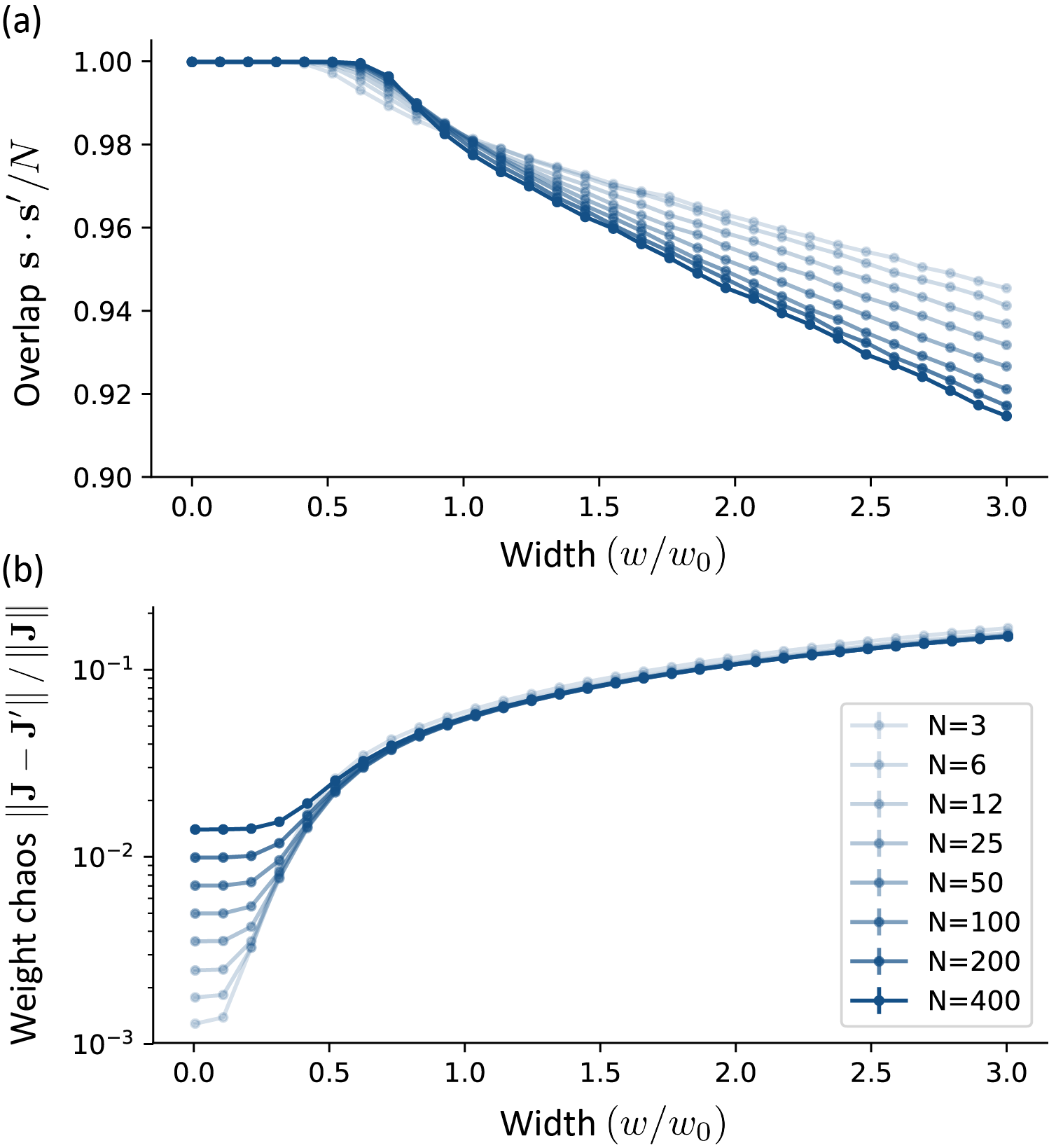}
    \caption{The effect of weight chaos in experimental parameter regimes, using an ensemble placement uncertainty of 1~$\mu$m and $1/\sqrt{M}$ fluctuations in atom number. The drift of metastable states between different realizations of noise weights is plotted in (a) as a function of the width $w$ of the ensemble distribution and for different system sizes $N$ (see legend in panel b). The states $\mathbf{s}$ and $\mathbf{s}^\prime$ refer to metastable states of two realizations $\mathbf{J}$ and $\mathbf{J}^\prime$, as described in the text. b) The direct change in the $\mathbf{J}$ connectivity matrix is plotted as a function of the width.}
    \label{fig:chaos}
\end{figure}

The effect of weight chaos is numerically explored in Fig.~\ref{fig:chaos} under experimentally realistic parameters. We consider an uncertainty of 1~$\mu$m in the placement of spin ensembles and fix the total number of atoms in the cavity at $\sum_{i=1}^N M = 10^6$ independent of the number of ensembles considered. We then present two different measures of weight chaos and its effect, as a function of the ensemble distribution width $w$. First, we quantify how metastable states drift between realizations.  To do so, we take a particular $\mathbf{J}$ matrix and find a metastable state $\mathbf{s}$ by allowing a random initial state to relax via SD. We then realize a weight matrix $\mathbf{J}^\prime$ that is nominally the same as $\mathbf{J}$ but has added to it  a particular realization of position and atom number noise. The state $\mathbf{s}$ is then allowed to again relax via SD to a potentially new state $\mathbf{s}^\prime$ to see if it has drifted. The overlap $\mathbf{s}\cdot\mathbf{s}'/N$ quantifies this drift and is shown in Fig.~\ref{fig:chaos}(a) for many realizations. In the second measure, we directly quantify the change in the weights as a percentage of their original values via the quantity $\begin{Vmatrix}\mathbf{J}-\mathbf{J}^\prime \end{Vmatrix}/\begin{Vmatrix}\mathbf{J} \end{Vmatrix}$, where $\begin{Vmatrix}\cdot \end{Vmatrix}$ denotes the Hilbert-Schmidt norm of the matrix.

Figure~\ref{fig:chaos} demonstrates that metastable states are robust against drift for widths less than approximately $w_0$, which includes the ferromagnetic and associative memory regimes of the confocal connectivity. Greater than 98\% overlap between experimental realizations are expected. The overlap begins to drop for widths beyond $w_0$ as the connectivity transitions to a spin glass, because the nonlocal connectivity becomes increasingly sensitive to position at larger radii. The overlap decreases with increasing system sizes in this regime up to the maximum $N=400$ ensembles probed in simulations. Atom number fluctuations set a floor on weight chaos that is independent of $w$.  This is visible in Fig.~\ref{fig:chaos}(b) for small $w$. The noise floor rises with increasing $N$ since $M$ must decrease in order to fix the total atom number.  A few percent deviation in $J_{ij}$ is acceptable for heuristic solvers, replacing one `good-enough' solution with another.  Position uncertainty will likely be smaller than 1~$\mu$m, so these results present a conservative picture.

\subsection{Experimental implementation}
We now present practical details of how the CCQED associative memory can be initialized in a particular (distorted) memory state. One may initialize the spin ensembles via longitudinal pumping to an arbitrary (encoded) $\tilde{\mathbf{x}}$, where $\tilde{x}_i=\pm1$ is the magnetization of the $i$th ensemble. As seen in Eq.~\eqref{eqn:HopfieldFromCavity}, the spins can be subjected to a field $\spinfield_i$ created by longitudinally pumping the cavity.  Physically, this is possible because in a confocal cavity, resonant light with any transverse profile (up to the resolving capability of the cavity mirrors) is supported due to the degeneracy of all TEM$_{\mathsf{lm}}$ modes of good parity.  If the longitudinal pumping is sufficiently strong, i.e., $\longfield_i > J_{ij}$, then the energy of spin flips $\dEi$ will be dominated by this field.
By choosing $\mathrm{sgn}(\longfield_i)=\tilde{x}_i$, 
SD dynamics will evolve the spin ensembles into state $\mathbf{x}$. The fact that only the pattern of signs matter for large fields presents a natural implementation of the threshold operation employed in the encoding  $\enc(\mathbf{x})$. The longitudinal pump is turned off once the state has been initialized, thereby allowing the system to evolve to the nearby metastable state determined by the connectivity $J_{ij}$ alone.

Many schemes would suffice for implementing the encoding and decoding transformations mentioned above.  The transformation can be accomplished most simply electronically. In this scheme, a new longitudinal field is optically constructed based on the electronically transformed state. Similarly, after imaging the cavity light, the decoder matrix would be applied electronically to yield the final output state from the experiment. Alternatively, an all optical implementation can be achieved using spatial light modulators or other methods~\cite{Miller:2015jp} if coherence in the field must be preserved. 


\section{Discussion}\label{sec:conclusion}

We presented a practicable quantum-optical system that can serve as a neural network for associative memory.  It does so by exploiting the natural spin coupling and superradiant dynamics provided by a CCQED system.   The physical system lacks the direct correspondence between memories and the connectivity matrix that makes simple the learning of patterns in the idealized Hopfield model. Nevertheless, we find a straightforward pattern storage scheme for encoding memories in this physical system that allows $N$ memories to be stored with a robust pattern completion process that tolerates an extensive number of initial errors. Remarkably, the emergent SD dynamics of the physical system extend the effective basins of attraction to enable robust recall. In terms of memory capacity, the system significantly outperforms Hebbian learning.

The multimode cavity QED platform is distinctive among annealing-based optimization schemes for its nonequilibrium dynamics.  In ``traditional'' annealing approaches, the system either undergoes classical equilibrium dynamics in a potential landscape~\cite{aarts:1988simulated}, or unitary quantum dynamics remaining close to the ground state~\cite{Hauke:2019vy}.  In the CCQED system, the optimization occurs by the intrinsic dynamics of a driven-dissipative system. By contrast, in equilibrium contexts, any coupling to the outside world disrupts the system:  dissipation must be overcome in these approaches, not embraced as in the present scheme.

Confocal cavity QED can be compared to other gain-based optimization schemes. These include weak-coupling, optics-based ``coherent Ising machines.'' These have explored the notion that optimization problems can be heuristically solved using the dissipative coupling between multiple optical parametric oscillators (OPOs)~\cite{McMahon:2016fy,Inagaki:2016eb,yamamoto2017coherent,hamerly2019experimental}. The steady-state pattern of the oscillator phases that develops at threshold manifests the solution to this problem.  This occurs because the coupling of OPOs is chosen so that the pattern of oscillation with lowest threshold corresponds to the solution of the optimization problem. As such, as one increases the pumping, this lowest threshold solution appears first.  Similar ideas have also been explored in coupled polariton condensates~\cite{Berloff:2017gs} and lasers~\cite{Pal:2017cd}. 

There are, however, crucial limitations to these weakly coupled optical gain-based optimizers:  the output spin states correspond to coherent photon states, which only emerge as one crosses the threshold. This means that near threshold, the spins are ``soft,'' allowing defects to form~\cite{hamerly2016topological}. In addition, because photon number is not conserved, photon intensity in each node can change, leading to dynamic changes in the $J_{ij}$ weights defining the problem~\cite{Kalinin:2018br}. This requires the addition of classical feedback to control these populations~\cite{kalinin2018networks,Leleu:2019bk}\footnote{Other, more complicated entanglement-preserving feedback schemes have been proposed~\cite{Yanagimoto:2019tu}.}. By contrast, the multimode cavity QED system has a nearly constant number of atoms per ensemble, so the connectivity matrix $J_{ij}$ remains stable.  The spin variables are well defined far above the superradiant threshold:  the superradiant enhancement of the dynamics forces ensembles to flip as one collective spin. We also point out that while the fully quantum limit of the CCQED system consists of a single strongly coupled atom at each position and so the state space becomes that of spin $1/2$ particles, in contrast, the photonic or polaritonic coherent Ising machines possess (continuous) bosonic Hilbert spaces.

An important challenge for future work is to extend the present treatment into the fully quantum realm. This could be done by suppressing the dephasing we used to favor classical states, tuning pump strength near to the superradiant phase transition, and by considering few spins per ensemble. As noted earlier, currently we require a large number of atoms per ensemble to overcome heating; an alternative would be to design a configuration with enhanced single-atom cooperativity. While the cooperativity of a single $^{87}$Rb atom coupled to TEM$_{00}$ mode is $C=g_0^2/2\kappa\gamma=2.5$~\cite{Kollar:2014us}, our system should already benefit from an enhanced cooperativity for an atom coupled to many degenerate modes with nonzero field amplitude at its location. The single-atom cooperativity is enhanced by the number of degenerate modes, due to the formation of a `supermode' with larger  electric field at the atom's position~\cite{Kollar2017sm,Vaidya2018}.  The confocal cooperativity is therefore far greater than unity; preliminary measurements suggest it is approximately 50~\cite{Vaidya2018}. If achieved, this would allow quantum entanglement to play a role in the dynamics~\cite{Kastoryano:2011hr}, enabling the exploration of the behavior of a fully quantum nonequilibrium neural network. Whether and how entanglement and quantum critical dynamics at the superradiant phase transition provide superior heuristic solution-finding capability remains unclear.  Specific questions of interest include how entanglement might affect memory capacity or retrieval fidelity.  We leave these questions for future exploration.  Beyond  quantum neuromorphic computing, such systems might be able to address quantum many-body problems~\cite{Carleo:2017cn}.

Finally, we note that our pattern storage and recall scheme provides a novel shift in paradigm from the traditional Hopfield framework of memory storage.  Indeed in this traditional framework, desired memories, as determined by network states driven by external stimuli (i.e., the patterns $\boldsymbol{\xi}^\mu$ in Sec.~\ref{sec:HModel}) are assumed to be directly stabilized by a particular choice of connectivity matrix $\mathbf J$.   For example, the Hebbian rule (Eq.~\eqref{heb}) and the pseudoinverse rule (Eq.~\eqref{pseudoheb}) are two traditional prescriptions for translating desired memories $\boldsymbol{\xi}^\mu$ into connection patterns $\boldsymbol{J}$ that engineer energy minima at or near the desired memories.  In contrast, the steepest descent dynamics in our system reveals that random choices of connectivity $\mathbf J$, whether originating from the CCQED system, a glassy Hopfield model beyond capacity, or remarkably, even the canonical SK spin glass, yield Ising systems possessing exponentially many local energy minima with extensive basin size.  Indeed, our storage scheme exploits this intrinsically occurring multiplicity of robust minima simply by mapping external stimuli $\boldsymbol{\xi}^\mu$ to them, rather than creating new minima close to the external stimuli. To date, this computational exploitation of the exponential multiplicity of local energy minima in glassy systems has not been possible because of their subextensive basin size under traditional 0TMH or Glauber dynamics.  By contrast, the intrinsic physical dynamics we have discovered in the CCQED system directly enables us to computationally harness the exponential multistability of glassy phases, by using an encoder to translate external stimuli to natively occurring internally generated patterns.  Further exploitation of these glassy states through \textit{nonlinear} encoders remains an intriguing direction for future work. 

Interestingly, in biological neural systems, multiple brain regions, like the hippocampus and the cerebellum, that are thought to implement associative memories, are synaptically downstream of other brain regions that are thought to recode sensory representations before memory storage; see, e.g.,~\cite{Ganguli2012-kq} for a review.  Such brain regions could play the role of an encoder that learns to translate sensory stimuli into natively occurring memory patterns, in analogy to our storage framework. Moreover, there has been some neurobiological evidence that natively occurring neural activity patterns observed before an experience are themselves used to encode that new experience~\cite{Dragoi2011-py,Dragoi2013-bn}, though there is debate about the prevalence of this observation~\cite{Silva2015-og}.  Of course, if neurobiological memory encoding does indeed select one of a large number of latent, preexisting stable activity patterns in order to encode a novel stimulus, it may be very difficult to experimentally observe such a stable activity pattern before presentation of the stimulus, due to limited availability of recording time.  Thus, our work suggests it may be worth explorations beyond the Hopfield framework for memory storage, in which the internal representations of memories are determined not only by the nature of sensory stimuli themselves, but also by the very the nature of natively occurring potential stable memory states that exist, but are not necessarily expressed, before the onset of the stimulus.

Thus overall, by combining the biological principles of neural computation with the intrinsic physical dynamics of a CCQED system, our work leads to enhanced associative memories with higher capacity and robustness, opens up a novel design space for exploring the computational capabilities of quantum-optical neural networks, and empowers the exploration of a novel paradigm for memory storage and recall, with implications both for how we might computationally harness the energy landscapes of glassy systems, as well as explore novel theoretical frameworks for the neurobiological underpinnings of memory formation. 


\begin{acknowledgments}
We thank Daniel Fisher and Hideo Mabuchi for stimulating conversations and acknowledge the Army Research Office for funding support. The authors wish to thank NTT Research for their financial support.  We would also like to thank the Stanford Research Computing Center for providing computational resources, especially access to the Sherlock cluster.  Y.G.~and B.M.~acknowledge funding from the Stanford Q-FARM Graduate Student Fellowship and the NSF Graduate Research Fellowship, respectively. J.K.~acknowledges support from the Leverhulme Trust (IAF-2014-025), and S.G.~acknowledges funding from the James S.~McDonnell and Simons Foundations and an NSF Career Award. 
\end{acknowledgments}


\appendix

\section{Raman coupling scheme and effective Hamiltonian}
\label{app:RamanScheme}

The effective multimode Dicke model in Eq.~\ref{eqn:multimodeDicke} may be engineered by coupling two internal states of $^{87}$Rb to the cavity fields as proposed in Ref.~\cite{Carmichael07}; see Fig.~\ref{fig:doubleRaman}.  The scheme has been demonstrated in the dispersive limit (yielding an effective Ising model) using a cavity capable of multimode (confocal) operation~\cite{kroeze2018spinor}.  The states $\ket{F = 2,m_F = -2}\equiv\ket{\uparrow}$ and $\ket{F = 1,m_F = -1}\equiv\ket{\downarrow}$ are coupled through two cavity-assisted two-photon Raman processes depicted in Fig.~\ref{fig:doubleRaman}. The two states $\ket{\uparrow}$ and $\ket{\downarrow}$ are separated in energy by $\omega_{\mathrm{HF}}$ due to hyperfine splitting and Zeeman shifts in the presence of a transversely oriented magnetic field. Typical experiments employ a magnetic field  of $\sim$2.8~G.  The hyperfine level splitting is $\omega_{\mathrm{HF}}\approx 6.8$~GHz. The cavity resonance frequency $\omega_C$ is detuned from the atomic excited state $5^{2}P_{3/2}$ by $\Delta_+$ and $\Delta_{-}$, respectively, for states $\ket{\uparrow}$ and $\ket{\downarrow}$. Two standing-wave transverse pump beams with Rabi frequencies $\Omega_{\pm}$ realize the two cavity-assisted Raman processes. The optical frequencies of the pump beams $\omega_\pm$ are given by
\begin{align}
\frac{1}{2} (\omega_+ - \omega_-) &= \omega_{\mathrm{HF}} - \transfield \nonumber \\
\frac{1}{2} (\omega_+ + \omega_-) &= \omega_C+\detune,
\end{align}
where $\transfield$ is the two-photon Raman detuning and $\detune$ is the detuning of the mean frequency of the pumps from the cavity resonance. 

To derive the model in Eq.~\eqref{eqn:multimodeDicke}, we extend the results and experimental setup in Ref.~\cite{kroeze2018spinor} to the present case of a multimode cavity of confocal configuration. The field profile of a single cavity transverse mode with mode index $m = (\mathsf{l},\mathsf{m})$ in a cavity of length $L$ is given by 
\begin{equation}
   \Phi_{m} = \Xi_m(\mathbf{r})\cos{\left[k\left(z+\frac{r^2}{2R(z)}\right)-\theta_{m}(z)\right]},
\end{equation}
with
\begin{align}
  \theta_{m}(z) &= \psi(z) + n_m \left[\psi(L/2) + \psi(z)\right] - \xi \nonumber \\
  \psi(z) &= \mathrm{arctan} \left( \frac{z}{z_R} \right) \\
  R(z) &= z + \frac{z^2_R}{z}.
  \label{eq:lightfieldphase}        
\end{align}
Here, $\xi$ is a phase offset fixed by the boundary condition that the light field vanishes at the two mirrors $z = L/2$, $n_m \equiv \mathsf{l}+\mathsf{m}$ is the sum of the two transverse indices, $k = 2 \pi/\lambda_{780~\mathrm{nm}}$  is the wavevector of the pump and cavity light, and $z_R = L/2$ is the Rayleigh range of a confocal cavity. Under the Raman coupling scheme, the light-matter interaction Hamiltonian for a multimode cavity can be obtained by summing up the contributions from the transverse modes. In the appropriate rotating frame~\cite{kroeze2018spinor}, the total Hamiltonian $H = H_{\uparrow} + H_{\downarrow} + H_{\mathrm{cavity}} + H_{\mathrm{Raman}}$ is given by
\begin{align}
H_{\uparrow} &= \int d^3 \mathbf{x} \hat{\psi}^{\dagger}_{\uparrow}(\mathbf{x}) \Big[ \frac{\hat{\mathbf{p}}^2}{2 m}  + V_{\mathrm{trap}}(\mathbf{x}) + \transfield \Big] \hat{\psi}_{\uparrow} (\mathbf{x}) \nonumber \\
H_{\downarrow} &= \int d^3 \mathbf{x} \hat{\psi}^{\dagger}_{\downarrow}(\mathbf{x}) \Big[ \frac{\hat{\mathbf{p}}^2}{2 m}  + V_{\mathrm{trap}}(\mathbf{x})\Big] \hat{\psi}_{\downarrow} (\mathbf{x}) \nonumber \\
H_{\mathrm{cavity}} &= - \sum_m \Delta_m a^{\dagger}_m a_m \nonumber \\
H_{\mathrm{Raman}} &= \sum_m \int d^3 \mathbf{x} \Big[ \eta^{-}_{m} (\mathbf{x}) \hat{\psi}^{\dagger}_{\uparrow} (\mathbf{x}) \hat{\psi}_{\downarrow} (\mathbf{x}) a^{\dagger}_m \nonumber \\ &\qquad\qquad+ \eta^{+}_{m} (\mathbf{x}) \hat{\psi}^{\dagger}_{\downarrow} (\mathbf{x}) \hat{\psi}_{\uparrow} (\mathbf{x}) a^{\dagger}_m  + \mathrm{h.c.}\Big],
\end{align}
where $\hat{\psi}^{\dagger}_{\uparrow,\downarrow}$ are the spinor wavefunction for the atomic states, $V_{\mathrm{trap}}$ is the trapping potential, and the cavity assisted Raman coupling strength has spatial dependence given by the cavity mode profile $\Phi_m$ and the standing-wave pump along the $x$-direction
\begin{equation}
   \eta^{\pm}_m (\mathbf{x}) = \frac{\sqrt{3} g_0 \Omega_{\pm}}{12 \Delta_{\pm}} \Phi_m (\mathbf{r}, z) \cos(k x). 
\end{equation}
The two Raman processes can be balanced by controlling the pump beam power such that
\begin{equation}
    \frac{\sqrt{3} g_0 \Omega_+}{12 \Delta_+} = \frac{\sqrt{3} g_0 \Omega_-}{12 \Delta_-} \equiv \eta.
\end{equation}
To remove the kinetic energy dependence and the spatial oscillation of the Raman coupling strength, we propose the following trapping scheme for $V_{\mathrm{trap}}(\mathbf{x})$: first introduce optical tweezers to tightly confine atoms in separate ensembles in different locations $\mathbf{r}_i$ in the cavity transverse plane but at the midpoint of the cavity axis $z = 0$; then superimpose on the optical tweezers a deep 2D optical lattice with periodicity of 780 nm along both the pump and cavity direction to add a potential $V_{\mathrm{lattice}} \propto \cos^2(k x/2) + \cos^2(k z/2)$. Under the combined trapping potential, the spinor wavefunction can be expanded as as sum of individual ensembles
\begin{align}
    \hat{\psi}_{\uparrow\downarrow} (\mathbf{x}) =& \sum_{i,\nu_i} Z(z) \sqrt{\rho(\mathbf{r}-\mathbf{r}_i)} \times \nonumber \\
    &\sqrt{[1 + \cos(kx)][1+\cos(kz)]}\hat{\psi}_{\uparrow\downarrow, \nu_i}, 
\end{align}
where $Z(z)$ and $\rho(\mathbf{r})$ describe the spatial confinement of an ensemble by the optical tweezers trap, $\nu_i$ indexes individual spins inside an ensemble indexed with $i$, and $\hat{\psi}_{\uparrow\downarrow, \nu_i}$ is the spin operator for a single atom $\nu_i$.

For each ensemble with $M$ atoms, we define collective spin operators
\begin{align}
    S^z_{i} &= \sum^{M}_{\nu_i= 1} (\hat{\psi}^{\dagger}_{\uparrow, \nu_i} \hat{\psi}_{\uparrow, \nu_i} - \hat{\psi}^{\dagger}_{\downarrow, \nu_i} \hat{\psi}_{\downarrow, \nu_i})/2 \nonumber \\
    S^x_{i} &= \sum^{M}_{\nu_i= 1} (\hat{\psi}^{\dagger}_{\uparrow, \nu_i} \hat{\psi}_{\downarrow, \nu_i} + \hat{\psi}^{\dagger}_{\downarrow, \nu_i} \hat{\psi}_{\uparrow, \nu_i})/2. 
\end{align}
Taking the assumption that the width $\sigma_{A}$ of the atom profile $Z(z)$ and $\rho(\mathbf{r})$ satisfies the condition $ \lambda_{780~\mathrm{nm}} \ll \sigma_{A} \ll z_R$, we can perform the spatial integral along $z$ and drop the oscillatory terms along $x$. The Raman coupling term becomes
\begin{equation}
    H_{\mathrm{Raman}} = \sum_m \sum^{N}_{i=1} g_{im}  S^x_{i} (a^{\dagger}_m + a_m),
\end{equation}
where the coupling strength between the $i$th ensemble and $m$th cavity mode $g_{im}$ is given by
\begin{equation}
g_{im} = \frac{\eta}{2} \int d^2 \mathbf{r}  ~\Xi_{m}(\mathbf{r}) \rho(\mathbf{r} - \mathbf{r}_i) \cos(n_m \pi/4),
\end{equation}
and we assume that different ensembles are separated in space with negligible overlap. 

Putting this together, the CCQED system can be modelled by a master equation:
\begin{equation}\label{mastereq}
  \partial_t \rho
  = - i [H_{CCQED}, \rho]
  + \sum_m \kappa \mathcal{L}[a_m \rho].
\end{equation}
Here, we account for cavity field loss at rate $\kappa$ by
Lindblad superoperators $\mathcal{L}[X]=X\rho X^\dag - \{X^\dag X,\rho \}/2$,  where $a_m$ is the annihilation operator for the $m$th cavity mode. The Hamiltonian of the CCQED system $H_{CCQED}$ is given by
\begin{multline}
    \label{eqn:HCCQED_Appendix}
    H_{CCQED} =  \sum_m \big[-\Delta_m a^\dag_m a_m+\longfield_m(a^\dag_m+a_m)\big] 
    \\+\transfield\sum_{i=1}^N S_i^z
    + \sum_{i=1}^N\sum_{m}g_{im}S_i^x(a^\dag_m+a_m).
\end{multline}
We will always choose transverse pump to be red detuned from the cavity modes so that $\Delta_m <0$.  The mode-space representation of a longitudinal pump term is $\longfield_m$. 

We will discuss the effects of additional classical and quantum noise terms in the Hamiltonian in appendices~\ref{app:classicalbath} and ~\ref{app:quantumbath}.

\section{Convolution of the nonlocal interaction with the finite extent of the spin ensemble}\label{app:nonlocal}
We now discuss how the finite spatial extent of the spin ensembles affects the connectivity. The effective nonlocal interaction between two ensembles is given by the convolution of the nonlocal interaction, proportional to $ \cos(2\mathbf{r}_i\cdot\mathbf{r}_j/\waist^2)$, with the ensemble density profiles $\rho_i(\mathbf{x})$:
\begin{equation}
    J_{ij}=\frac{-\tilde g_0^2\detune}{\pi(\detune^2 + \kappa^2)} \int \mathrm{d}\mathbf{x}\mathrm{d}\mathbf{y} \rho_i(\mathbf{x})\rho_j(\mathbf{y})\cos\left(2\frac{\mathbf{x}\cdot\mathbf{y}}{\waist^2} \right). 
\end{equation}
The ensemble densities may be approximated by Gaussian profiles centered at positions $\mathbf{r}_i$ of width $\sigma_A$. We then find the effective nonlocal interaction
\begin{multline}
    J_{ij} = \frac{-\tilde g_0^2\detune}{\pi(\detune^2 + \kappa^2)}\cos\left(2\frac{\waist^2\mathbf{r}_i\cdot\mathbf{r}_j}{\waist^4+4\sigma_A^4}\right) \\
    \times\frac{\waist^4}{(\waist^4+4\sigma_A^4)}\exp\left[\frac{-2\sigma_A^2(|\mathbf{r}_i|^2+|\mathbf{r}_j|^2)}{\waist^4+4\sigma_A^4} \right].
\end{multline}
The effect of this convolution is to rescale the term inside the cosine and dampen the amplitude of $J_{ij}$ for ensembles located far from the cavity center. The unconvolved nonlocal interaction is recovered in the limit of small $\sigma_A/\waist$. In current experiments, $\sigma_A$ is at least an order of magnitude smaller than $\waist$, thus we consider the unconvolved nonlocal interaction in this work.


\section{Derivation of confocal cavity QED connectivity probability distribution}
\label{app:random-coupling}

We now derive the probability distribution and correlations of the $J_{ij}$ matrices for the CCQED system with a Gaussian distribution of atomic positions. We recall that the photon-mediated interaction takes the form  $J_{ij}\propto\cos(2\mathbf{r}_i\cdot\mathbf{r}_j/w_0^2)$, where $w_0$ is the width of the TEM$_{00}$ mode. We  consider spin ensemble positions $\mathbf{r}_i$ to be randomly distributed according to a 2D isotropic Gaussian distribution, centered at the origin of the cavity transverse plane and with a standard deviation $w$. Explicitly, $\mathbf{r}_i=( r_{i,x}, r_{i,y})^T$, where $r_{i,k}$ is a Gaussian random variable with mean $0$ and variance $w^2$.  To find the probability distribution of $J_{ij}$, we proceed by first finding the probability distribution for $\mathbf{r}_i\cdot\mathbf{r}_j=r_{i,x}r_{j,x}+r_{i,y}r_{j,y}$,  the dot product of 2D Gaussian random variables. We start by writing the terms in the dot product as
\begin{equation}
    r_{i,x}r_{j,x} = \left(\frac{r_{i,x}+r_{j,x}}{2} \right)^2 -\left(\frac{r_{i,x}-r_{j,x}}{2} \right)^2.
\end{equation}
Since the variances of $r_{i,x}$ and $r_{j,x}$ are equal, their sum and difference are independent random variables with each of variance $2w^2$. This implies that
\begin{align}
    r_{i,x}r_{j,x} &\sim \left[\frac{\mathcal{N}(0,2w^2)}{2} \right]^2 -\left[\frac{\mathcal{N}(0,2w^2)}{2} \right]^2 \nonumber \\
    &\sim\frac{w^2}{2}\left[\mathcal{N}(0,1)^2-\mathcal{N}(0,1)^2 \right],
\end{align}
where, as above, $\sim$ indicates equality in distribution, and
each $\mathcal{N}(\mu,\sigma^2)$ denotes an independent Gaussian random variable with mean $\mu$ and variance $\sigma^2$.  Thus, the full dot product $\mathbf{r}_i\cdot\mathbf{r}_j$ is distributed as
\begin{align}
    \mathbf{r}_i\cdot\mathbf{r}_j &\sim \frac{w^2}{2}\Big[ \mathcal{N}(0,1)^2-\mathcal{N}(0,1)^2  +\mathcal{N}(0,1)^2-\mathcal{N}(0,1)^2  \Big] \nonumber \\
    & \sim \frac{w^2}{2} \Big[ \chi^2_2 - \chi^2_2 \Big],
\end{align}
where $\chi^2_\nu$ is a chi-square random variable with $\nu$ degrees of freedom. The difference of $\chi^2_\nu$ random variables produces a variance-gamma distribution. To see this, note that the moment generating function for the difference of $\chi^2_\nu$ distributions is
\begin{align}
    M_{\chi^2_\nu-\chi^2_\nu}(t)&=M_{\chi^2_\nu}(t)M_{\chi^2_\nu}(-t) \nonumber \\
    &=(1-2t)^{-\nu/2}(1+2t)^{-\nu/2} \nonumber \\
    &=\left(\frac{1/4}{1/4-t^2} \right)^{\nu/2},
\end{align}
while the moment generating function for the variance-gamma distribution with parameters $\mu,\alpha,\beta,\lambda$ is 
\begin{equation}
    M_{VG}(t)=e^{\mu t}\left(\frac{\alpha^2-\beta^2}{\alpha^2-(\beta+t)^2} \right)^\lambda.
\end{equation}
Thus, the moment generating functions coincide for $\mu=\beta=0$, $\alpha=1/2$, and $\lambda=\nu/2=1$. The random variable $\mathbf{r}_i\cdot\mathbf{r}_j$ is distributed like a variance-gamma random variable with the above parameters. Denoting the probability density of $\mathbf{r}_i\cdot\mathbf{r}_j$ as $\prob_{\mathbf{r}_i\cdot\mathbf{r}_j}(x)$, we find
\begin{equation}
    \prob_{\mathbf{r}_i\cdot\mathbf{r}_j}(x) = \frac{1}{2w^3}\sqrt{\frac{2|x|}{\pi}}K_{\frac{1}{2}}\left(|x|/w^2  \right),
\end{equation}
where $K_{\nu}(x)$ is a modified Bessel function of the second kind. The Bessel function simplifies for $\nu=1/2$ to the form $K_{1/2}(x) = \sqrt{\pi/2x}\,e^{-x}$, yielding the simplified expression
\begin{equation}
    \prob_{\mathbf{r}_i\cdot\mathbf{r}_j}(x) = \frac{1}{2w^2}\exp\left(\frac{-|x|}{w^2}\right).
\end{equation}

We now wish to find the probability density $\prob_{\cos}(x)$ for the cosine of the random variable $2\mathbf{r}_i\cdot\mathbf{r}_j/w_0^2$. Applying the general formula~\cite{riley2002mathematical} for taking functions of random variables allows us to write
\begin{equation}
    \prob_{\cos}(x) = \sum_{k,\pm}\prob_{\mathbf{r}_i\cdot\mathbf{r}_j}\left[\tfrac{w_0^2}{2}\cos^{-1}_{k,\pm}(x)\right]\left|\frac{\textrm{d}}{\textrm{d}x}\tfrac{w_0^2}{2}\cos^{-1}_k(x) \right|,
\end{equation}
where by $\cos_{k,\pm}^{-1}(x)$ we denote the infinite number of possible inverses of cosine; more precisely,
\begin{equation}
    \cos^{-1}_{k,\pm}(x) = \pm\arccos(x) + 2\pi k,\quad k\in\mathbb{Z},
\end{equation}
where $\arccos(x)\in[0,\pi]$ is the principal inverse of cosine. The derivative of $\cos^{-1}_{k,\pm}(x)$ with respect to $x$ appearing above is independent of $k$,
\begin{equation}
    \frac{\textrm{d}}{\textrm{d}x}\cos^{-1}_{k,\pm}(x) = \frac{\mp1}{\sqrt{1-x^2}}.
\end{equation}
Substituting in the derivative, we find
\begin{multline}
    \prob_{\cos}(x) = \\ \frac{w_0^2}{2w^2\sqrt{1-x^2}}\sum_{k=-\infty}^{\infty} \exp\left(-\tfrac{w_0^2}{2w^2}|\arccos(x)+2\pi k| \right).
\end{multline}
Using the fact that $|\arccos(x)|\leq \pi$, the infinite sum can be rearranged as a geometric series,
\begin{multline}
    \prob_{\cos}(x) =  \frac{w_0^2}{2w^2\sqrt{1-x^2}}\Bigg[\exp\left(-\frac{w_0^2\arccos(x)}{2w^2} \right) \\
    +2\cosh\left(\frac{w_0^2\arccos(x)}{2w^2} \right)\sum_{k=1}^\infty e^{-\pi kw_0^2/w^2} \Bigg].
\end{multline}
Evaluating the geometric series and rewriting in terms of hyperbolic functions finally results in
\begin{equation}
    \prob_{\mathrm{cos}}(x)=\frac{w_0^2\mathrm{csch}\left(\frac{\pi w_0^2}{2w^2}\right)}{2w^2\sqrt{1-x^2}}\cosh\left[\frac{w_0^2}{2w^2}(\pi-\arccos x)\right].
\end{equation}
This probability density is the main result of this  Appendix. 

We now present a number of quantities related to the distribution of couplings. First, the mean of the probability distribution can be computed in the usual way,
\begin{align}
    \mu_J = \int_{-1}^1 x \prob_{\cos}(x) dx = \frac{1}{1+4\big(\frac{w}{w_0}\big)^4},
\end{align}
as well as the variance:
\begin{align}
    \sigma_J^2&= \nonumber \int_{-1}^1 dx (x-\mu_J)^2  \prob_{\cos}(x) \\
    &=\frac{16\big(\frac{w}{w_0}\big)^8\left[5+8\big(\frac{w}{w_0}\big)^4\right]}{\left[1+4\big(\frac{w}{w_0}\big)^4\right]^2\left[1+16\big(\frac{w}{w_0}\big)^4\right]}.
\end{align}

We may also find the correlation between couplings as a function of $w$. This is important, for instance, in comparing the connectivity to that of an SK spin glass in which the couplings are uncorrelated.  We first use a trigonometric identity to write
\begin{multline}
    J_{ij}J_{jk} = \cos\left(2\frac{\mathbf{r}_i\cdot\mathbf{r}_j}{w_0^2}\right) \cos\left(2\frac{\mathbf{r}_j\cdot\mathbf{r}_k}{w_0^2}\right) \\
    = \frac{1}{2}\left[\cos\left(\frac{2}{w_0^2}\mathbf{r}_j\cdot(\mathbf{r}_i+\mathbf{r}_k) \right)+\cos\left(\frac{2}{w_0^2}\mathbf{r}_j\cdot(\mathbf{r}_i-\mathbf{r}_k) \right) \right].
\end{multline}
Since $\mathbf{r}_i$ and $\mathbf{r}_k$ are both Gaussian random variables with equal variances, their sum and difference are uncorrelated Gaussian random variables with twice the original variance. Thus, the two terms have the same expected value, and we find
\begin{align}
    \left<J_{ij}J_{jk}\right> &=  \left<\cos\left(\frac{2}{w_0^2}\mathbf{r}_j\cdot(\mathbf{r}_i+\mathbf{r}_k) \right)\right> \nonumber \\
    &= \left<\cos\left(\frac{2}{w_0^2}\mathcal{N}^2(0,w^2)\cdot\mathcal{N}^2(0,2w^2) \right)\right>,
\end{align}
where $\left< \cdot \right>$ denotes an expectation over the random variables, and
$\mathcal{N}^2(\mu,\sigma^2)$ denotes an isotropic two-dimensional Gaussian random variable with mean $\mu$ and variance $\sigma^2$. We can rewrite this in the symmetric form
\begin{equation}
    \left< J_{ij}J_{jk} \right> = \left<\cos\left[\frac{2}{w_0^2}\mathcal{N}^2(0,\sqrt{2}w^2)\cdot\mathcal{N}^2(0,\sqrt{2}w^2) \right]\right>.
\end{equation}
This expression now matches the expression for the expected value of the original $\prob_{\cos}(x)$ distribution, but with $w^2$ rescaled by $\sqrt{2}$. Thus, we have already computed this quantity and can extract 
\begin{equation}
   \left<J_{ij}J_{jk}\right> = \frac{1}{1+8(w/w_0)^4},   
\end{equation}
exhibiting full correlation at $w=0$ and vanishing correlation at $w \to \infty$.

Finally, to determine the probability of generating negative couplings as a function of $w$, we integrate the probability density to yield the cumulative distribution function
\begin{equation}
    \prob_{\cos<X}=
    \mathrm{csch}\left(\frac{\pi w_0^2}{2w^2}\right)\sinh\left[\frac{w_0^2}{2w^2}(\pi-\arccos X)\right].
\end{equation}
Evaluating the cumulative distribution at $x=0$, the probability of finding a negative coupling term is
\begin{equation}
    \prob_{\cos<0}=
    \frac{1}{2\cosh(w_0^2\pi/4w^2)},
\end{equation}
which approaches $0$ for small $w$ and $1/2$ for large $w$. Thus, in the limit of large $w$, we see that the two-point correlations decay and that negative couplings become equally likely.  These are characteristics typical of a spin glass.
 
\section{Spin-flip rates and decoherence}\label{app:classicalbath}

A classical noise term inserted into the Hamiltonian induces decoherence.  This allows us to restrict our analysis to the classical subspace with definite $S^x$. We now derive the form of the spin-flip rate equations in the presence of classical noise.  We start from the Hamiltonian $H=H_{CCQED} + H_N$
with $H_{CCQED}$ defined in Eq.~\ref{eqn:HCCQED_Appendix},
and the noise term
   $ H_N = \noise(t)\sum_i S_i^x$.
We may note that the only term in $H$ which takes one outside the classical subspace is the transverse field term $\transfield\sum_i S_i^z$. For the classical limit to hold, the effects of this term should be small compared to the decoherence rate.  As discussed in Sec.~\ref{sec:SD}, the noise term will arise naturally from the Raman pumping lasers, and can also be deliberately enhanced to ensure one remains in the classical limit.

\subsection{Decay of coherences}

To pinpoint the effect of the noise on coherences, we will rotate into the interaction picture with respect to $H_N$. The time evolution in the interaction picture is 
\begin{equation}
\rho(t) = e^{ -i \int_0^t dt \noise(t)\sum_i S_i^x} \rho_0(t)  e^{ i \int_0^t dt \noise(t)\sum_i S_i^x},
\end{equation}
where $\rho_0(t)=e^{-iH_{CCQED}t}\rho_0 e^{iH_{CCQED}t}$. To analyze the time dependence of specific elements in the density matrix, we denote total $S^x$ states of the full system by vectors $\mathbf{s}=(s_1^x,s_2^x,\cdots,s_N^x)$, where $s_i^x$ denotes the $S^x$ eigenvalue of the $i$th spin ensemble. The density matrix elements may then be indexed as $\rho_{\mathbf{ss}'}$ for two $S^x$ states $\mathbf{s}$ and $\mathbf{s}'$. As a notational tool, we denote the total difference in $S^x$ eigenvalues between $\mathbf{s}$ and $\mathbf{s}'$ as $\Delta s = \sum_i (s^x_i-{s^\prime}^x_i)$. If the noise correlations are faster than the spin dynamics due to the transverse field $\transfield$, then the time evolution for a general matrix element $\rho_{\mathbf{ss}'}$ is found by averaging over realizations of the stochastic noise source:
\begin{equation}\label{eqn:noiseavg}
\rho_{\mathbf{ss}'}(t)=\left \langle \exp\left( \int_0^t dt\, i \Delta s\, \noise(t) \right)\right \rangle_{\noise(t)} (\rho_0(t))_{\mathbf{ss}'}.
\end{equation}
The average may be performed using the cumulant expansion of the noise source. We assume that $\noise(t)$ is a zero-mean, Gaussian signal such that only the second cumulant is nonzero. The noise-averaged term appearing in Eqn.~\eqref{eqn:noiseavg} is then given by
\begin{equation}
\label{eqn:decoRate}
\Big<\ldots\Big>=
\exp\left[ \frac{(i\Delta s)^2}{2}\iint_0^t dt' dt'' \,\left \langle \noise(t'),\noise(t'') \right \rangle \right].
\end{equation}
We will further take the noise field to be stationary so that the correlation function is time translation invariant, $\left\langle \noise(t'),\noise(t'') \right\rangle = \left\langle \noise(0),\noise(t''-t') \right\rangle$. In this case, the noise correlations are defined by the spectral density, $J_c(\omega)$, using the expression
\begin{equation}
\left\langle \noise(0),\noise(t) \right\rangle = \int_{-\infty}^\infty \frac{d\omega}{2\pi} e^{-i\omega  t}J_c(\omega).  
\end{equation}
Inserting this into Eq.~\eqref{eqn:decoRate}, we find that
\begin{equation}\label{eqn:decoherence}
\Big<\ldots\Big>=
 \exp\left[ -\frac{(\Delta s)^2}{\pi}\int_{-\infty}^\infty d\omega \, \frac{J_c(\omega) }{\omega^2}\sin^2\left(\tfrac{\omega t}{2}\right)  \right].
\end{equation}
This expression allows us to evaluate the decoherence induced by classical noise sources with different spectral densities $J_c(\omega)$.

We now consider the analytically tractable case of a noise source with an ohmic spectral density $J_c(\omega)=2\alpha\omega e^{-\omega/\cutoff}$ for $\omega\geq0$, where $\alpha$ is a dimensionless constant and $\cutoff$ is an exponential cutoff frequency for the bath. Evaluating Eq.~\eqref{eqn:decoherence} using this spectral density, we find that coherences decay with a functional form
\begin{equation}
\frac{\partial}{\partial t}\left|\frac{\rho_{\mathbf{s}\mathbf{s}'}(t)}{\rho_{\mathbf{s}\mathbf{s}'}(0)}\right| = (1+t^2\cutoff^2)^{-\alpha (\Delta s)^2/2\pi}.
\end{equation}
This is close to a power law in form.  We can now extract the timescale over which the coherences decay. The above function is unity at $t=0$ and decays to half of its initial value at time
\begin{equation}
    t_{1/2} = \frac{1}{\cutoff}\sqrt{2^{2\pi/\alpha (\Delta s)^2}-1}. 
\end{equation}
Thus, if one considers a coherent superposition of two states that have a difference $\Delta s$ between their eigenvalues of $S^x$, the rate at which that coherence decays increases with $\Delta s$. As such, coherence decays faster for ensembles with larger numbers of atoms. The classical limit is valid if $\transfield t_{1/2} \ll 1$. This is readily attainable in an experimentally relevant parameter regime, in which $\detune$ is on the order of MHz, $\cutoff$ is on the order of kHz, and $\alpha$ is on the order of 10.

\subsection{Spin-flip rate in presence of an ohmic bath}
We now examine the form of the spin-flip rate function that arises when coupled to an ohmic noise source. 
In Eq.~\eqref{eqn:spinFlipRates}, we gave the general expression for the spin-flip rate in terms of the noise correlation function.  Using an ohmic spectral density, this becomes
\begin{multline}\label{eqn:ohmic_rate}
K_i(\dE) =  h(\dE) 
\\+\frac{\transfield^2}{8} e^{- J_{ii}/|\detune|}\sum_{n=1}^\infty \frac{(J_{ii}/|\detune|)^n}{n!}\frac{n\kappa}{(\dE-n\detune)^2+n^2\kappa^2},
\end{multline}
where $h(\dE)$ is a symmetric peaked function. In the absence of  noise, $h(\dE)=\delta(\dE)$, reflecting the fact that perfectly degenerate spin flips occur at a fast rate because they do not need to  trade energy with cavity modes. Realistically, this delta function is broadened to a finite peak in the presence of even a small amount of noise. As the peak is symmetric, it induces heating since it encourages energy lowering and raising spin-flips at equal rates:  we should tune the noise strength to keep the width of this function smaller than $\detune$.  We now compute the form of this function in the presence of ohmic noise to leading order in $\alpha\cutoff^2/\detune^2$. This is well-justified in the parameter regime stated above. The expression is
\begin{equation}
    h(\dE) = \frac{\sqrt{2\pi}\transfield^2 e^{- J_{ii}/|\detune|}}{8\Gamma(\alpha)2^{\alpha}\sqrt{\cutoff|\dE|}}\left(\frac{|\dE|}{\cutoff}\right)^{\alpha}K_{\alpha-\frac{1}{2}}\left( \frac{|\dE|}{\cutoff}\right),
\end{equation}
where $K_{\alpha-\frac{1}{2}}\left( x\right)$ is the modified Bessel function of the second kind, which has exponentially decaying tails. 

\begin{figure}[t!]
	\centering
	\includegraphics[width=\columnwidth]{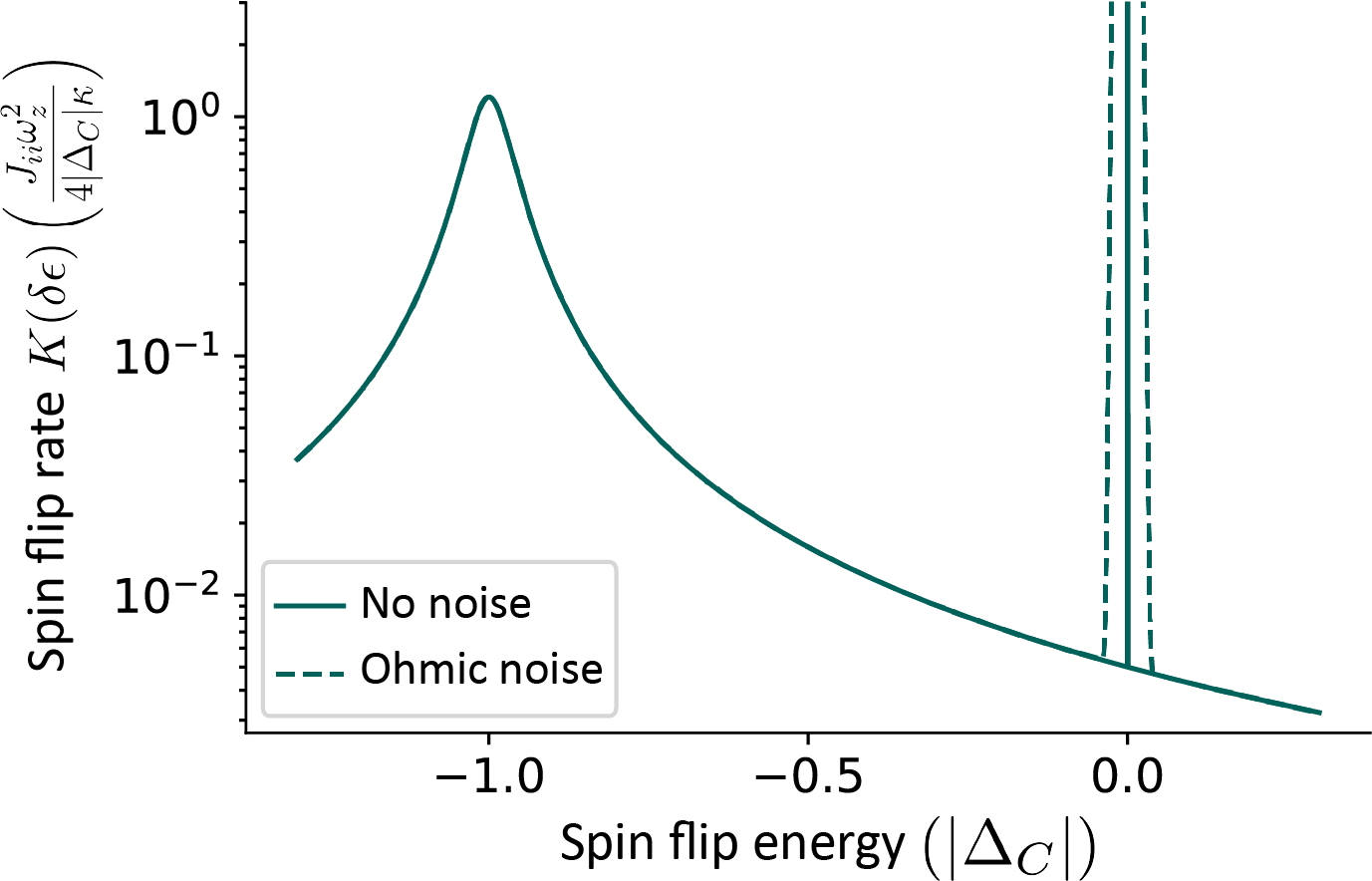}
	\caption{The spin-flip rate function for an ohmic noise source. The Lorentzian peak centered at $\detune = -2$~MHz has a width set by $\kappa=150$~kHz and provides cooling, while the  peak at zero causes heating if the noise source broadens it too much. Plotted is the spin-flip rate function with no noise source (solid line), in which the central peak is a zero-width delta function, and with a classical ohmic noise sufficient to completely suppress superpositions of $S^x$ states of up to $M=10^4$ spins per ensemble (dashed line). The rate functions overlap away from the point of zero spin-flip energy. The central peak is broadened, but still remains small compared to the typical spin-flip energy $\alt\detune$; as such, it does not interfere with the cooling dynamics.}
	\label{fig:ohmicnoise}
\end{figure}

Figure~\ref{fig:ohmicnoise} plots the spin-flip rate function both for the zero noise case and for noise strong enough to fully suppress superpositions in ensembles of up to $M=10^4$ spins. In the latter case, we use $\alpha=20$ and $\cutoff=2\pi\times4$~kHz. The symmetric peak broadens, but remains significantly smaller than the typical spin-flip energy scale $\detune$. In Sec.~\ref{sec:unravelling}, we verified that such a peak is sufficiently narrow so as not to interfere with the SD dynamics that arise naturally in the cavity. Thus, ohmic noise can be employed to fully suppress coherence while preserving cavity dynamics.


\section{Spin-flip rates with a quantum bath}
\label{app:quantumbath}

This section provides an alternative derivation of spin-flip rates, where we consider dephasing to arise from a quantum bath instead of the classical bath discussed in Appendix~\ref{app:classicalbath}; see also Ref.~\cite{Fiorelli:2020kd}. We start from a model $H=H_{CCQED}+H_\gamma$, where $H_\gamma$ is a quantum noise source:
\begin{equation}
  H_\gamma =  \sum_{i,k} \left[
    \sigma_i^x \xi_k (B^\dagger_{ik} + B^{}_{ik})
    + \nu_k B^{\dagger}_{ik} B^{}_{ik}
  \right].
\end{equation}
Here, $B_{ik}$ is a bosonic annihilation operator for the $k$th bath mode coupled to the ensemble at site $i$.  We assume independent but identical baths for each site $i$, so the behavior of this source is fully determined by its spectral function $J_Q(\omega) = \sum_k \xi_k^2\delta(\omega-\nu_k)$. We will discuss below how the result of this model connects to that of classical noise by considering a high-temperature limit.

As before, we will consider the transverse field term perturbatively.  It will be useful to define longitudinal
and transverse parts of the Hamiltonian:
\begin{equation}
    H_T = \transfield \sum_{i=1}^N S^z_i, \qquad
    H_L = H_{CCQED}-H_T.
\end{equation}
We derive a spin-only master equation containing the spin-flip rates by treating the term $H_T$ perturbatively using the Bloch-Redfield approach~\cite{Breuer2002}. Crucial to this procedure is the observation that $H_0 = H_{L} + H_\gamma$ can be diagonalized by a unitary transformation that we will write in two parts as $U=U_F U_\gamma$, where $U_F$ and $U_\gamma$ are chosen to diagonalize $H_{L}$ and $H_\gamma$, respectively. The former is written
\begin{equation}
  U_F= \exp\left[ \sum_m F_m \left(
      \frac{a^\dagger_m}{\Delta_m + i\kappa} - \mathrm{H.c.}\right) \right],
\end{equation}
where we defined $F_m = \longfield_m + \sum_i g_{im} S_i^x$.  This transformation commutes with $S^x_i$, and thus with $H_\gamma$. As discussed in Sec.~\ref{sec:SD}, this transform leads to the replacement $a_m \to a_m + F_m/(\Delta_m + i \kappa)$, which also modifies the Lindblad dissipation term describing cavity loss.  Generically,
\begin{equation}
   \kappa \mathcal{L}[a_m+\alpha]=
   \kappa \mathcal{L}[a_m] + \kappa
   \left[\alpha^\ast a_m - \alpha a^\dag_m, \rho\right],
\end{equation}
with the second term shifting the Hamiltonian.
Incorporating this contribution of the Lindblad term into the transformed $\tilde{H}_{L}$, we find
\begin{equation}
  \tilde{H}_{L} 
  = -\sum_m \left[ \Delta_m a^\dagger_m a_m  
    - \frac{\Delta_m F_m^2}{\Delta_m^2 + \kappa^2} \right].
\end{equation}
This expression is now diagonal, with eigenstates defined by classical states of $S^x_i$ and photon number states. We note that the second term is the Hopfield model in a longitudinal field:
\begin{align}
  \ham &= 
  \sum_m
  \frac{\Delta_m F_m^2}{\Delta_m^2 + \kappa^2} \nonumber \\
  &= 
  -\sum_{i,j=1}^N J_{ij} S^x_i S^x_j
  - \sum_{i=1}^N \spinfield_i S^x_i,
\end{align}
with connectivity and longitudinal fields:
\begin{equation}
  J_{ij} = -\sum_m\frac{\Delta_m g_{im} g_{jm}}{\Delta_m^2 + \kappa^2} ,
  \quad
  \spinfield_i = -2 \sum_m \frac{\longfield_m g_{im} \Delta_m}{\Delta_m^2 + \kappa^2}.
\end{equation}
The term $H_\gamma$ may be similarly diagonalized by the transformation
\begin{equation}
  U_\gamma = \exp\left[ -\sum_{ik} S^x_i \frac{\xi_{k}}{\nu_k} 
  (B^\dagger_{ik} - B^{}_{ik} ) \right].
\end{equation}
This commutes with $H_{L}$ and  transforms $H_\gamma$ to
\begin{equation}
  \tilde{H}_\gamma= \sum_{i,k} \nu_k B^{\dagger}_{ik} B^{}_{ik}.
\end{equation}

Now that $\tilde{H}_0 = \tilde H_{L}+\tilde H_{\gamma}$ is diagonal in the transformed frame, we may implement the standard Bloch-Redfield procedure in which the (transformed) term $\tilde{H}_T $ is treated perturbatively. We first determine the form of $\tilde{H}_T= U^\dag H_T U$. Working in the interaction picture of $\tilde{H}_0$, we find
\begin{equation}
  \label{eqn:transHT}
  \tilde{H}_T(t) = \frac{\transfield}{2} \sum_i 
  \left(
    i S^-_i(t) D_{F,i}(t) D_{\gamma,i}(t) 
    + \mathrm{H.c.}\right).
\end{equation}
We now examine the factors that appear in this expression. First, note the raising and lowering operators take the unusual form $S^\pm = S^y \pm i S^z$ because the $S^x$ spin states are the eigenstates of $H_0$. The interaction picture time dependence of these operators is
\begin{equation}
    S_i^\pm(t) = e^{-i\ham t}S_i^\pm e^{-i\ham t}. 
\end{equation}
This operator takes a simple form in the $S^x$ subspace because $\ham$ is diagonal in this subspace. Specifically, consider $S^x$ states denoted $\ket{\mathbf{s}}=\ket{s_1^x,s_2^x,\cdots,s_N^x}$, where $s_i^x$ denotes the $S_i^x$ eigenvalue for the $i$th spin ensemble. The time dependence of the matrix elements is then
\begin{equation}
    \bra{\mathbf{s}^\prime}S_i^\pm(t)\ket{\mathbf{s}} = e^{i(E_{\mathbf{s}^\prime}-E_{\mathbf{s}})t}\bra{\mathbf{s}^\prime}S_i^\pm\ket{\mathbf{s}},
\end{equation}
where $E_{\mathbf{s}}$ is the energy eigenvalue of the state $\ket{\mathbf{s}}$ of $\ham$. The difference in energy appearing in the exponential, which results from flipping a single spin in the $i$th ensemble, can be written
\begin{equation}
    \dEi^\pm = 
    E_{\mathbf{s}^\prime}-E_{\mathbf{s}}
    =
    -J_{ii}\mp2\sum_{i=1}^N J_{ij}s_j^x.
\end{equation}
The time dependence of the raising and lowering operators therefore takes the simple form $S_i^\pm(t) = e^{i\dEi^\pm t}S_i^\pm$. The other factors in Eq.~\eqref{eqn:transHT} are polaronic operators describing how the spins are dressed by the photons  
\begin{align}
  D_{F,i}
  &=
  \exp\left[
     \sum_m g_{im} 
    \left(
      \frac{a^\dagger_m }{\Delta_m + i\kappa} - 
      \frac{a^{}_m }{\Delta_m - i\kappa}
    \right)
  \right],
  \\
  D_{\gamma,i}
  &=
  \exp\left[
     -\sum_k \frac{\xi_k}{\nu_k} (B^\dagger_{ik} - B^{}_{ik} )
  \right].
\end{align}
Their time-dependent forms involve the interaction-picture time dependence of $a_m, B_{ik}$.

With the full forms of $\tilde H_0$ and $\tilde H_T(t)$ now specified, the standard Bloch-Redfield procedure (in the Markovian approximation) begins by writing:
\begin{equation}
  \partial_t \tilde{\rho}
  =
  - \int^t dt^\prime  \mathrm{Tr}_{\mathrm{bath}}
  \left[\tilde{H}_T(t), \left[\tilde{H}_T(t^\prime), \rho(t) \right] \right].
\end{equation}
We include both the cavity modes and noise operators $B_{ik}$ in tracing over the ``bath" degrees of freedom above, yielding a master equation for the spin degrees of freedom only.  Note that tracing out the cavity modes at this stage preserves the vital physics describing the spin-flip dynamics, as the diagonalized Hamiltonian $H_0$ includes the full cavity-mediated interaction between spins. From this point on, we assume the confocal limit, $\Delta_m=\detune$.

The Bloch-Redfield procedure eventually yields an expression that may be written in the interaction picture as
\begin{multline}
  \partial_t \rho = 
  -i [H_{\mathrm{Lamb}}, \rho] \\
  + \sum_i 
  \Big( 
    K_i(\dEi^-) \mathcal{L}[S^-_i]
    +  
    K_i(\dEi^+) \mathcal{L}[S^+_i]  
  \Big),
\end{multline}
where the spin-flip rates 
are $K_i(\dE)=\Re[\Gamma_i(\dE)]$ and 
\begin{multline}\label{rates}
  \Gamma_i(\dE)
  = 
  \frac{\transfield^2}{8}\,  
  \int_0^\infty 
  d \tau \exp\Bigg[ -i \dE\tau-C(\tau) \\
  - \frac{{J}_{ii}}{\detune}\big(1- e^{-(\kappa-i\detune)\tau}\big)\Bigg].
\end{multline}
The Lamb shift $H_{\mathrm{Lamb}}$ term is
\begin{equation}
  H_{\mathrm{Lamb}} = 2\sum_i
    \left[L_i(\dEi) + L_i(-\dEi) \right] S^+_i S^-_i,
\end{equation}
where $L_i(\dE)=\Im[\Gamma_i(\dE)]$.

Since $S_i^+S_i^-$ is diagonal in the $S^x$ basis, $H_{\mathrm{Lamb}}$ generates no spin-flip dynamics. For the quantum bath, $C(\tau)$ takes the form
\begin{multline}
 C_Q(\tau)  = \\ 
  \int d\omega \frac{J_Q(\omega)}{\omega^2}
  \left[
    \coth\left({\beta \omega}/{2}\right)
    (1-\cos(\omega \tau))
    +
    i \sin(\omega \tau)
  \right],
\end{multline}
where $J_Q(\omega)$ is the spectral density of the quantum bath. We can recover the results of Appendix~\ref{app:classicalbath} for a classical bath by taking the limit $\beta\to0$ while rescaling the spectral density to keep the integrand finite.  This corresponds to suppressing quantum noise while maintaining a finite level of thermal (classical) noise. Formally, we keep
$J_C(\omega)=J_Q(\omega)\coth(\beta \omega/2) \simeq J_Q(\omega)(2 k_B T/\omega)$ finite while $J_Q(\omega) \to 0$.
This effectively removes the imaginary term from $C(\tau)$, giving
\begin{equation}
C_C(\tau) = \int \frac{d\omega}{2\pi} \frac{J_C(\omega)}{\omega^2}\sin^2\left(\frac{\omega \tau}{2}\right).
\end{equation}

We conclude this section by presenting the case in which  $J_Q(\omega)$ describes an ohmic quantum bath,
\begin{equation}
    J_Q(\omega)=\frac{\alpha_Q}{4}\omega e^{-\omega/\cutoff}.
\end{equation} 
We use a slightly different normalization than in the classical case for algebraic convenience.  Note also, following the previous paragraph, that the high-temperature classical limit of this bath would be classical white noise, rather than classical ohmic noise.
The function $C_Q(\tau)$ can be calculated exactly in the zero-temperature bath limit:
\begin{equation}
    C_Q(\tau) = \frac{\alpha_Q}{4}\ln(1+i\cutoff\tau).
\end{equation}
The integrals in the spin-flip rate can then be evaluated to give the exact expression
\begin{multline}
    K_i(\dE) = 
    \frac{\transfield^2}{8} e^{-{J}_{ii}/|\detune|} 
    \Bigg\{
    \frac{\theta  (-\dE)\pi }{\Gamma(\alpha_Q)\cutoff}\exp\left[\frac{-|\dE|}{\cutoff}\right]\left(\frac{|\dE|}{\cutoff}\right)^{\alpha_Q-1}  \\
    + \sum_{n=1}^\infty \frac{({J}_{ii}/|\detune|)^n}{n!}\mathrm{Im}\left[\phi\big(\alpha_Q,\tfrac{1}{\cutoff}(\dE-n\detune-in\kappa)\big)\right]
    \Bigg\},
\end{multline}
where we have introduced the function $\phi(s,z)=z^{s-1}e^z\Gamma(1-s,z)$.  Here, $\Gamma(s,z)$ is the upper incomplete gamma function and $\theta(x)$ denotes the Heaviside step function. The function $\phi(s,z)$ asymptotically approaches $1/z$  for large $|z|$. Thus, we can explicitly calculate the rate to be 
\begin{multline}\label{eqn:qBathRates}
    K_i(\dE) = 
    \frac{\transfield^2}{8} e^{-{J}_{ii}/|\detune|}\\
   \left( h_Q(\dE)+ \sum_{n=1}^\infty \frac{({J}_{ii}/|\detune|)^n}{n!}\frac{n\kappa}{(\dE-n\detune)^2+n^2\kappa^2} \right)
\end{multline}
in the large detuning limit defined by $|\detune|\gg\dE,\cutoff$.  The bath-dependent term $h_Q(\dE)$ is given by
\begin{equation}
    h_Q(\dE) = \frac{\theta  (-\dE)\pi}{\Gamma(\alpha_Q)\cutoff}\exp\left[\frac{-|\dE|}{\cutoff}\right]\left(\frac{|\dE|}{\cutoff}\right)^{\alpha_Q-1}.
\end{equation}
The function $h_Q(\dE)$ dominates the rate function for $\dE<0$, but vanishes for $\dE>0$. The cavity parameters $\detune$ and $\kappa$ drop out of the bath term, showing that the dynamics are predominantly determined by the bath structure. 

\begin{figure}[t!]
    \centering
    \includegraphics[width=\columnwidth]{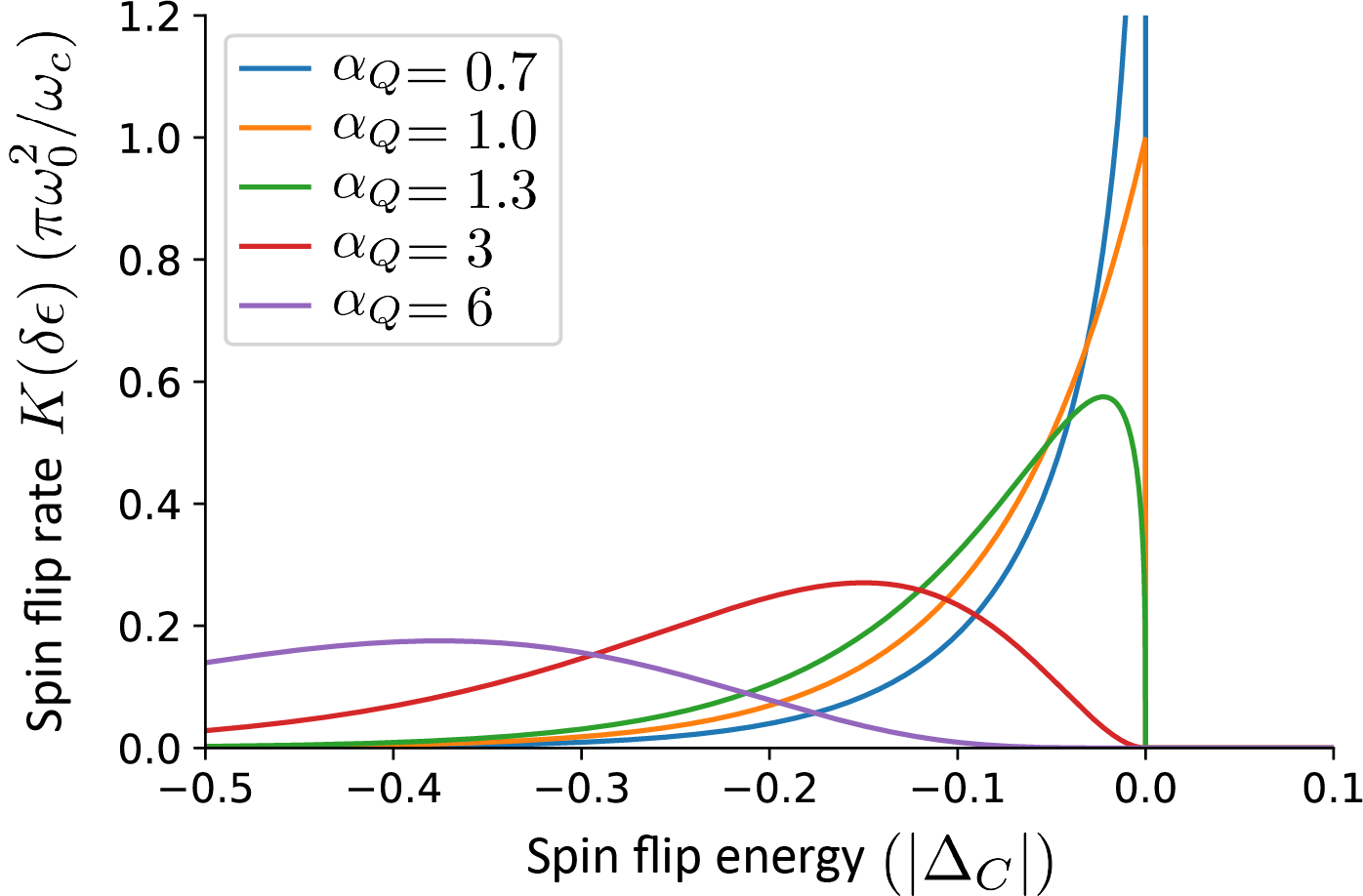} \caption{The spin-flip rate versus energy for an ohmic quantum bath for various bath coupling  strengths $\alpha_Q$.  Three regimes  of dynamics are apparent.  The spin-flip rate diverges for $\alpha_Q<1$ near $\dE=0$, giving rise to dynamics that perform the smallest energy-lowering spin flips possible.  For $\alpha_Q \simeq 1$, the spin-flip rate is roughly flat for negative $\dE$, matching 0TMH dynamics. Finally, for $\alpha_Q>1$, the spin-flip rate rises sharply as $\dE$ approaches zero. This describes dynamics akin to SD, in which the most energy-lowering spin flip has the highest spin-flip rate.}
    \label{fig:q-bath-rates}
\end{figure}

Figure~\ref{fig:q-bath-rates} plots $K_i(\dE)$ for several values of the bath coupling strength $\alpha_Q$. The expression has the property that for $\alpha_Q>1$ and negative $\dE$, $h_Q(\dE)$ approaches zero at $|\dE|=0$, attains a maximum at $|\dE|=(\alpha_Q-1)\cutoff$, and then falls off exponentially for larger $|\dE|$. This means that for energy lowering processes with $|\dE|<(\alpha_Q-1)\cutoff$, spin flips that lower the energy by a greater amount have a higher spin-flip rate; this is like SD.  For $\alpha_Q\approx1$, the spin-flip rate function is nearly flat while $\dE<0$ and yields 0TMH-like dynamics. Last, $h_Q(\dE)$ diverges at $\dE=0$ for $\alpha_Q<1$. This results in the spins preferring to perform trivial spin flips that lower the energy by only small amounts rather than making substantial energy lowering flips. Such dynamics would allow the spins to fluctuate as much as possible before settling to a local minimum in the long-time limit and is undesirable for associative memory.

 
\section{Derivation of mean-field dynamics}
\label{app:mean-field}

The deterministic ensemble dynamics of Sec.~\ref{sec:ensembles} are now derived from the stochastic spin dynamics of Sec.~\ref{sec:unravelling}.
This is a specific example of deriving mean-field equations from a continuous time Markov chain~\cite{Kolesnichenko2014}. As such, we first formalize the structure of the Markov chain. Each individual spin can either be in the up or down state, and so we define the occupancy vector $\mathbf{x}_{i,j}$ to denote the current state of the $i$th spin in ensemble $j$, where $\mathbf{x}_{i,j}=(1,0)$ for the up state and $\mathbf{x}_{i,j}=(0,1)$ for the down state. These define the microscopic degrees--of--freedom in the Markov chain.  We average over the microscopic degrees--of--freedom to produce a macroscopic description, $\mathbf{x}_j = \sum_{i=1}^M \mathbf{x}_{i,j}/M$.
The ensemble occupancy vectors can be written as $\mathbf{x}_j = (x_j^+,x_j^-)$, where $x_j^+$ $(x_j^-)$ is the fraction of spins in the $j$th ensemble that are up (down).  

To obtain the mean-field equations of motion, we construct the generator matrix describing the transition rates between different occupancy states. We must specify the spin-flip rate experienced \textit{per spin} rather than the total ensemble rate. We thus normalize by the number of spins that are available to flip in the given direction, $M x_i^\pm=2S x_i^\pm$, in terms of the ensemble populations. After normalization, the generator matrix is  
\begin{widetext}
\begingroup
\renewcommand*{\arraystretch}{2.2}
\begin{equation}
    Q_i = \frac{1}{2S}\begin{pmatrix}
    -\dfrac{S(S+1)-s^x_i(s^x_i-1)}{x^+_i}K_i(\dEi^-) & \dfrac{S(S+1)-s^x_i(s^x_i-1)}{x^+_i}K_i(\dEi^-) \\
    \dfrac{S(S+1)-s^x_i(s^x_i+1)}{x^-_i}K_i(\dEi^+) & -\dfrac{S(S+1)-s^x_i(s^x_i+1)}{x^-_i
    }K_i(\dEi^+)
  \end{pmatrix},
\end{equation}
where $S=M/2$ and the variable $s^x_i=S(x_i^+-x_i^-)$ is the average spin state of ensemble $i$.
The entries of the generator matrix are defined such that the off-diagonal elements give the transition rates for the up $\to$ down and down $\to$ up transitions, while the two diagonal elements give the total rate at which the spins exit the up or down state. 

The mean-field limit, exact for $S\to\infty$, describes the time evolution of the ensemble occupancy vectors in terms of a differential equation involving the generator matrix. Explicitly, under the only constraint that the rate functions be Lipschitz continuous, the limit differential equation for the ensemble occupancy vectors is
\begin{equation}
    \lim_{S\to\infty}\frac{d}{dt}\mathbf{x}_i = \mathbf{x}_i \cdot Q_i 
    = \frac{1}{2S}\begin{pmatrix}
    - K_i(\dEi^-)\big[S(S+1)-s^x_i(s^x_i-1)\big]+ K_i(\dEi^+)\big[S(S+1)-s^x_i(s^x_i+1)\big] \\
    K_i(\dEi^-)\big[S(S+1)-s^x_i(s^x_i-1)\big] - K_i(\dEi^+)\big[S(S+1)-s^x_i(s^x_i+1)\big]
  \end{pmatrix}^T.
\end{equation}
\endgroup
\end{widetext}
We may rewrite this equation as an ordinary differential equation in terms of the ensemble magnetizations $m_i=x_i^+-x_i^-=s^x_i/S$.  This is
\begin{multline}
\label{eqn:MFODEApp}
    \lim_{S\to\infty}\frac{d}{dt} m_i = S\left[K_i(\dEi^+)-K_i(\dEi^-)\right](1-m_i^2)\\  +(1-m_i)K_i(\dEi^+)
    -(1+m_i)K_i(\dEi^-).
\end{multline}
Noting that the sign of $K_i(\dEi^+)-K_i(\dEi^-)$ depends on whether the ensemble is aligned with its local field $\localfield_i=\sum_j J_{ij}s^x_j$, we arrive at the expression for the large ensemble equation of motion Eq.~\eqref{eqn:MFODE} presented in the main text. Referring back to Fig.~\ref{fig:ensemble_dynamics}, the mean-field equations match the full stochastic unraveling quite well for the chosen parameters.


%

\end{document}